\documentclass[12pt]{article}
\pdfoutput=1
\usepackage{bbm}
\usepackage{color}
\usepackage{authblk}
\usepackage{latexsym}
\usepackage{epsfig,amssymb,euscript, mathrsfs,cite}
\usepackage{amsmath}
\usepackage{mathrsfs} 
\usepackage{enumerate}
\definecolor{MyDarkBlue}{rgb}{0.15,0.15,0.45}
\usepackage[linktocpage=true]{hyperref}
\hypersetup{
colorlinks=true,
citecolor=MyDarkBlue,
linkcolor=MyDarkBlue,
urlcolor=MyDarkBlue,
pdfauthor={},
pdftitle={},
pdfsubject={hep-th}
}

\usepackage{mathtools}

\newsavebox{\ns}
\newsavebox{\dbrane}
\newsavebox{\dbshort}

\def\be{\begin{equation}}
\def\ee{\end{equation}}
\def\bea{\begin{eqnarray}}
\def\eea{\end{eqnarray}}

\newcommand{\nn}{\notag \\}

\def\cO{{\mathcal O}}
\def\eq#1 { \begin{equation} #1 \end{equation} }

\newcommand{\dd}{\mathrm{d}}

\newcommand{\tr}{\mathrm{tr}}

\newlength{\sswidth}

\numberwithin{equation}{section}       

\begin{document}

\begin{titlepage}

\vfill

\begin{flushright}
Imperial/TP/2021/JG/01\\
ICCUB-21-XXX
\end{flushright}

\vfill

\begin{center}
   \baselineskip=16pt
   {\Large\bf A new family of $AdS_4$ S-folds\\ in type IIB string theory}
  \vskip 1.5cm
Igal Arav$^1$, K. C. Matthew Cheung$^2$, Jerome P. Gauntlett$^2$\\
Matthew M. Roberts$^2$ and Christopher Rosen$^3$\\
     \vskip .6cm     
                          \begin{small}
                                \textit{$^1$Institute for Theoretical Physics, University of Amsterdam,\\
                                Science Park 904, PO Box 94485,\\
                                1090 GL Amsterdam, The Netherlands}
        \end{small}\\
        \begin{small}\vskip .3cm
      \textit{$^2$Blackett Laboratory, 
  Imperial College\\ London, SW7 2AZ, U.K.}
        \end{small}\\
             \begin{small}\vskip .3cm
      \textit{$^3$Departament de F\'isica Qu\'antica i Astrof\'isica and Institut de Ci\`encies del Cosmos (ICC), \\
      Universitat de Barcelona, Mart\'i Franqu\`es 1, ES-08028, \\Barcelona, Spain}
        \end{small}\\
                       \end{center}
\vfill

\begin{center}
\textbf{Abstract}
\end{center}
\begin{quote}
We construct infinite new classes of $AdS_4\times S^1\times S^5$ solutions of type IIB string theory
which have non-trivial $SL(2,\mathbb{Z})$ monodromy along the $S^1$ direction. The solutions are 
supersymmetric and holographically dual, generically, to $\mathcal{N}=1$ SCFTs in $d=3$. 
The solutions are first constructed as $AdS_4\times \mathbb{R}$ solutions in $D=5$ $SO(6)$ gauged supergravity and then uplifted to $D=10$.
Unlike the known $AdS_4\times \mathbb{R}$ S-fold solutions, there is no continuous symmetry associated with the $\mathbb{R}$ direction.
The solutions all arise as limiting cases of Janus solutions of $d=4$, $\mathcal{N}=4$ SYM theory which are supported both by a different value of the coupling constant on either side of the
interface, as well as by fermion and boson mass deformations.
As special cases, the construction recovers three known S-fold constructions, preserving $\mathcal{N}=1,2$ and 4 supersymmetry, as well as
a recently constructed $\mathcal{N}=1$ $AdS_4\times S^1\times S^5$ solution (not S-folded). We also present some novel
``one-sided Janus" solutions that are non-singular.

\end{quote}

\vfill

\end{titlepage}

\tableofcontents

\newpage

\section{Introduction}\label{sec:intro}

The landscape of non-geometric solutions of string/M-theory which are associated with the AdS/CFT correspondence is still largely unexplored territory.
By definition, such solutions are patched together using duality symmetries and hence they are not ordinary solutions of the
low-energy supergravity approximation. Nevertheless, in favourable situations one can still utilise supergravity constructions
to obtain valuable insights.

Within the context of type IIB string theory, which is the focus of this paper, we can consider S-folds i.e. non-geometric solutions 
that are patched together using the $SL(2,\mathbb{Z})$ symmetry. For AdS/CFT applications we 
are interested in solutions of type IIB supergravity of the form $AdS\times M$ with, in general, the axion-dilaton, the three-forms and
the self dual five-form all active on $M$. The S-fold construction implies that $M$ will have monodromies in $SL(2,\mathbb{Z})$, which act on
the axion-dilaton and the three-forms. If these monodromies involve contractible loops in $M$ then, in general, one is led to the presence of 
brane singularities and regions where the supergravity approximation breaks down. However, one can hope to make further progress if
the solutions lie within the context of F-theory as in the $AdS_3$ solutions discussed in \cite{Couzens:2017way,Couzens:2017nnr}, for example.

We can also consider $AdS\times M$ solutions of type IIB supergravity where the $SL(2,\mathbb{Z})$ monodromies do not involve contractible loops.
In this case, provided that the fields are all varying slowly on $M$, we can expect the type IIB supergravity approximation to be valid, and
that such solutions do indeed correspond to dual CFTs. 
Examples of such solutions were presented in \cite{Inverso:2016eet} and further discussed in \cite{Assel:2018vtq}: 
the spacetime is of the
form $AdS_4\times S^1\times S^5$ with non-trivial $SL(2,\mathbb{Z})$ monodromy just around the $S^1$ direction. The solutions
preserve the supersymmetry associated with $\mathcal{N}=4$ SCFTs in $d=3$, and we shall refer to them as $\mathcal{N}=4$ S-folds.
These solutions can be constructed as a certain limit of a class of
$\mathcal{N}=4$ Janus solutions \cite{DHoker:2006qeo} which describe $\mathcal{N}=4$, $d=3$ superconformal interfaces of 
$d=4$, $\mathcal{N}=4$ SYM theory. 
Using this perspective, and the results of \cite{Gaiotto:2008sa,Gaiotto:2008sd}, a specific conjecture for the SCFT dual to these $\mathcal{N}=4$ S-folds was given in \cite{Assel:2018vtq}.

$\mathcal{N}=1$ and $\mathcal{N}=2$ S-fold solutions of the form $AdS_4\times S^1\times S^5$ 
have also been constructed in \cite{Guarino:2019oct,Guarino:2020gfe,Bobev:2020fon}. In particular, it was
shown in \cite{Bobev:2020fon} how they can be obtained as limiting solutions of $\mathcal{N}=1$ \cite{Clark:2005te,DHoker:2006vfr,Suh:2011xc} 
and $\mathcal{N}=2$ \cite{DHoker:2006qeo} Janus solutions, also describing interfaces
of $d=4$, $\mathcal{N}=4$ SYM theory.
Furthermore, the $\mathcal{N}=1$ $AdS_4\times S^1\times S^5$ S-folds have been generalised to $\mathcal{N}=1$
$AdS_4\times S^1\times SE_5$ S-folds, where $SE_5$ is an arbitrary five-dimensional Sasaki-Einstein manifold \cite{Bobev:2019jbi}.

In the Janus solutions that are used to construct the S-folds just mentioned \cite{Assel:2018vtq,Bobev:2020fon}, 
the complex gauge coupling $\tau$ of 
$\mathcal{N}=4$ SYM theory takes different values on either side of the interface. It was recently pointed out that
this is not necessarily the case and it is possible to have interfaces in $\mathcal{N}=4$ SYM with the same value
of $\tau$ on either side of the interface which are supported by spatially dependent fermion and boson mass deformations, while preserving
$d=3$ conformal symmetry \cite{Arav:2020obl}. The associated supersymmetric Janus solutions of type IIB supergravity which
are holographically dual to such interfaces were also constructed in \cite{Arav:2020obl} by first constructing them in $D=5$ $SO(6)$ gauged supergravity. 
Furthermore, there is a particularly interesting $AdS_4\times\mathbb{R}$ solution that can be obtained
as a limit of this class of Janus solutions which is periodic in the $\mathbb{R}$ direction and uplifts to give a smooth\footnote{As far as we are aware this
is the first example of a supersymmetric $AdS_4\times M_6$ solution of type IIB supergravity, with compact $M_6$ that is smooth i.e. without sources.}
$AdS_4\times {S^1}\times S^5$ solution of type IIB supergravity (i.e. with no S-folding) \cite{Arav:2020obl}.

The constructions of \cite{Arav:2020obl} can be immediately generalised to give Janus solutions which have spatially dependent masses 
and varying $\tau$. It is therefore natural to ask 
if there are limiting classes of such Janus solutions which can be utilised to construct new S-fold solutions and/or periodic solutions.
While we have not found any more periodic solutions, we have found infinite new classes of $AdS_4\times\mathbb{R}$ solutions
of $D=5$ $SO(6)$ gauged supergravity that give rise to infinite
new classes of S-fold solutions of the form $AdS_4\times S^1\times S^5$, generically
preserving $\mathcal{N}=1$ supersymmetry in $d=3$. 

Our new construction will utilise various consistent sub-truncations of $D=5$ $SO(6)$ gauged supergravity all lying within the 10-scalar 
truncation of \cite{Bobev:2016nua} which, not surprisingly, just keeps 10 of the 42 scalars as well as the metric. 
One of these scalars, the $D=5$ dilaton $\varphi$, which for the vacuum $AdS_5$ solutions
is dual to the coupling constant of $\mathcal{N}=4$ SYM theory, plays a privileged role
as we expand upon below\footnote{We note that, in general, 
the type IIB dilaton of the uplifted solutions is not the same as the $D=5$ dilaton, as explained in appendix \ref{upliftform}.}.
Within this truncation we numerically construct families of $AdS_4\times\mathbb{R}$ solutions that
arise as certain limits of Janus solutions with $\mathcal{N}=4$ SYM on either side of the interface. We then uplift these to obtain $AdS_4\times\mathbb{R}\times S^5$
of type IIB supergravity, using the results of \cite{Lee:2014mla,Baguet:2015sma}. 
Additional $AdS_4\times\mathbb{R}\times S^5$ solutions in $D=10$ can then be generated
using $SL(2,\mathbb{R})$ transformations. Finally, within this larger family of solutions of type IIB supergravity one 
can find discrete examples where we can S-fold leading to supersymmetric
 $AdS_4\times {S^1}\times S^5$ S-fold solutions 
of type IIB string theory. 

The $D=5$ metric for the solutions we discuss in this paper are all of the form
\begin{align}\label{confgauge}
ds^2=e^{2A(r)}[ds^2(AdS_4)-dr^2]\,,
\end{align}
with all of the $D=5$ scalar fields just a function of the radial coordinate. The ansatz therefore preserves $d=3$ conformal invariance.
The $D=5$ solutions associated with the known $\mathcal{N}=1,2$ and $4$ S-folds 
are all direct products of the form $AdS_4\times\mathbb{R}$ with constant warp factor $A$ and with all of the
$D=5$ scalars constant, except for the $D=5$ dilaton, $\varphi$, which varies linearly in the radial coordinate. 

\begin{figure}[h!]
\centering
{
\includegraphics[scale=0.3]{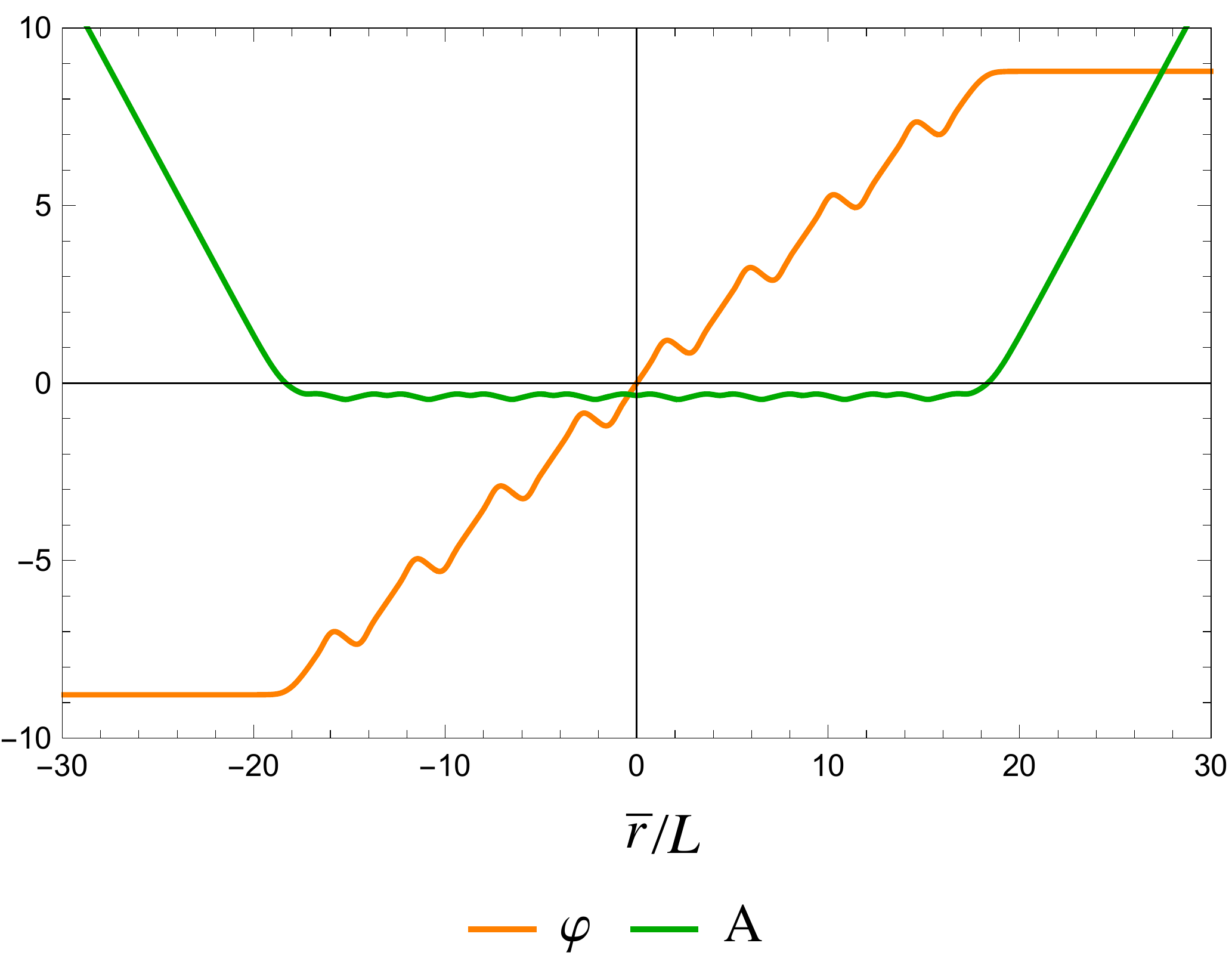}\qquad\qquad
\includegraphics[scale=0.3]{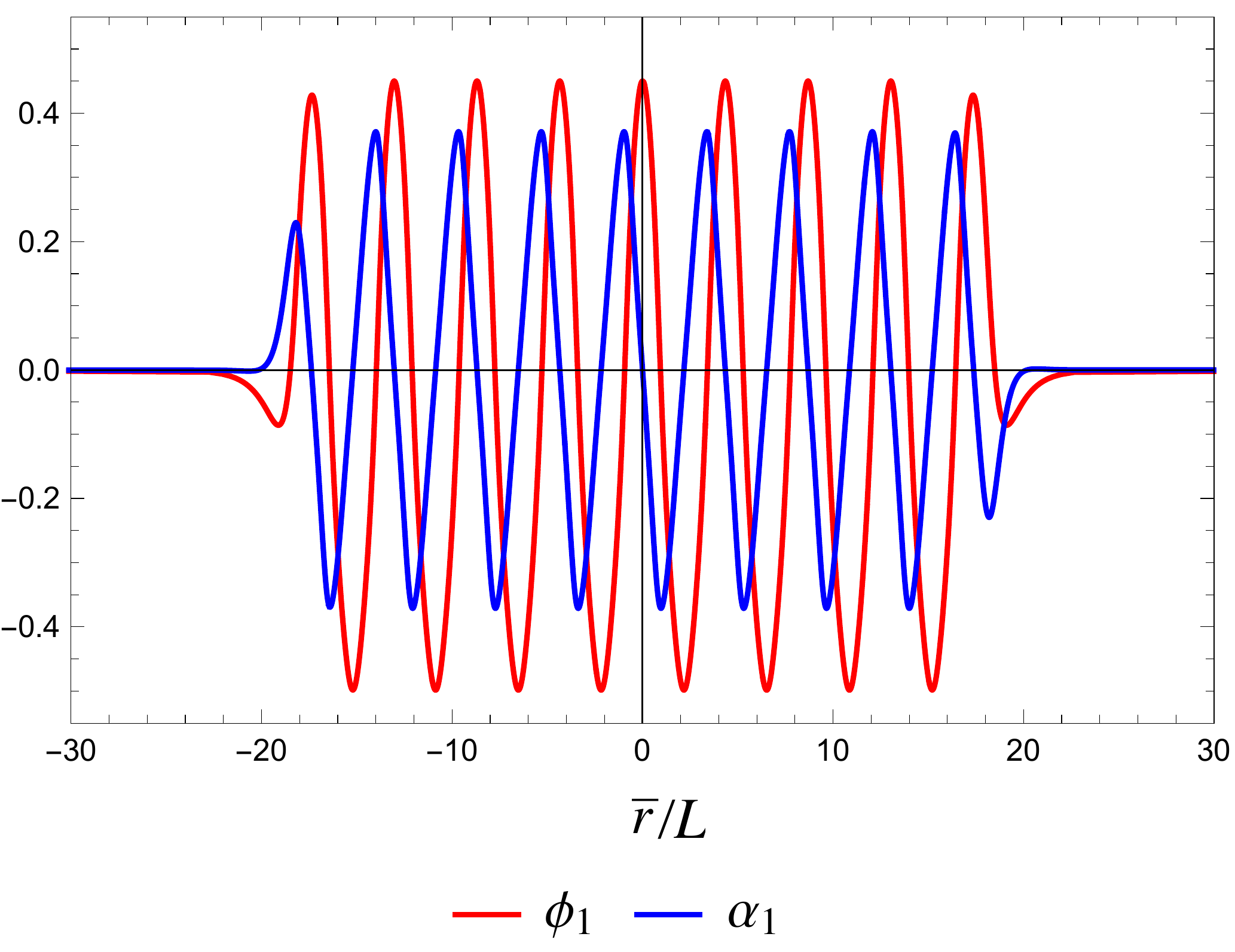}
}
\caption{A $D=5$ Janus solution that is approaching the new $AdS_4\times\mathbb{R}$ solutions for the $SO(3)$ invariant model. 
As $\bar r\to\pm \infty$, the solution is approaching $AdS_5$ on either side of the interface:
the warp factor is behaving as $A\to\pm \bar r/L$, the $D=5$ dilaton is approaching two different constants
$\varphi\to\varphi_\pm$, while the remaining scalar fields $\phi_1$, $\alpha_1$ and $\phi_4$ (not displayed) are going to zero. In the 
intermediate regime we see the build up of a periodic structure for the warp factor and the scalar fields, with $\varphi$ having, in addition, a dependence linear in $\bar r$ i.e. $\varphi$ is a ``linear plus periodic" (LPP) function. In the new limiting
$AdS_4\times\mathbb{R}$ solutions the intermediate structure extends all the way out to infinity.
Note that we have used the proper distance radial coordinate $\bar r$ given in \eqref{gaugechoices}.}\label{fig1}
\end{figure}

The new $AdS_4\times\mathbb{R}$ solutions involve several novel features. First, the metric on $AdS_4\times\mathbb{R}$ is no longer a direct product but a warped product,
since the warp factor now has non-trivial dependence on the radial direction. Secondly, and importantly, the warp factor $A(r)$ and all of the $D=5$ scalars are now periodic in the $\mathbb{R}$ direction, with the same period $\Delta r$, except for $\varphi$ which is now a
``linear plus periodic" (LPP) function of $r$.
Thus, unlike the known $AdS_4\times\mathbb{R}$ S-fold solutions,
the metric no longer admits a Killing vector associated with translations in the $\mathbb{R}$ direction and, furthermore, 
the solution is no longer invariant under the continuous symmetry consisting of translating along the $\mathbb{R}$ direction combined with a suitable dilaton shift. Thirdly, and as a consequence of the latter, we do not believe that the new solutions can be constructed in the maximally supersymmetric $D=4$ gauged supergravity theory which can be used to construct the known S-fold solutions \cite{Inverso:2016eet,Guarino:2019oct,Guarino:2020gfe}. This is simply because the $D=4$ theory is expected to arise after carrying out a Scherk-Schwarz dimensional reduction of maximal $D=5$ gauged supergravity on the $\mathbb{R}$ direction and this reduction requires such a continuous symmetry.
In figure \ref{fig1} we have illustrated how the new solutions arise as limiting cases
of Janus solutions of $\mathcal{N}=4$ SYM which, generically, have the $\mathcal{N}=4$ SYM coupling taking different
values on either side of the interface, as well as additional fermion and boson mass deformations.

The plan of the paper is as follows. We begin in section \ref{tenscalar} by discussing the 10-scalar truncation of maximal $D=5$ 
$SO(6)$ gauged supergravity given in \cite{Bobev:2016nua} as well as various sub-truncations. 
In section \ref{conssfolds} we discuss the general framework for constructing the new $AdS_4\times\mathbb{R}$ 
solutions in $D=5$ and the procedure for then obtaining $AdS_4\times {S^1}\times S^5$ S-folds solutions of type IIB string theory.

In sections \ref{so3model} and 
\ref{su2model} we discuss in more detail the constructions for two particular sub-truncations: an $SO(3)\subset SU(3)\subset SO(6)$ invariant model involving four scalar
fields and an $SU(2)\subset SU(3)\subset SO(6)$ invariant model involving five scalar fields.
The $SO(3)$ invariant model, called the $\mathcal{N}=1^*$ equal mass model in \cite{Arav:2020obl},  
includes the $AdS_4\times\mathbb{R}$ solutions associated with
the known $\mathcal{N}=1$ and $\mathcal{N}=4$ S-fold solutions as well as the periodic $AdS_4\times\mathbb{R}$ solution found in \cite{Arav:2020obl}. We note that figure
\ref{fig1} is associated with this model.
The $SU(2)$ invariant model includes the $AdS_4\times\mathbb{R}$ solutions associated with
the known $\mathcal{N}=2$ S-fold solutions and it also includes 
those associated with the known $\mathcal{N}=1$ S-fold solutions. In both truncations, our new family of S-fold solutions includes the previous known solutions. Furthermore,
in both cases one can identify the existence of some of our new family of solutions by a perturbative construction about the known $\mathcal{N}=1$ S-fold solution (but, interestingly, not around the $\mathcal{N}=2,4$ solutions).

In section \ref{onesided} we briefly discuss some novel ``one-sided Janus" solutions which
approach the $AdS_5$ vacuum on one side and either a known S-fold solution, an LPP dilaton
solution or the periodic $D=5$ solution of \cite{Arav:2020obl} on the other. Unlike other one-sided Janus solutions, they are non-singular.
In the case that it approaches the $\mathcal{N}=4$ $AdS_4\times\mathbb{R}$ S-fold solution
we are able to construct the solution analytically and we show how, after uplifting to type IIB supergravity, it fits into the general class of $AdS_4\times S^2\times S^2\times\Sigma$
solutions preserving $\mathcal{N}=4$ supersymmetry that were studied in 
\cite{DHoker:2007zhm,DHoker:2007hhe} (see also \cite{Aharony:2011yc,Assel:2011xz,Assel:2012cj} for some later developments).
We also discuss how the solution is related to solutions describing D3-branes ending on D5-branes.
In appendix \ref{upliftform} we have included some useful results concerning how to uplift solutions of the 10-scalar model in $D=5$ to type IIB supergravity.
In appendix \ref{appren}, prompted by the analysis in section \ref{onesided}, we refine the holographic renormalisation analysis 
for the 10-scalar truncation of \cite{Arav:2020obl} in a way that is consistent with the preservation of additional supersymmetry in the boundary theory.

\section{The 10-scalar model}\label{tenscalar}
We are interested in a truncation of $\mathcal{N}=8$, $SO(6)$ gauged supergravity in $D=5$, discussed in \cite{Bobev:2016nua},
that involves the metric and ten scalar fields which parametrise the coset
\begin{align}\label{tensctrunc}
\mathcal{M}_{10} =  SO(1,1)\times SO(1,1) \times \Big[\frac{SU(1,1)}{U(1)} \Big]^4 \,.
\end{align}
The $SO(1,1)\times SO(1,1)$ is parametrised by two scalars $\beta_1,\beta_2$ while the remaining eight scalars of this truncation, parametrising four copies of the Poincar\'e disc, can be packaged into
four complex scalar fields $z^A$ via
\begin{align}\label{zedintermsofscs}
     z^1 &= \tanh \Big[ \frac{1}{2} \big( \alpha_1 + \alpha_2 + \alpha_3 + \varphi 
       -i \phi_1 -i \phi_2 -i \phi_3 +i \phi_4 \big) \Big] \,,\nn
            z^2 &= \tanh \Big[ \frac{1}{2} \big( \alpha_1 - \alpha_2 + \alpha_3 - \varphi 
       -i \phi_1 +i \phi_2 -i \phi_3 -i\phi_4 \big) \Big] \,,\nn
     z^3 &= \tanh \Big[ \frac{1}{2} \big( \alpha_1 + \alpha_2 - \alpha_3 - \varphi 
       -i\phi_1 -i\phi_2 +i \phi_3 -i \phi_4 \big) \Big] \,,\nn
     z^4 &= \tanh \Big[ \frac{1}{2} \big( \alpha_1 - \alpha_2 - \alpha_3 + \varphi 
       -i \phi_1 +i \phi_2 +i\phi_3 +i\phi_4 \big) \Big] \,.
     \end{align}
Schematically, these 10 scalar fields are dual to the following Hermitian operators in $\mathcal{N}=4$ SYM theory:
\begin{align}\label{opfieldmapz}
\Delta=4:\qquad \qquad \varphi&\quad \leftrightarrow \quad \tr F_{\mu\nu} F^{\mu\nu},\,, \nn
\Delta=3:\qquad \qquad  \phi_i &\quad \leftrightarrow \quad \tr(\chi_i\chi_i+\text{cubic in $Z_i$})+h.c.\,, \qquad i=1, 2, 3 \,, \nn
   \phi_4&\quad \leftrightarrow \quad \tr(\lambda \lambda+\text{cubic in $Z_i$})+h.c.\,, \nn
\Delta=2:\qquad \qquad  \alpha_i &\quad \leftrightarrow \quad \tr(Z_i^2)+h.c.\,, \qquad\qquad\qquad\qquad i=1, 2, 3\,,\nn
\beta_1 &\quad \leftrightarrow \quad \tr(|Z_1|^2+|Z_2|^2 -2 |Z_3|^2 )\,,\nn
   \beta_2 &\quad \leftrightarrow \quad \tr(|Z_1|^2-|Z_2|^2)\,.
   \end{align}
The operators of $d=4$, $\mathcal{N}=4$ SYM appearing on the right hand side of \eqref{opfieldmapz} have been written in an $\mathcal{N}=1$ language, with
$Z_i$ and $\chi_i$ the bosonic and fermionic components of the associated three chiral superfields $\Phi_i$ while
$\lambda$ is the gaugino of the vector multiplet. Thus, the $D=5$ dilaton $\varphi$ is dual to the coupling constant
of $\mathcal{N}=4$ SYM theory, while $\phi_i,\phi_4$ are fermionic mass terms and $\alpha_i$, $\beta_1$,
$\beta_2$ are bosonic mass terms.

The action is given by
\begin{align}\label{bulkaction}
S_{Bulk}=\frac{1}{4\pi G_{(5)}}\int d^5 x\sqrt{|g|}\Big[-\frac{1}{4}R +3(\partial\beta_1)^2+(\partial\beta_2)^2+ \frac{1}{2}\mathcal{K}_{A\bar{B}}\partial_{\mu}z^{A}\partial^{\mu}\bar{z}^{\bar{B}} - \mathcal{P}\Big]\,,
\end{align}
and we work with a $(+----)$ signature convention. 
Here $\mathcal{K}$ is the K\"ahler potential given by
\begin{align}\label{kpot}
\mathcal{K}=-\sum_{A=1}^4\log(1-z^A\bar z^A)\,.
\end{align}
The scalar potential ${\cal P}$ can be conveniently derived from a superpotential-like quantity
\begin{align}\label{superpotlike}
\mathcal{W} \equiv ~&\frac{1}{L}e^{2\beta_1+2\beta_2}\left(1+z^1z^2+z^1z^3+z^1z^4+z^2z^3+z^2z^4+z^3z^4+z^1z^2z^3z^4\right)\nn
+ &\frac{1}{L}e^{2\beta_1-2\beta_2}\left(1-z^1z^2+z^1z^3-z^1z^4-z^2z^3+z^2z^4-z^3z^4+z^1z^2z^3z^4\right)\nn
 +&\frac{1}{L}e^{-4\beta_1}\left(1+z^1z^2-z^1z^3-z^1z^4-z^2z^3-z^2z^4+z^3z^4+z^1z^2z^3z^4\right)\,,
\end{align}
via
\begin{align}\label{peeform}
   \mathcal{P} = \frac{1}{8}e^{\mathcal{K}}\left[\frac{1}{6}\partial_{\beta_1}\mathcal{W}\partial_{\beta_1}\overline{\mathcal{W}}
     +\frac{1}{2}\partial_{\beta_2}\mathcal{W}\partial_{\beta_2}\overline{\mathcal{W}}+\mathcal{K}^{\bar{B} A}    
      \nabla_{A}\mathcal{W}\nabla_{\bar{B}}\overline{\mathcal{W}} -\frac{8}{3}\mathcal{W}\overline{\mathcal{W}}\right]\,,
 \end{align}
where $\mathcal{K}^{\bar{B}A}$ is the inverse of ${\cal K}_{A \bar B}$ and 
$ \nabla_{A}\mathcal{W}\equiv \partial_A\mathcal{W}+\partial_A \mathcal{K}\mathcal{W}$.

The model is invariant under $\mathbb{Z}_2\times S_4$ discrete symmetries which, importantly, leave $\mathcal{W}$ invariant. First, it is invariant under
the $\mathbb{Z}_2$ symmetry
\begin{align}\label{genz2}
z^A\to -z^A\,,\quad\Leftrightarrow\quad \{\phi_i,\phi_4,\alpha_i,\varphi\}\to- \{\phi_i,\phi_4,\alpha_i,\varphi\}\,.
\end{align}
Second, it is invariant under an $S_3$ permutation symmetry which acts on $(-z^2,-z^3,z^4)$ as well as $\beta_1,\beta_2$
and is generated by two elements:
\begin{align}\label{gens31}
\{z^3 &\leftrightarrow -z^4
\quad\Leftrightarrow\quad \phi_1 \leftrightarrow \phi_3\,,\alpha_1 \leftrightarrow \alpha_3\}\,,\quad
\beta_1 \rightarrow -\frac{1}{2}(\beta_1 + \beta_2)\,,\quad \beta_2 \rightarrow \frac{1}{2}(\beta_2 - 3 \beta_1)\,,\nn
\{z^2 &\leftrightarrow -z^4  \quad\Leftrightarrow\quad \phi_1 \leftrightarrow \phi_2\,,\alpha_1 \leftrightarrow \alpha_2\}\,,\qquad
\beta_2 \rightarrow - \beta_2\,.
\end{align}
There is also an invariance under the interchange of pairs of the $z^A$:
\begin{align}\label{gens33}
z^1 &\leftrightarrow z^4,\quad  -z^2 \leftrightarrow -z^3\,,
\, &\Leftrightarrow\quad  (\phi_2,\phi_3)\to -(\phi_2,\phi_3)\,,\quad(\alpha_2,\alpha_3)\to -(\alpha_2,\alpha_3)\,,\nn
z^1 &\leftrightarrow - z^2,\quad  -z^3 \leftrightarrow z^4\,,
\, &\Leftrightarrow\quad  (\phi_1,\phi_3)\to -(\phi_1,\phi_3)\,,\quad(\alpha_1,\alpha_3)\to -(\alpha_1,\alpha_3)\,,\nn
z^1 &\leftrightarrow -z^3,\quad  -z^2 \leftrightarrow z^4\,,
\, &\Leftrightarrow\quad  (\phi_1,\phi_2)\to -(\phi_1,\phi_2)\,,\quad(\alpha_1,\alpha_2)\to -(\alpha_1,\alpha_2)\,.
\end{align}
Together \eqref{genz2}-\eqref{gens33} generate 
$\mathbb{Z}_2\times S_4$ as observed in \cite{Bobev:2020ttg}. We also note that 
\eqref{gens31}, \eqref{gens33} are discrete subgroups of the $SO(6)$ R-symmetry while 
\eqref{genz2} is part of the $SL(2,\mathbb{R})$ symmetry of $D=5$ gauged supergravity.

The model is also invariant under shifts of the dilaton
\begin{align}\label{dilshift}
\varphi\to\varphi+c\,.
\end{align}
For later use, we note that this shift symmetry is generated by the following holomorphic Killing vector
\begin{align}
l =\frac{1}{2} \sum_{A=1}^4 {(-1)^{s(A)}} \left( 1- (z^A)^2 \right) \frac{\partial}{\partial z^A}\,,
\end{align}
where $s(A)=0$ for $A=1,4$ and $s(A)=1$ for $A=2,3$. Furthermore, if we define
\begin{equation}
\widetilde{\mathcal{K}} \equiv \mathcal{K} + \log \mathcal{W} + \log \overline{\mathcal{W}} \,,
\end{equation}
we have 
\begin{align}\label{ksymcon}
l^A \partial_A \widetilde{\mathcal{K}} + l^{\bar{A}} \partial_{\bar{A}} \widetilde{\mathcal{K}} = 0\,,
\end{align}
and the corresponding moment map 
$\mu=\mu(z^A,\bar z^A)$, satisfying
\begin{equation}\label{mommapcon}
\mu = i l^A \partial_A \widetilde{\mathcal{K}} = \mathcal{K}^{A\bar{B}} \partial_{\bar{B}} \mu \,  \partial_{\bar{A}}\widetilde{\mathcal{K}} \,,
\end{equation}
is given by 
\begin{equation}\label{mommap1}
\mu = -\frac{i}{2} \sum_{A=1}^4 (-1)^{s(A)} \frac{z^A - \bar{z}^A}{1- z^A \bar{z}^A} \,.
\end{equation}
In terms of the fields given in \eqref{zedintermsofscs} we find that the moment map only depends on $\phi_i$, $\phi_4$ and takes the form
\begin{align}\label{mommap2}
\mu = \frac{1}{2} & \left[   \tan( -\phi_1 - \phi_2 - \phi_3 + \phi_4 )  
 - \tan( -\phi_1 + \phi_2 - \phi_3 - \phi_4 ) \right. 
 \nn
 & \left.
 - \tan( -\phi_1 - \phi_2 + \phi_3 - \phi_4 ) 
 + \tan( -\phi_1 + \phi_2 + \phi_3 + \phi_4 ) 
 \right] \,.
\end{align}
Expanding about $\phi_i=0$ we have to lowest order $\mu\sim 2\phi_4$.

The 10-scalar truncation is not a supergravity theory. However, the conditions for a solution of the 10-scalar
model to preserve a preferred supersymmetry as a solution of $D=5$ $SO(6)$ gauged supergravity were written down in 
\cite{Bobev:2016nua} and also used in \cite{Arav:2020obl}. 
These preferred supersymmetry transformations are left invariant under the $\mathbb{Z}_2\times S_4$ discrete symmetries \eqref{genz2}-\eqref{gens33}. The equations of motion of the 10-scalar model are also invariant under additional discrete symmetries, given in appendix \ref{appren}, which transform the supercharges of the maximal $D=5$ gauged supergravity theory into each other and do not preserve the preferred supersymmetries that we focus on in this paper.
Here we use exactly the same conventions as \cite{Arav:2020obl}.

\begin{figure}[h!]
\centering
{\includegraphics[scale=0.65]{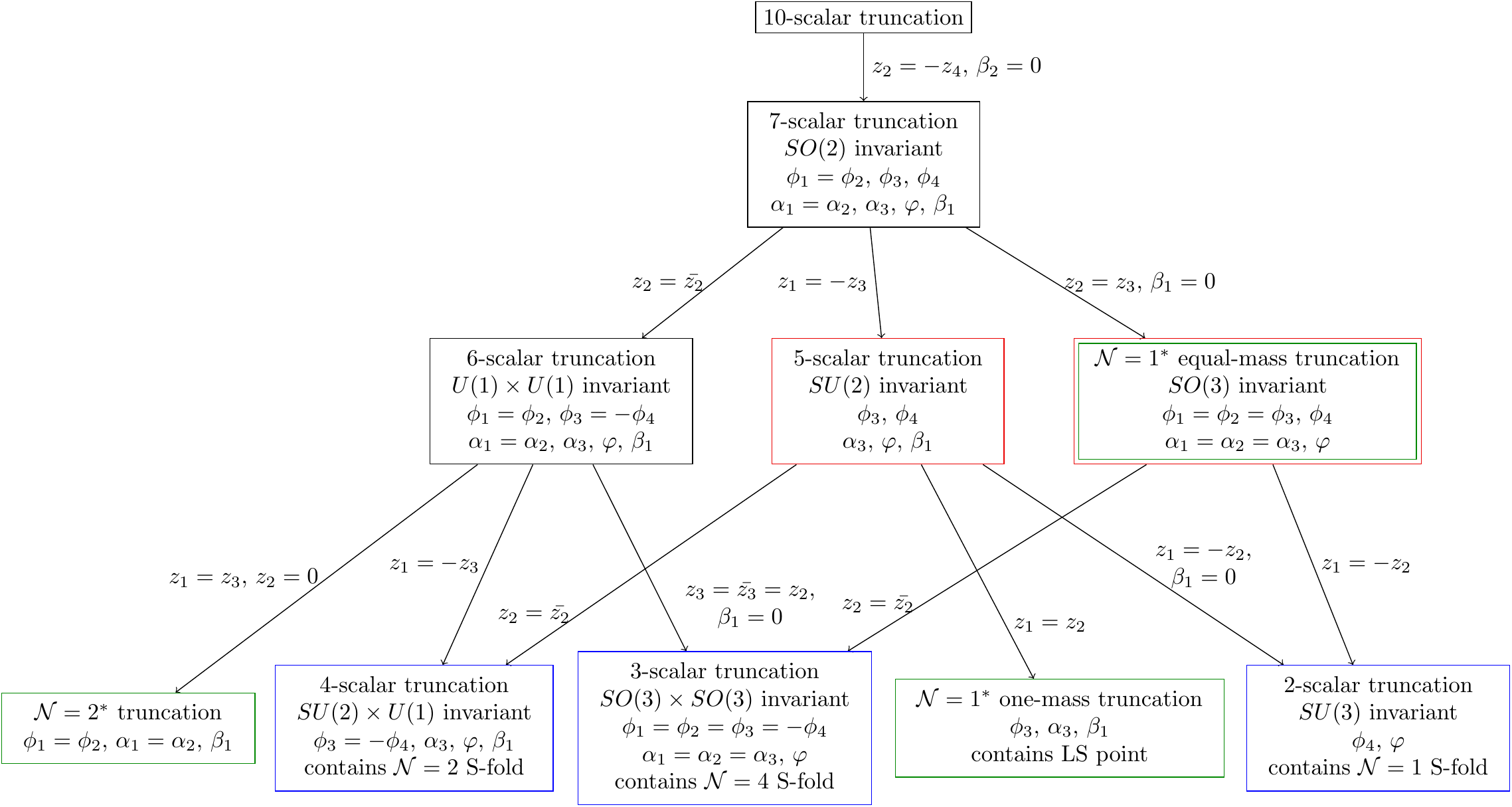}}
\caption{Various sub-truncations of the ten scalar model. 
In this paper we focus on the $\mathcal{N}=1^*$ equal mass, $SO(3)$ invariant
truncation and the 5-scalar $SU(2)$ invariant truncation, marked by red boxes, as well as their associated sub-truncations in the bottom line. 
The boxes with the blue outline are truncations that contain known $AdS_4\times\mathbb{R}$ S-fold solutions discussed in
\cite{Bobev:2020fon}. 
The boxes with the green outline are truncations
which were used in \cite{Arav:2020obl}.}\label{truncdiag}
\end{figure}

There are a number of different consistent sub-truncations of the 10-scalar model which were also discussed in \cite{Bobev:2016nua}, that we
summarise in figure \ref{truncdiag}. The figure also displays where one can find
the three known $AdS_4\times\mathbb{R}$ solutions with a linear $D=5$ dilaton $\varphi$ which are
associated with S-folds preserving $\mathcal{N}=1,2$ and $4$ supersymmetry, as well as the symmetry subgroup of $SO(6)$ that is preserved by the truncation.
These sub-truncations preserve various subsets of the $\mathbb{Z}_2\times S_4$ discrete symmetries given in \eqref{genz2}-\eqref{gens33}.
All of the sub-truncations preserve the $\mathbb{Z}_2$ symmetry \eqref{genz2} as well as shifts of the dilaton \eqref{dilshift} when the dilaton $\varphi$ is present in the truncation.
In this paper we will be mostly interested in two cases: the $\mathcal{N}=1^*$ equal mass, $SO(3)$ invariant model, with $SO(3)\subset SU(3)\subset SU(4)$ 
and involving four scalar fields; and
the 5-scalar $SU(2)$ invariant model, with $SU(2)\subset SU(3)\subset SU(4)$.
While the $SO(3)$ invariant model does not preserve any additional symmetries, the $SU(2)$ model
preserves a further $\mathbb{Z}_2$ that is contained in \eqref{gens33}.


\section{Constructing S-folds}\label{conssfolds}
The construction of the S-fold solutions starts with solutions of $D=5$ supergravity. These are then uplifted to type IIB, where additional
solutions are generated using the $SL(2,\mathbb{R})$ symmetry of type IIB supergravity. Finally, the S-folding procedure, using the $SL(2,\mathbb{Z})$ symmetry of type IIB string theory, is made.

\subsection{Ansatz in $D=5$}
We consider solutions of $D=5$ supergravity of the form
\begin{align}\label{metjanus}
ds^2=e^{2A} ds^2(AdS_4)-N^2dr^2\,,
\end{align}
where $ds^2(AdS_4)$ is the metric on $AdS_4$, which we take to have unit radius,
and $A$, $N$ as well as the scalar fields $\beta_1,\beta_2,z^A$ are
functions of $r$ only. Clearly this ansatz preserves $d=3$ conformal invariance. There is still some freedom in choosing the radial coordinate.
In this paper we will either use the ``conformal gauge" with $N=e^{{A}}$, as in \eqref{confgauge}, 
or the ``proper distance gauge" with $N=1$
\begin{align}\label{gaugechoices}
\text{conformal gauge:}\qquad &N=e^{{A}}\,,\qquad \text{radial coordinate: $r$} \,,\nn
\text{proper distance gauge:}\qquad &N=1\,,\qquad \,\,\,\text{radial coordinate: $\bar r$} \,,
\end{align}
with $d\bar r=e^{A} dr$.

We are interested in supersymmetric configurations which, generically, are associated with $\mathcal{N}=1$ supersymmetry in  $d=3$
(i.e. two Poincar\'e plus two superconformal supercharges). As shown in \cite{Arav:2020obl}, 
we obtain such solutions provided that we satisfy
the following\footnote{With essentially no loss of generality, the parameter $\kappa=\pm 1$ appearing in \cite{Arav:2020obl}, 
which fixes the projections on the Killing spinors, has been set to $\kappa=+1$.} BPS equations (in the conformal gauge), 
\begin{align}
\label{eq:RewrittenJanusBPSAeq}
\partial_r {A} - {i}&=2B_r, \nn
\partial_r B_r &= 2 \mathcal{F} B_r \bar{B}_r\,,
\end{align}
where $\mathcal{F}$ is a real quantity just depending on $\mathcal{W}$, $\mathcal{K}$ given by
\begin{equation}\label{ceffdef}
\mathcal{F} \equiv 1-\frac{3}{2}\frac{1}{|\mathcal{W}|^2}\nabla_A\mathcal{W}\mathcal{K}^{A\bar B}\nabla_{\bar B}\bar{\mathcal{W}}
-\frac{1}{4}|\partial_{\beta_1}\log \mathcal{W}|^2
-\frac{3}{4}|\partial_{\beta_2}\log \mathcal{W}|^2\,,
\end{equation}
as well as
\begin{align}
\label{eq:RewrittenJanusBPSzeq}
\partial_r z^A &= -3 \mathcal{K}^{A\bar{B}} \frac{\nabla_{\bar{B}}\overline{\mathcal{W}}}{\overline{\mathcal{W}}} \bar{B}_r \,, \nn
\partial_r \beta_1 &= - \frac{1}{2}  \partial_{\beta_1}\log\overline{\mathcal{W}} \bar{B}_r \, ,\nn
\partial_r \beta_2 &= - \frac{3}{2} \partial_{\beta_2}\log\overline{\mathcal{W}} \bar{B}_r \, .
\end{align}

In these equations the quantity $B_r$ is defined as $B_r\equiv \frac{1}{6}e^{i\xi+A+\mathcal{K}/2}{\mathcal{W}}$
where $\xi(r)$ is a phase that appears in the Killing spinors. It is helpful to point out that the BPS equations are left invariant under
the transformation\footnote{In general, the transformation $z^A\to \bar z^A$ by itself, which can be obtained by combining (B.2) with (2.8)-(2.10), is a symmetry of the equations of motion for the 10-scalar model but also acts on the preferred supersymmetries.}
\begin{align}\label{z2twosymgm1}
r\to-r,\qquad z^A\to \bar z^A,\qquad\xi\to -\xi+\pi\,.
\end{align}
The BPS equations are also invariant under the discrete $\mathbb{Z}_2\times S_4$ symmetries in
\eqref{genz2}-\eqref{gens33} and this will also be the case for any of the sub-truncations in figure \ref{truncdiag} for which they are still present.
Additional general aspects of the space of solutions to these BPS equations were discussed in section 5 of \cite{Arav:2020obl}.

It will also be useful to notice that the dilaton shift symmetry \eqref{dilshift} of the 10-scalar model
gives rise to a conserved quantity for the BPS equations.
Specifically, using \eqref{mommapcon} one can check that an integral of motion for the BPS equations is given by 
\begin{equation}\label{eq:IntegOfMotion}
\mathcal{E} \equiv \frac{1}{L^3}e^{3 A} \mu(z,\bar{z}) \,,
\end{equation}
where the moment map was given in \eqref{mommap1} or \eqref{mommap2}.
This result can be derived via the Noether procedure as follows.  The Killing vector $l^A$
generating the symmetry \eqref{dilshift}, gives rise to a conserved current for the full equations of motion. For our radial ansatz
we deduce that the radial component of this current, given by
\begin{equation}
\mathcal{E}  \propto \sqrt{g} g^{rr} \left( \mathcal{K}_{A\bar{B}} \partial_r \bar{z}^{\bar{B}} l^A + \mathcal{K}_{B\bar{A}} \partial_r z^{B}  l^{\bar{A}} \right)\,,
\end{equation}  
is a conserved quantity, independent of $r$.
Using the BPS equations we then obtain
\begin{align}
\mathcal{E} &\propto e^{3A} \left( \partial_A \widetilde{\mathcal{K}} B_r l^A + \partial_{\bar{A}} \widetilde{\mathcal{K}} \bar{B}_r l^{\bar{A}} \right) \,,\nn
&= e^{3A} \left[ ( l^A \partial_A \widetilde{\mathcal{K}} + l^{\bar{A}} \partial_{\bar{A}} \widetilde{\mathcal{K}}  )\operatorname{Re}(B_r) 
- \frac{i}{2} \left( l^A \partial_A \widetilde{\mathcal{K}} - l^{\bar{A}} \partial_{\bar{A}} \widetilde{\mathcal{K}} \right) \right]\,,\nn
&= -{} e^{3 A} \left( i l^A \partial_A \widetilde{\mathcal{K}} \right)
= -{} e^{3 A} \mu\,.
\end{align}
where to get to the second line we wrote 
$B_r = \operatorname{Re}(B_r) - \frac{i }{2}$, and to get to the third line
we used \eqref{ksymcon} and \eqref{mommapcon}.

\subsection{Janus solutions}
We now briefly summarise some aspects of the Janus solutions constructed in \cite{Arav:2020obl}. We first recall
that the $AdS_5$ vacuum solution, dual to $d=4$, $\mathcal{N}=4$ SYM, has a warp factor given by
\begin{align}\label{ads5vac}
e^{A}={L}\cosh\frac{\bar r}{L}\,,
\end{align}
with all of the scalars vanishing, $z^A=0$.

Janus solutions, describing superconformal interfaces of $d=4$, $\mathcal{N}=4$ SYM, can be obtained by solving the BPS equations 
and imposing boundary conditions so that they approach the $AdS_5$ vacuum solution \eqref{ads5vac} at $\bar r=\pm\infty$, with suitable falloffs for the scalar fields, associated with supersymmetric sources for the dual operators.
A detailed analysis of holographic renormalisation for such Janus solutions was carried out in \cite{Arav:2020obl}
(using the proper distance gauge). 
The focus in \cite{Arav:2020obl} was to construct Janus solutions that are dual to interfaces of $\mathcal{N}=4$ SYM that are supported
by fermion and boson masses that have a non-trivial spatial dependence on the direction transverse to the interface. These solutions were constructed
within the following truncations, shown in green boxes in figure \ref{truncdiag}: the $\mathcal{N}=2^*$ truncation (three scalar fields), 
the $\mathcal{N}=1^*$ one-mass truncation (three scalar fields) and the $\mathcal{N}=1^*$ equal-mass, $SO(3)$ invariant truncation (four scalar fields).

Within the Janus solutions of the $\mathcal{N}=1^*$ equal-mass, $SO(3)$ invariant truncation (green and red box in figure \ref{truncdiag})
a special limiting $AdS_4\times\mathbb{R}$ solution was found with
the warp factor $A$ and all of the scalar fields periodic in the $\mathbb{R}$ direction. As such, this solution can be compactified on the $\mathbb{R}$ direction and after uplifting to type IIB, one obtains 
a regular $AdS_4\times S^1\times S^5$ solution (without S-folding). In the sequel we will present
new $AdS_4\times\mathbb{R}$ solutions which are no longer periodic in the $\mathbb{R}$ direction that
can also be found as limiting classes of Janus solutions. In the new solutions the $D=5$ dilaton, $\varphi$, is a LPP function
while the remaining scalars and warp factor are periodic in the $\mathbb{R}$ direction;
an illustration is given in figure \ref{fig1}.
All of our new S-fold solutions arise as limits of $D=5$ Janus solutions with $\varphi_{(s)}$, which parametrises the source for the
operator dual to $\varphi$, taking different values on either side of the interface. In other words the Janus solutions are 
interfaces of $d=4$, $\mathcal{N}=4$ SYM with the coupling constant taking different values on either side of the interface.

It will also be helpful to recall that for the $\mathcal{N}=1^*$ one-mass truncation, in addition to the $AdS_5$ vacuum solution dual to 
$d=4$, $\mathcal{N}=4$ SYM, there are also two other $AdS_5$ solutions, LS$^\pm$, which are both dual to the Leigh-Strassler
$\mathcal{N}=1$ SCFT. In \cite{Arav:2020obl,Arav:2020asu} interesting limiting solutions of the Janus solutions associated with
interfaces involving the LS SCFT were found. In particular we found solutions dual to an RG interface with $\mathcal{N}=4$ SYM on one side of the interface
and the LS theory on the other, as well as Janus solutions with the LS theory on either side of the interface. In this paper we also construct
solutions within the 5-scalar $SU(2)$ truncation in figure \ref{truncdiag} (red box), which contain the LS$^\pm$ fixed points. 
In addition to the new LPP solutions we also find limiting Janus solutions that involve Janus interfaces for the LS$^\pm$ fixed points themselves i.e. solutions with
LS$^\pm$ on either side of the interface with a linear $D=5$ dilaton.

Finally, as somewhat of an aside, we note that the conserved quantity $\mathcal{E}$ given in \eqref{eq:IntegOfMotion} implies a constraint amongst the sources and expectation values of operators of $\mathcal{N}=4$ SYM theory for the Janus configurations. Following the 
holographic renormalisation carried out in \cite{Arav:2020obl}, which used the proper distance gauge,
the expansion at, say, the ${\bar r}\to\infty$ end of the interface is given by
\begin{align}
\phi_i&=\phi_{i,(s)}e^{-{\bar r}/L}+\cdots+{\phi}_{i,(v)}e^{-3{\bar r}/L}+\cdots\,,\quad
\alpha_i=\alpha_{i,(s)}\frac{{\bar r}}{L}e^{-2{\bar r}/L}+{\alpha}_{i,(v)}e^{-2{\bar r}/L}+\cdots\,,\nn
\beta_i&=\beta_{i,(s)}\frac{{\bar r}}{L}e^{-2{\bar r}/L}+{\beta}_{i,(v)}e^{-2{\bar r}/L}+\cdots\,,\qquad
\varphi=\varphi_{(s)}+\cdots+{\varphi}_{(v)}e^{-4{\bar r}/L}+\cdots\,,\nn
A&=\frac{{\bar r}}{L}+\cdots+{A_{(v)}}e^{-4{\bar r}/L}+\cdots\,.
\end{align}
Here $\phi_{i,(s)}, \alpha_{i,(s)},...$ give the source terms of the dual operators, while ${\phi}_{i,(v)}, {\alpha}_{i,(v)},...$ can be used
to obtain the expectation values, explicitly given in \cite{Arav:2020obl}.
Using this expansion as well the conditions on sources and expectation values imposed by the BPS conditions, we find that
the integral of motion is given by
\begin{align}
\mathcal{E}=\frac{1}{L^3}(2\phi_{4,(v)}-4\phi_{1,(s)}\phi_{2,(s)}\phi_{3,(s)})\,.
\end{align}

\subsection{$AdS_4\times\mathbb{R}$ solutions and S-folds }\label{sfoldproc}
Our principal interest in this paper concerns a new class of solutions to the BPS equations 
of the form (in conformal gauge):
\begin{align}\label{sfoldansatz}
ds^2&=e^{2{A}}[ds^2(AdS_4)-dr^2]\,,\nn
\varphi&=k r+ f(r)\,,
\end{align}
where $k$ is a constant and $A,f$ and all other scalars satisfy 
\begin{align}
A(r)=A(r+\Delta r)\,,\quad f(r)=f(r+\Delta r)\,, \quad z^A(r)=z^A(r+\Delta r)\,.
\end{align}

Notice that, in general, the $D=5$ dilaton $\varphi$ is an LPP function, while the warp factor and the remaining scalar fields
are all periodic functions of $r$, with period $\Delta r$. Over one period $\varphi$ changes by an amount $\Delta\varphi$ given by
\begin{align}\label{varphidef}
\Delta\varphi\equiv \varphi(r+\Delta r)-\varphi(r)=k\Delta r\,.
\end{align}
Although we have defined $\Delta\varphi$ in the conformal gauge, importantly (and unlike $k,\Delta r$) it is invariant under coordinate changes\footnote{After integrating we can write $\rho=c r +H(r)$ with $H(r+\Delta r)=H(r)$ and $H$ having no zero mode. Inverting this, we can write $r=(1/c)\rho+\tilde H(\rho)$ with $\tilde H(\rho+\Delta\rho)=\tilde H(\rho)$, where $\Delta\rho=c\Delta r$. In this gauge we can then write $\varphi=(k/c)\rho+\tilde f(\rho)$ with
$\tilde f(\rho+\Delta\rho)=\tilde f(\rho)$ and $\Delta\varphi=k\Delta r$.}
of the form $r\to \rho$ with $d\rho=G(r) dr$ where $G(r)$ is a periodic function, $G(r+\Delta r)=G(r)$. 
We can also define the proper distance of a period $\Delta {\bar r}$, which is given by
\begin{align}
\Delta {\bar r}=\int_0^{\Delta r} e^{A} dr\,.
\end{align}

For the special case when $k=0$, when $\varphi$ is also periodic, these solutions are periodic in the $r$ direction and we can then immediately compactify the
radial direction to obtain an $AdS_4\times S^1$ solution. In this case, if we identify after just one period, $\Delta {\bar r}$ is the 
length of the $S^1$.
 We presented one such solution in 
\cite{Arav:2020obl} and this will appear in our new constructions. For this purely periodic
solution the period of the warp factor is half of that of the scalar fields.
Another special case is when $k\ne 0$ and $f=0$, so that $\varphi$ is purely linear in $r$, 
as well as ${A}$ and all
other scalar fields being constant. These $AdS_4\times\mathbb{R}$ solutions are associated with the known $AdS_4$ S-fold solutions:
we can periodically identify the radial direction after uplifting to type IIB supergravity and making a suitable identification with an $SL(2,\mathbb{Z})$ transformation, as we outline in more generality below.

We now continue with the more general class of LPP solutions of the form \eqref{sfoldansatz} with both $k\ne 0$ and $f\ne 0$ and show that these too can give rise to new classes of $AdS_4$ S-fold solutions. 
We begin by noting, as explained in appendix \ref{upliftform} (see also \cite{Bobev:2020fon}), that the dilaton-shift 
symmetry \eqref{dilshift} of the $D=5$ theory, $\varphi\to \varphi+c$, acts as a specific $SL(2,\mathbb{R})$
transformation in $D=10$. If the type IIB dilaton, $\Phi$ and axion $C_0$ are parametrised
as
\begin{equation}
m_{\alpha\beta} = 
\begin{pmatrix}
e^{\Phi}C_0\,^2+e^{-\Phi} & -e^{\Phi}C_0\\
 -e^{\Phi}C_0 & e^{\Phi}
\end{pmatrix},
\end{equation}
then the transformation is given by $m\to ({\mathcal S}^{-1})^Tm{\mathcal S}^{-1}$ where ${\mathcal S} \in SL(2,\mathbb{R})$, in the hyperbolic conjugacy class, is given by
\begin{align}\label{lambdaexp}
{\mathcal S}(c)=\begin{pmatrix}
e^{c} & 0\\
 0& e^{-c}
\end{pmatrix},
\end{align}
Equivalently, we have 
$\Phi \to \Phi+2c$ and $C_0 \to e^{-2c}C_0$. 

To carry out the S-fold procedure, we next note that starting from the uplifted $D=5$ solutions we can obtain a family of
uplifted type IIB solutions after acting with a general element $P\in SL(2,\mathbb{R})$. For example, the axion and dilaton in this larger
family will be of the form $\tilde m(\varphi)=(P^{-1})^Tm(\varphi) P^{-1}$, where we have included the dependence on the $D=5$ dilaton for emphasis.
Within this larger family of type IIB solutions we then look for solutions
that we can periodically identify along the radial direction with period $q\Delta r$ i.e. $q \in\mathbb{N}$ times the fundamental period $\Delta r$, up to the action of an 
${\cal M}\in SL(2,\mathbb{Z})$ transformation. Recalling that as we translate by $\Delta r$ in the radial direction in the conformal gauge \eqref{sfoldansatz} we have
$\varphi\to \varphi+\Delta\varphi$, and hence we require that
\begin{align}\label{mpcond}
\tilde m(\varphi+q\Delta\varphi)=({\cal M}^{-1})^T\tilde m(\varphi){\cal M}^{-1}\,,
\end{align}
which can be achieved provided that $P\in SL(2,\mathbb{R})$ is such that 
\begin{align}\label{emmcond}
{\cal M}=\pm P{\mathcal S}(q\Delta\varphi) P^{-1}\,.
\end{align}

The different S-folded solutions which can be obtained in this way are labelled by the conjugacy classes of ${\cal M}$ in $SL(2,\mathbb{Z})$.
A discussion of such classes can be found in\cite{DeWolfe:1998eu,DeWolfe:1998pr} (see also \cite{Dabholkar:2002sy}). 
For any conjugacy class ${\cal M}$, we have that $-{\cal M}$ and $\pm{\cal M}^{-1}$ also represent conjugacy classes. Clearly from the form
of ${\mathcal S}$ in \eqref{lambdaexp} we must be in the hyperbolic conjugacy class with $|Tr({\cal M})|>2$. 
We have the following possibilities for ${\cal M}$ (as well as the conjugacy classes  $-{\cal M}$ and $\pm{\cal M}^{-1}$): we can have
\begin{align}\label{genm}
{\cal M}=\begin{pmatrix}
n & 1 \\
-1 & 0
\end{pmatrix}\,,\qquad n\ge 3\,,
\end{align}
with trace $n$, as well as ``sporadic cases" ${\cal M}(t)$ of trace $t$. For example for $3\le t\le 12$ the complete list is given 
by\footnote{Note that writing $\mathcal{M}_n$ for the matrix in \eqref{genm}, we can also write
${\cal M}(8)^{-1}=-\mathcal{M}_{2}\mathcal{M}_{-3}$,
${\cal M}(10)^{-1}=-\mathcal{M}_{4}\cdot \mathcal{M}_{-2}$ and
${\cal M}(12)^{-1}=-\mathcal{M}_{2}\cdot \mathcal{M}_{-5}$.}
\begin{align}
{\cal M}(8)=\left(\begin{matrix}
1 & 2 \\
3 & 7
\end{matrix}\right)\,,\qquad
{\cal M}(10)=\left(\begin{matrix}
1 & 4 \\
2 & 9
\end{matrix}\right)\,,\qquad
{\cal M}(12)=\left(\begin{matrix}
1 & 2 \\
5 & 11
\end{matrix}\right)\,.
\end{align}
For these cases, in order to find solutions to \eqref{mpcond} (focussing on the upper sign in \eqref{emmcond}) we must have 
\begin{align}\label{keycond}
q\Delta\varphi = \mathrm{arccosh}\frac{n}{2}\,, \qquad \mathrm{for} \qquad n\ge 3, \quad q\ge 1\,.
\end{align}
For example, for the S-folds that are identified using ${\cal M}$ in $SL(2,\mathbb{Z})$ given in \eqref{genm} we have
\begin{align}
P=
\begin{pmatrix}
1 & -\frac{1}{\sqrt{n^2-4}} \\
\frac{1}{2}(-n+\sqrt{n^2-4}) & \frac{1}{2}(1+\frac{n}{\sqrt{n^2-4}})
\end{pmatrix}\,.
\end{align}

Interestingly, the S-folding procedure preserves the supersymmetry as we now explain.
If we translate the $D=5$ solution by $\Delta r$ then we have $\varphi\to\varphi+\Delta\varphi$.
Such a shift in the dilaton is equivalently obtained by carrying out a K\"ahler transformation
$\mathcal{K}\to \mathcal{K}+f+\bar f$ and $\mathcal{W}\to e^{-f}\mathcal{W}$ with $f=f(z^A)$. Under this transformation
the preserved supersymmetries, a symplectic Majorana pair, transform as 
$\varepsilon_1\to e^{(f-\bar f)/4}\varepsilon_1$ and $\varepsilon_2\to e^{-(f-\bar f)/4}\varepsilon_2$ as noted in \cite{Arav:2020obl}.
Now, as we explained above, this transformation is implemented on the bosonic fields as an element of 
${\mathcal S} \in SL(2,\mathbb{R})$. In appendix \ref{upliftform} we show that this is also true for the preserved supersymmetries.
Thus, as we translate by $\Delta r$, the solution and the preserved supersymmetries get transformed by the
same element of $SL(2,\mathbb{R})$. This will also be true after uplifting to $D=10$ and hence, after conjugating by $P\in SL(2,\mathbb{R})$,
the S-fold procedure will not break any supersymmetry.


\subsection{Free energy of the S-folds}
The $AdS_4\times S^1\times S^5$ S-fold solutions of the kind we have just described should be dual, in general, to $\mathcal{N}=1$ SCFTs in $d=3$.
A key observable is $ \mathcal{F}_{S^3}$, the free energy of the SCFT on $S^3$. This can be calculated holographically after a dimensional reduction
on $S^1\times S^5$ to a four-dimensional theory of gravity and then evaluating the
regularised on-shell action for the $AdS_4$ vacuum solution of this theory. With a four-dimensional theory that has an 
$AdS_4$ vacuum solution with unit radius we have
 \begin{align}
 \mathcal{F}_{S^3}=\frac{\pi}{2G_{(4)}}\,.
 \end{align}
Here $G_{(4)}$ is the four-dimensional Newton's constant which can be obtained from the five-dimensional Newton's constant
via 
\begin{align}
\frac{1}{G_{(4)}}=  \frac{1}{G_{(5)}}\int_0^{q\Delta r} dr e^{3A}\,.
\end{align}
Here we remind the reader that the radial coordinate, $r$, is associated with the $D=5$ conformal gauge, as in \eqref{sfoldansatz}, and also that in 
the construction of the S-fold solution we made the S-fold identification after going along $q$ periods of the periodic functions.
Recalling that the $AdS_5$ vacuum with radius $L$ solves the equations of motion and is dual to $d=4$, $\mathcal{N}=4$SYM with gauge group $SU(N)$, we have the standard result
          \begin{align} \label{stdresult}
\frac{1}{16 \pi G_{(5)}}=\frac{N^2}{8 \pi^2 L^3}\,.
\end{align}
Putting this together we get our final formula for the free energy:
 \begin{align}\label{fenexp}
 \mathcal{F}_{S^3}&=\frac{N^2}{L^3}q\int_0^{\Delta r}  dr e^{3A}\,,\nn
 &=\frac{N^2}{L^3}  \frac{\mathrm{arccosh}\frac{n}{2}}{\Delta\varphi}\int_0^{\Delta r}  dr e^{3A}\,.
 \end{align}
The first expression is valid for all solutions, including the periodic solution (for which it is natural to take $q=1$), while
the second expression is valid for the S-folded solutions. In the special case of the known $\mathcal{N}=1,2,4$ S-folds which have
a purely linear $D=5$ dilaton (i.e. $\varphi=kr$ in \eqref{sfoldansatz})
and $A$ is constant, we can rewrite this as
 \begin{align}\label{freeensfold}
 \mathcal{F}_{S^3}
 &=\frac{N^2}{L^3}\frac{e^{3A}}{k}\mathrm{arccosh}\frac{n}{2}\,.
 \end{align}

Finally, following the arguments in \cite{Assel:2018vtq}, at fixed $n$ the type IIB supergravity approximation should be valid in the large $N$ limit 
since higher derivative corrections will be suppressed by terms of order $1/\sqrt{N}$.

\section{$\mathcal{N}=1^*$ equal mass, $SO(3)$ invariant model}\label{so3model}
This model is obtained from the 10-scalar model by setting
$z^2=z^3=-z^4$, or equivalently 
$\alpha_1=\alpha_2=\alpha_3$ and $\phi_1=\phi_2=\phi_3$, as well as
$\beta_1=\beta_2=0$. This four-scalar model is parametrised by the two complex fields
\begin{align}
z^1=\tanh\big[\frac{1}{2}\big(3\alpha_1+\varphi-3i\phi_1+i\phi_4\big)\big],\quad 
z^2=\tanh\big[\frac{1}{2}\big(\alpha_1-\varphi-i\phi_1-i\phi_4\big)\big]\,.
\end{align}
The integral of motion \eqref{eq:IntegOfMotion}
for this truncation is given by
\begin{align}\label{ethismodel}
\mathcal{E}=\frac{1}{L^3}e^{3A}\frac{1}{2}[-\tan(3\phi_1-\phi_4)+3\tan(\phi_1+\phi_4)]\,.
\end{align}

This model has two further sub-truncations as illustrated in figure \ref{truncdiag}, and in particular contains
the known $\mathcal{N}=1$ and $\mathcal{N}=4$ $AdS_4\times\mathbb{R}$ S-fold solutions. Firstly, if 
 we set $z^1=-z^2$, equivalently, $\alpha_1=\phi_1=0$, then we obtain a two-scalar $SU(3)$ invariant model depending on $\varphi,\phi_4$ that
 overlaps\footnote{They consider a model with four
scalars: $(\varphi,\chi,c,\omega)$. One should set $c=\omega=0$ and then identify
$\sin \phi_4=\tanh\chi$ as well as $g=2/L$.} 
with the truncation considered
in the context of $\mathcal{N}=1$ S-folds in section 4 of \cite{Bobev:2020fon}. 

The $\mathcal{N}=1$ $AdS_4\times\mathbb{R}$ S-fold solution is given (in conformal gauge) by
\begin{align}\label{nonesfold}
\varphi&=\frac{\sqrt{5}}{2}r,\quad
\phi_4=\cos^{-1}\sqrt{\frac{5}{6}}\,,\quad
e^{A}=\frac{5L}{6}\,,\qquad
\alpha_1=\phi_1=0\,,
\end{align}
and we have $\mathcal{E}=\frac{25\sqrt{5}}{108}$. There is another $\mathcal{N}=1$ S-fold solution
obtained from the symmetry \eqref{genz2}, with opposite sign for $\mathcal{E}$.
The free energy of these solutions can be obtained from \eqref{freeensfold}
and is given by
\begin{align}
\mathcal{F}_{S^3}=\frac{25\sqrt{5}}{108}\mathrm{arccosh}\frac{n}{2}{N^2}\,.
\end{align}
in agreement with \cite{Bobev:2020fon}.

On the other hand if we further set $z^2=\bar z^2$, or equivalently $\phi_1=-\phi_4$,
then we obtain a three-scalar $SO(3)\times SO(3)$ invariant model depending on $\alpha_1,\phi_1,\varphi$ that overlaps\footnote{They consider a model with five
scalars: $(\varphi,\chi,\alpha,c,\omega)$. One should set $c=\omega=0$ and then identify $\alpha_1=\alpha$ and
$\sin 4\phi_1=-\tanh 4\chi$. We also note that setting 
$z^2=\bar z^2$ in the BPS equations \eqref{eq:RewrittenJanusBPSzeq} leads to an additional 
algebraic reality constraint. The compatibility of imposing this constraint with the BPS equations 
can be verified
as in section 5 of \cite{Arav:2020obl} for a similar issue associated with the reality of the scalar fields $\beta_1,\beta_2$.}
with the truncation considered
in the context of $\mathcal{N}=4$ S-folds in section 2 of \cite{Bobev:2020fon}. 
The $\mathcal{N}=4$ S-fold solution is given (in conformal gauge) by
\begin{align}\label{nfoursfold}
\varphi&=\frac{1}{\sqrt{2}}r\,,\quad 
\phi_1=-\phi_4=-\frac{1}{2}\cot^{-1}\sqrt{2}\,,\quad
e^{A}=\frac{L}{\sqrt{2}}\,,\qquad
\alpha_1=0\,,
\end{align}
and has $\mathcal{E}=\frac{1}{2}$.
Again there is another $\mathcal{N}=4$ S-fold solution
obtained from the symmetry \eqref{genz2}, with opposite sign for $\mathcal{E}$.
From  \eqref{freeensfold} the free energy of these solutions is given by
\begin{align}
\mathcal{F}_{S^3}=\frac{1}{2}\mathrm{arccosh}\frac{n}{2}{N^2}\,.
\end{align}
in agreement with \cite{Assel:2018vtq,Bobev:2020fon}.

The model also contains a single periodic $AdS_4\times\mathbb{R}$ solution that was found numerically in \cite{Arav:2020obl} which has $\mathcal{E}=0$. 
In this solution the warp factor $e^A$ and all the scalar fields, including $\varphi$, are periodic in the radial direction.
Thus, it can immediately be compactified to give an $AdS_5\times S^1$ solution of $D=5$ supergravity and then uplifted to
an  $AdS_4\times S^1\times S^5$ solution of type IIB using the results of appendix \ref{upliftform}. From the numerical results we can calculate the 
free energy \eqref{fenexp} and we find 
\begin{align}
\mathcal{F}_{S^3}\approx q \times 1.90107{N^2}\,,
\end{align}
where $q$ is the number of periods over which we have compactified.

The periodic solution was found as a limiting case of a class of Janus solutions in \cite{Arav:2020obl}. The focus in
\cite{Arav:2020obl} was Janus solutions that approach the $\mathcal{N}=4$ SYM vacuum with
the same value of $\varphi_{(s)}$ on either side of the interface, corresponding to the same value of $\tau$ of $\mathcal{N}=4$ SYM
on either side of the interface. It is straightforward to generalise these Janus solutions to allow $\varphi_{(s)}$ to take different values on
either side of the interface. As already noted, taking limits of these solutions leads to new families of $AdS_4\times\mathbb{R}$ solutions
with $\varphi$ an LPP function of the radial coordinate, $r$, which parametrises $\mathbb{R}$. Before summarising these new solutions, all found numerically, we discuss how some of the new
family of solutions can also be seen by perturbing the $AdS_4\times\mathbb{R}$ solution associated with the
$\mathcal{N}=1$ S-fold solution.

\subsection{Periodic perturbation about the $\mathcal{N}=1$ S-fold}
Within the $\mathcal{N}=1^*$ equal mass model, we consider linearised perturbations of the BPS equations
about the $AdS_4\times\mathbb{R}$ solution \eqref{nonesfold}, associated with the $\mathcal{N}=1$ S-fold. 
There are zero modes associated with shifts of $\varphi$, ${A}$
and there is also a freedom to shift the coordinate $r$. There are two linearised modes that depend exponentially on $r$.
Of most interest is that there is also a linearised periodic mode of the form 
\begin{align}
\delta\alpha_1&= \sin\frac{\sqrt{5} r}{3}\,,\qquad
\delta\phi_1=-\sqrt{5} \cos\frac{\sqrt{5} r}{3}\,.
\end{align}

With a little effort we can use this periodic mode to construct a perturbative expansion in a parameter $\epsilon$, that takes the form\
\begin{align}\label{pertsolone}
\alpha_1&= \sum_{m,p=1}^\infty
a_{m,p}^{(\alpha_1)}\epsilon^m \sin{}pKr\,,\qquad\qquad\qquad
\phi_1= \phi_1^{zm}(\epsilon)+\sum_{m,p=1}^\infty
a_{m,p}^{(\phi_1)}\epsilon^m \cos{}pKr\,,\nn
\phi_4&=\phi_4^{zm}(\epsilon)+\sum_{m,p=1}^\infty
a_{m,p}^{(\phi_4)}\epsilon^m \cos{}pKr\,,\qquad
\varphi= k(\epsilon)r +\sum_{m,p=1}^\infty
a_{m,p}^{(\varphi)}\epsilon^m \sin{}pKr
\,,\nn
{A}&= A^{zm}(\epsilon)+\sum_{m,p=1}^\infty
a_{m,p}^{({A})}\epsilon^m \cos{}pKr\,,
\end{align}
where all functions are periodic in the radial direction with period $\Delta r\equiv \frac{2\pi}{K}$, with $\varphi$ 
having an extra linear piece, and hence an LPP function, exactly as in \eqref{sfoldansatz}-\eqref{varphidef}.
The wavenumber $K$ is itself given by the following series in $\epsilon$:
\begin{align}
K\equiv \frac{2\pi}{\Delta r}&=\frac{\sqrt{5}}{3}-\frac{184 \sqrt{5}}{13}\epsilon ^2-\frac{2155938 \sqrt{5}}{2197}\epsilon ^4-\frac{1193970682204}{1856465 \sqrt{5}}\epsilon ^6+\cdots\,,
\end{align}
which we notice is decreasing as we move away from the $\mathcal{N}=1$ S-fold solution.
Interestingly, we notice that $\alpha_1$ has vanishing zero mode in this expansion, while the zero modes of the remaining periodic functions are explicitly given by
\begin{align}
\phi_1^{zm}&=-5 \sqrt{5} \epsilon ^2-\frac{9431 \sqrt{5}}{26}\epsilon ^4-\frac{6269904259}{26364 \sqrt{5}}\epsilon ^6+\cdots\,,\nn
\phi_4^{zm}&=\cos^{-1}\sqrt{\frac{5}{6}}-\sqrt{5} \epsilon ^2-\frac{61645 \sqrt{5}}{676} \epsilon ^4-\frac{110249429617}{1713660 \sqrt{5}} \epsilon ^6+\cdots\,,\nn
A^{zm}&= \log \frac{5 L}{6}-3 \epsilon ^2-\frac{102177}{338}  \epsilon ^4-\frac{60279560187}{1428050}\epsilon ^6+\cdots\,.
\end{align}
In addition the slope of $\varphi$ takes the form
\begin{align}
k=\frac{\sqrt{5} }{2}-\frac{9\sqrt{5} }{2} \epsilon ^2-\frac{513855 \sqrt{5}}{1352} \epsilon ^4-\frac{295876107351 }{1142440 \sqrt{5}}\epsilon ^6+\cdots\,.
\end{align}
Furthermore we also have $\Delta\varphi\equiv k\Delta r$ is given by
\begin{align}\label{pertepsilon}
\Delta\varphi=3 \pi+\frac{1305 \pi}{13}  \epsilon ^2+\frac{95032143 \pi}{8788} \epsilon ^4+\frac{11893037855571 \pi}{7425860}  \epsilon ^6+\cdots\,.
\end{align}
The integral of motion \eqref{ethismodel} is given by
\begin{align}
\mathcal{E} =\frac{25 \sqrt{5}}{108}\big(1-6 \epsilon ^2-\frac{14598}{169}\epsilon ^4-\frac{1590041883}{142805}\epsilon ^6+\cdots\big)\,.
\end{align}

One finds that all of the expansion parameters $a^{(*)}_{m,p}$ appearing in \eqref{pertsolone} are only non-zero when $m+p$ is even.
This implies the following property of the perturbative solution under a half period shift in the radial coordinate. 
Specifically, let $\Psi=\{A,\alpha_1,\phi_1, \phi_4\}$ denote the periodic functions
so that the whole solution is specified by $\Psi(\epsilon,r)$ and $\varphi(\epsilon,r)$. 
We then find
\begin{align}\label{revepsign}
\Psi(\epsilon,r+\pi/K)=\Psi(-\epsilon,r),\qquad
\varphi(\epsilon,r+\pi/K)=\varphi(-\epsilon,r) + \text{constant}\,,
\end{align}
where the constant can be removed by \eqref{dilshift}. 
This means that changing the sign of $\epsilon$ gives, essentially, the same solution
(i.e. up to a shift in the radial direction plus a shift of $\varphi$).

Finally, after uplifting to type IIB, using the results of appendix \ref {upliftform}, and carrying out the S-fold procedure, as described in section \ref{sfoldproc}, we obtain new S-folds of type IIB provided that we can solve
\eqref{keycond}.
The free energy for the S-folded solutions can then be obtained from \eqref{fenexp} and is given by
 \begin{align}\label{freeone}
 \mathcal{F}_{S^3}
 &=\frac{25\sqrt{5}}{108}\left(1-\frac{1305}{13}\epsilon^4 -\frac{26414316}{13^3}\epsilon^6+\dots\right)\mathrm{arccosh}\frac{n}{2}N^2\,.
 \end{align}
To solve \eqref{keycond} we first note that $2\cosh 3\pi\sim 12391.6$. Thus, the smallest value of $n$ that can be reached in
\eqref{keycond} is $n=12392$, which occurs for $q=1$ and $\epsilon\sim 0.0003$. 
There are additional branches of solutions, labelled by $q$,
which, for a given $n$, have smaller values of $\epsilon$. Thus, we can find S-fold solutions
with arbitrarily small $\epsilon$. We also note that while these $AdS_4\times\mathbb{R}$ solutions are perturbatively
connected with the $\mathcal{N}=1$ $AdS_4\times\mathbb{R}$ S-fold solution, they are not as S-folds of type IIB string theory. This is clear when we recall that for the latter
we can solve \eqref{keycond} for any $n\ge 3$ by suitably adjusting the period $\Delta r$ over which we S-fold, while for the perturbative 
solutions, as just noted, we have $n\ge 12392$.

The $\mathcal{N}=1^*$ equal mass, $SO(3)$ invariant truncation we are considering also contains the known 
$\mathcal{N}=4$ $AdS_4\times\mathbb{R}$ S-fold solution \eqref{nfoursfold}. If we consider the linearised perturbations of the BPS equations
about this solution we again find zero modes associated with shifts of $\varphi$, ${A}$
and there is also a freedom to shift the coordinate $r$. The remaining modes all depend exponentially  on the radial coordinate.
In particular, there is no longer a linearised periodic mode and this feature will manifest itself in the family of new solutions we
discuss in the next section.

\subsection{New S-fold solutions}
The new $AdS_4\times \mathbb{R}$ solutions, with $\varphi$ a LPP function,  can be constructed as limiting cases of Janus solutions.
A convenient way to numerically solve the BPS equations \eqref{eq:RewrittenJanusBPSAeq}-\eqref{eq:RewrittenJanusBPSzeq} is to set 
initial conditions for the scalar fields at a turning point of the metric warp function,
$A$, which corresponds to $\text{Re}(B_r)=0$ along with the values of the scalar fields at the turning points.
Some general comments concerning this procedure were made in sections 5 and 6 of \cite{Arav:2020obl}. 

In more detail we consider Janus solutions with the turning point of $A$ located at $r=r_{tp}$. Since 
the BPS equations are unchanged by shifting the radial coordinate by a constant, we can take $r_{tp}=0$.
We can also use the shift symmetry \eqref{dilshift} to
choose $\varphi(r_{tp})=0$. We can then 
focus\footnote{If we relax the condition that the initial data is invariant under the $\mathbb{Z}_2$ symmetry, then we do not find any LPP solutions of the type we are interested in for constructing S-folds. Instead we find some interesting ``one-sided" Janus solutions that 
we discuss in section \ref{onesided}.
We also note that the general periodic perturbative solution
\eqref{pertsolone} did not assume invariance under the $\mathbb{Z}_2$ symmetry, yet it is in fact invariant.}
on solutions that are invariant under the $\mathbb{Z}_2$ symmetry, obtained by
combining \eqref{genz2} and \eqref{z2twosymgm1}, 
\begin{align}\label{z2twosymgm12}
r\to-r,\qquad z^A\to -\bar z^A,\qquad\xi\to -\xi+\pi\,.
\end{align}
This implies that $\phi_i,\phi_4$ are even functions of $r$ and $\alpha_i,\varphi$ are odd functions. In particular, at the turning point
we can take $\alpha_i(r_{tp})=0$ as part of our initial value data. For the $SO(3)$ invariant model, these Janus solutions are therefore fixed
by the values of $\phi_1(r_{tp})$ and $\phi_4(r_{tp})$. 
By suitably tuning the values of the scalar field at the turning points we are able to construct the limiting cases of solutions associated with the S-folds.

\begin{figure}[h!]
\centering
{
\includegraphics[scale=0.75]{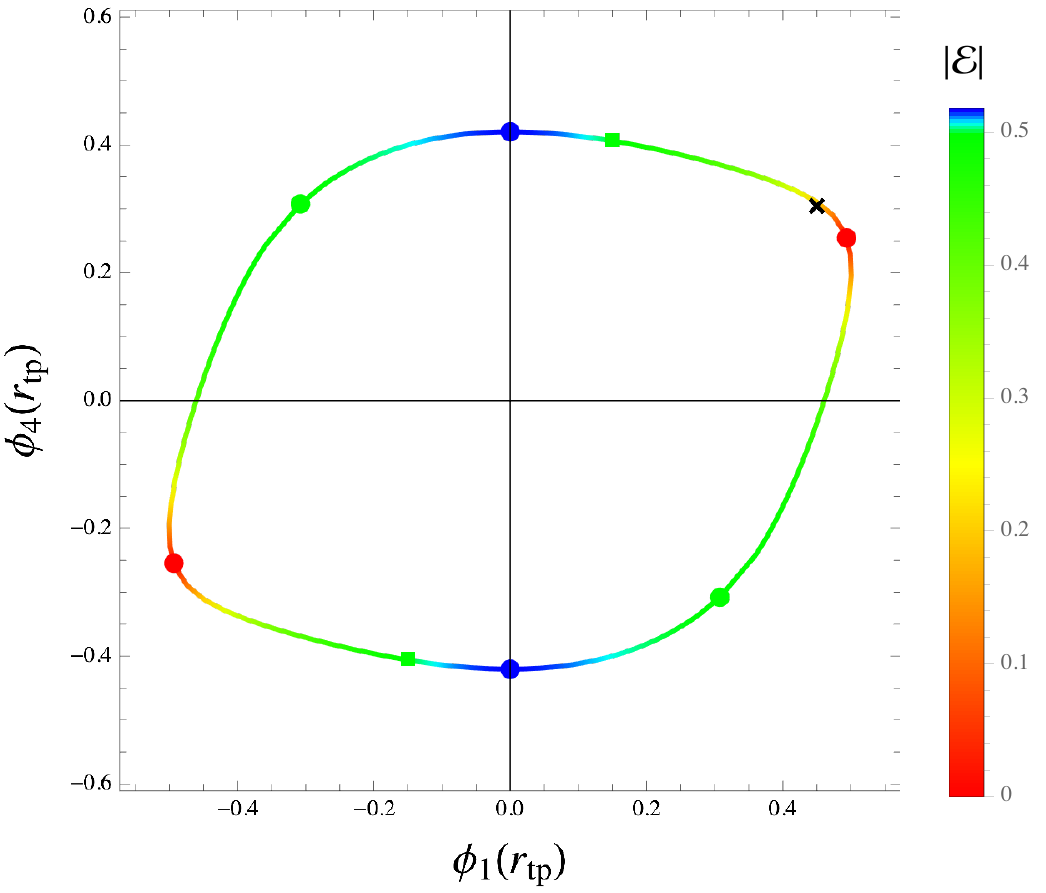}}
\caption{Turning point initial data for the $AdS_4\times\mathbb{R}$ solutions of the $\mathcal{N}=1^*$ equal mass $SO(3)$ invariant model.
Red dots correspond to the exactly periodic solution, blue dots correspond to the $\mathcal{N}=1$ linear dilaton solutions, green dots to the $\mathcal{N}=4$ linear dilaton solutions and green squares to the bounce solutions. The remaining points correspond to $AdS_4\times\mathbb{R}$ solutions with $\varphi$ a LPP function of $r$. All points inside the curve correspond to Janus solutions of $\mathcal{N}=4$ SYM theory 
(the black cross is the Janus solution in figure \ref{fig1}),
while points outside the curve have singularities. 
Points on the curve with the same colour represent the same solution, up to shifts of $\varphi$
and the discrete symmetry \eqref{genz2}.
}\label{fig:equalmass_phi_tp}
\end{figure}

The space of solutions that we have found in this way is summarised by the coloured curve in figure \ref{fig:equalmass_phi_tp}, with the
colour giving the value of $|\mathcal{E}|$, given by \eqref{ethismodel}.
If one starts with turning point data that lies anywhere within the curve, one obtains a Janus solution of $\mathcal{N}=4$ SYM theory 
with fermion and boson masses and a coupling constant that varies as one crosses the interface. For example, the Janus solution depicted in figure
\ref{fig1} corresponds to the black cross inside the curve in figure \ref{fig:equalmass_phi_tp}.
On the other hand if one starts outside the curve then one finds that the solution becomes singular on both sides of the interface as
in the solutions discussed in \cite{Arav:2020obl}, for example.

Observe that the figure is
symmetric under changing the signs of both $\phi_1(r_{tp})$ and $\phi_4(r_{tp})$, as a result of the symmetry \eqref{genz2}. 
The associated $AdS_4\times\mathbb{R}$ solutions obtained by this symmetry, 
which is a discrete $R$-symmetry combined with an $S$-duality transformation for the associated Janus solutions, are physically equivalent. The value of $\mathcal{E}$ is positive for the upper part of the curve between the two red dots and negative for the lower part.
We next point out that the blue dots correspond to the two
$\mathcal{N}=1$ $AdS_4\times\mathbb{R}$ S-fold solutions, with $\varphi$ a linear function of $r$, as in \eqref{nonesfold}. The red dots 
correspond to the fully periodic $AdS_4\times\mathbb{R}$ solution found in \cite{Arav:2020obl}. 
We will come back to the green dots and squares in a moment.
The remaining points on the curve 
all correspond to $AdS_4\times\mathbb{R}$ solutions with $\varphi$ an LPP function of $r$. 
Also, if one starts at the $\mathcal{N}=1$ S-fold solution at the top of the curve, then one can match on to the perturbative family of solutions that we constructed in the previous subsection and there is a similar story for the $\mathcal{N}=1$ S-fold solution at the bottom of the curve.

Points on the curve with the same colour have the same value of $|\mathcal{E}|$ and represent, essentially, the same solution, up to dilaton shifts \eqref{dilshift} and the discrete symmetry \eqref{genz2} if $\mathcal{E}$ has the opposite sign.
Indeed if we move to the right from the blue dot at the top all the way to the red dot at the right, the LPP solutions (all of which have $\mathcal{E}$ positive)
are essentially the same as those as one moves to the left; although the turning point data at $r=r_{tp}$ is different, the data of one of the solutions at $r=r_{tp}$ 
agrees with the turning point data of the other solution at $r=r_{tp}+\Delta r/2$, after making a suitable shift of $\varphi$ using 
\eqref{dilshift}. One can explicitly check this feature analytically for the perturbative solution 
\eqref{pertsolone}. We also note that this feature is consistent with the fact that there is just a single 
periodic solution which has the property that if one uses \eqref{dilshift} to have no zero mode for $\varphi$, then the solution
is invariant under a half period shift combined with a $\mathbb{Z}_2$ 
symmetry transformation \eqref{genz2}.

\begin{figure}[h!]
\centering
{
\includegraphics[scale=0.5]{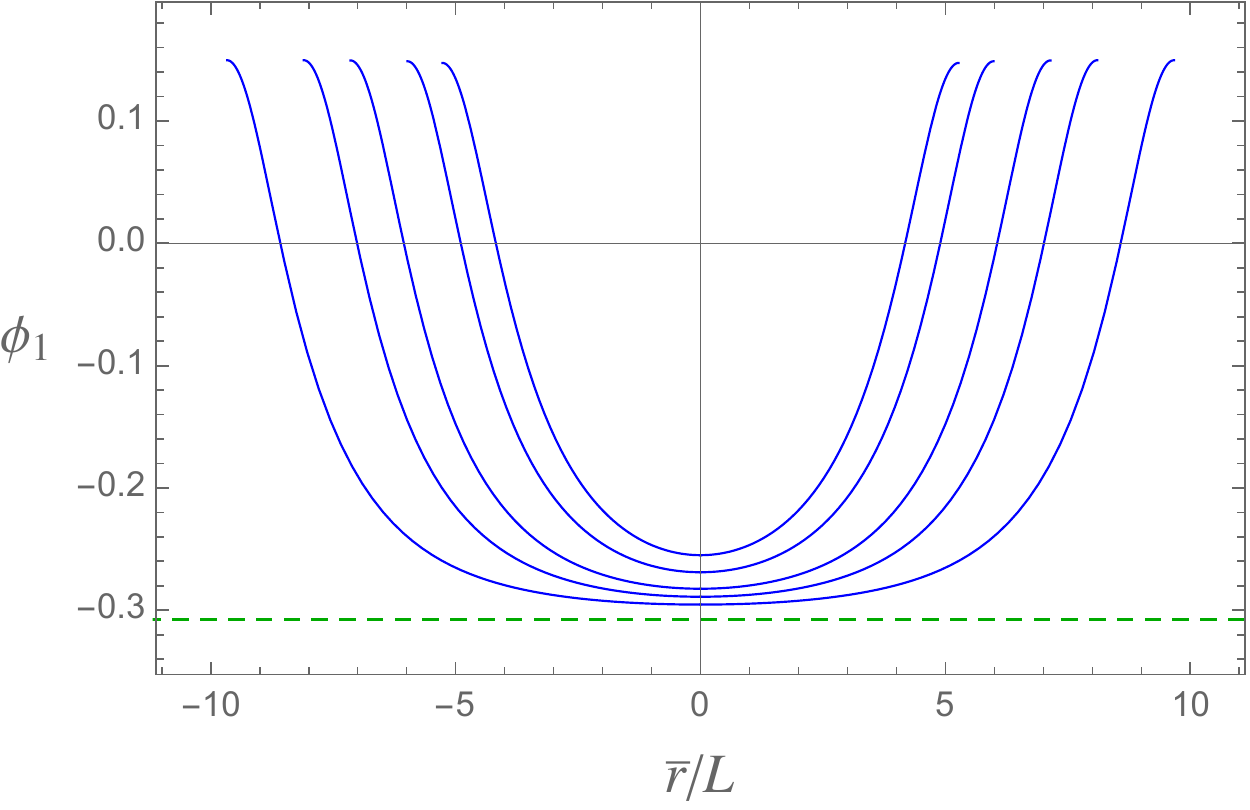}\qquad
\includegraphics[scale=0.5]{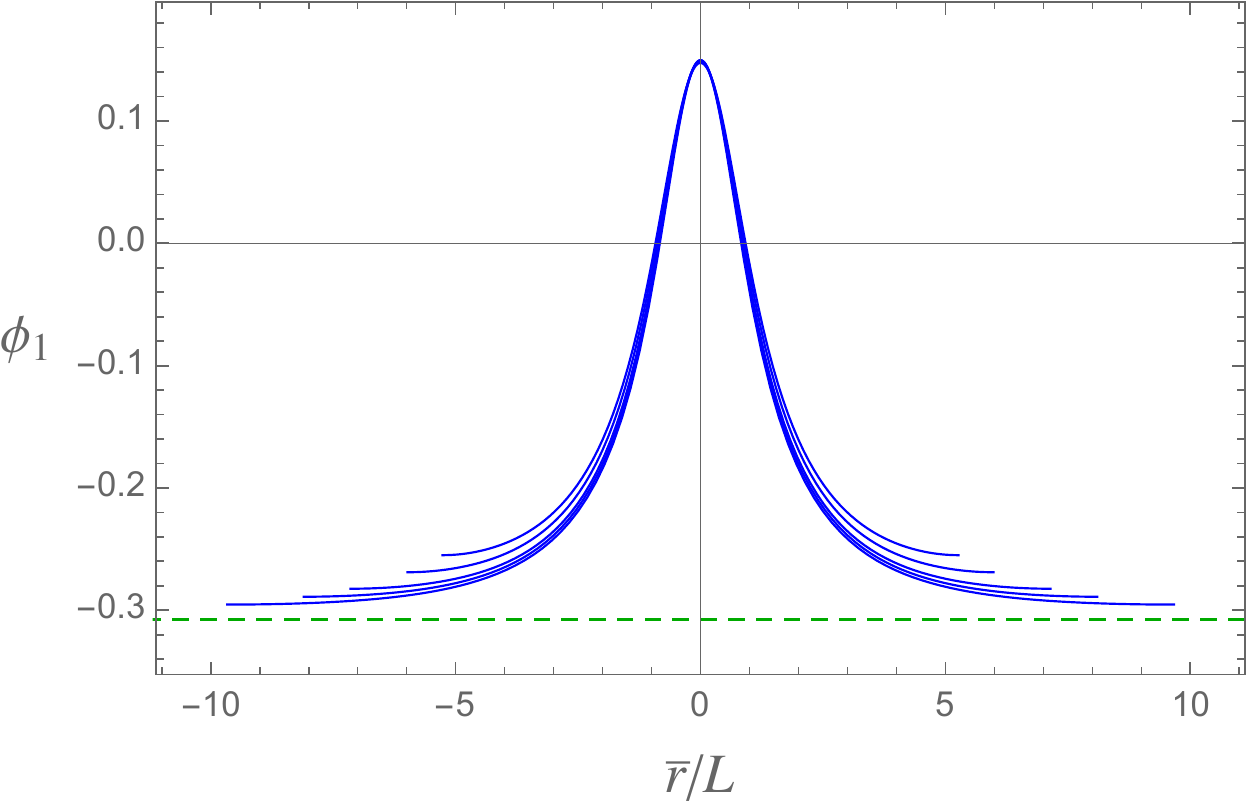}
}
\caption{Family of LPP solutions for the $\mathcal{N}=1^*$ equal mass $SO(3)$ invariant model with turning point data illustrating the approach to the green dots and
squares 
as in figure \ref{fig:equalmass_phi_tp}, with $\mathcal{|E|}=1/2$. The figures display just the periodic behaviour of $\phi_1$ for clarity and just one period. 
The left panel shows that the limiting solutions associated with the green dots degenerate into the $\mathcal{N}=4$ linear dilaton solution,
marked with a dashed green line.
The right panel shows
the limiting solution associated with the green square becomes a bounce solution which approaches the $\mathcal{N}=4$ linear dilaton solution, at both $ \bar r\to \pm \infty$, with a kink in $\phi_1$ centred at $\bar r=0$. 
}\label{fig:equalmass_F_T_varphideg}
\end{figure}

We now return to the green dots and squares in figure \ref{fig:equalmass_phi_tp}. 
The green dots, located at $\mathcal{|E|}=1/2$
represent the $\mathcal{N}=4$ linear dilaton solutions given in \eqref{nfoursfold}, while the green squares represent ``bounce" solutions that
involve those solutions, as we now explain.
We first consider the limiting class of the LPP solutions as we move along the coloured curve
in figure \ref{fig:equalmass_phi_tp} towards the upper green dot to the left. To illustrate, in the left panel of figure \ref{fig:equalmass_F_T_varphideg} we
have displayed the behaviour of one of the periodic functions, $\phi_1(r)$, as one approaches the critical initial data 
associated with the green dot, which has $\phi_1(r_{tp})=-1/2\cot^{-1}\sqrt{2}\sim -0.308$. The figure shows that
in this limit, the solution simply degenerates into the $\mathcal{N}=4$ linear dilaton solution \eqref{nfoursfold} for all values of $r$.
In the right panel of figure \ref{fig:equalmass_F_T_varphideg}  we have also displayed the approach to the upper green square to the right. In this case the
solution develops a region that approaches the $\mathcal{N}=4$ linear dilaton solution \eqref{nfoursfold} as one moves away from $r=0$ in either direction. Exactly at the initial values associated with the green square the solution will no longer be an LPP solution but degenerates into a ``bounce solution" which approaches the $\mathcal{N}=4$ linear dilaton solution \eqref{nfoursfold} at both $\bar r/L\to \pm \infty$, with a kink in the middle. 
We also see that these degenerations of the LPP solutions split the whole family of solutions
into two branches of LPP solutions: one that includes the 
perturbative solutions built using the $\mathcal{N}=1$ linear dilaton solutions and another that contains the periodic solution.

\begin{figure}[h!]
\centering
{
\includegraphics[scale=0.75]{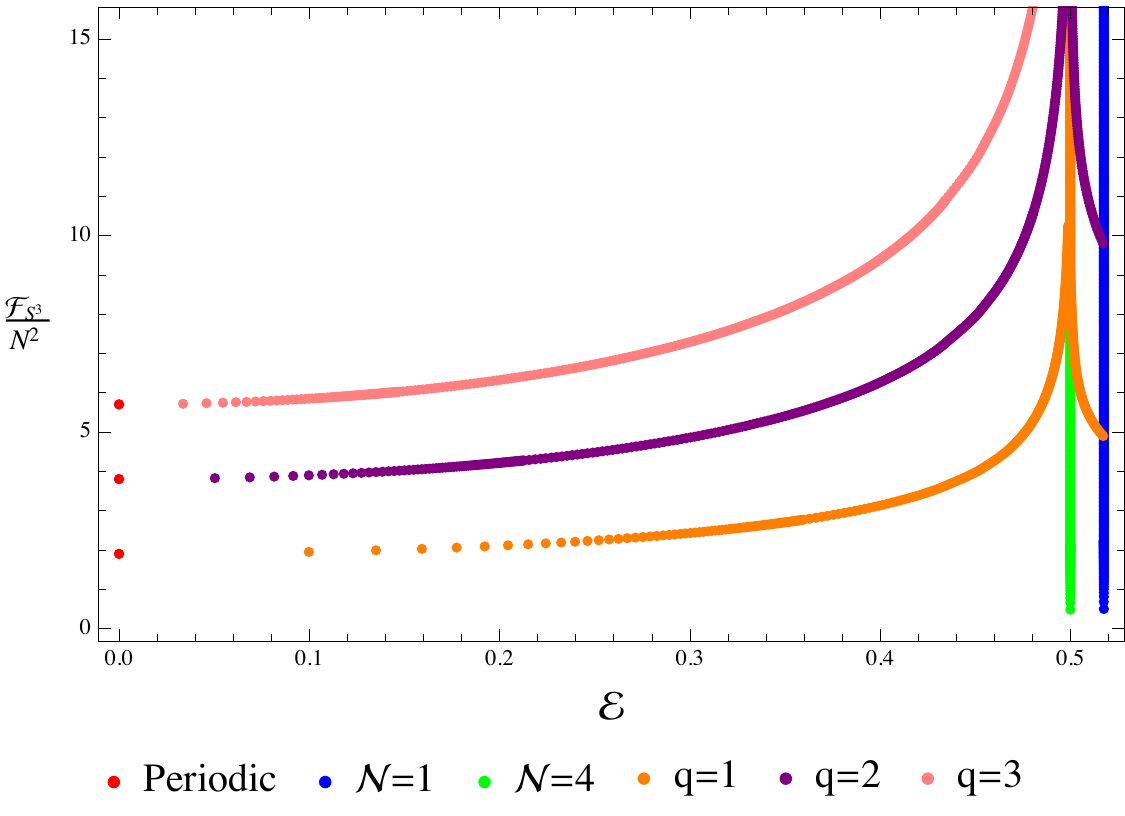}
}
\caption{Plot of the discrete S-folded solutions and the associated free energy of the dual field theory, $\mathcal{F}_{S^3}$, for the $\mathcal{N}=1^*$ equal mass $SO(3)$ invariant model as in figure \ref{fig:equalmass_phi_tp}. The discrete points rapidly become indistinguishable from continuous lines. }\label{fig:equalmass_F_T_varphi}
\end{figure}

In order to obtain S-fold solutions of type IIB string theory we also need to impose the quantisation condition \eqref{keycond}. 
In figure \ref{fig:equalmass_F_T_varphi} we have plotted some of these discrete solutions as well as $\mathcal{F}_{S^3}$ given in
\eqref{fenexp}. 
The discrete set of vertical points coloured blue and green correspond to the $\mathcal{N}=1$ and 
$\mathcal{N}=4$ S-fold solutions with linear dilatons, respectively, and $n$ increasing from 3 to infinity as one goes up;
for these S-folds we can obtain all values $n\ge3$ by suitably adjusting the period
$\Delta r$ over which we S-fold. 
The red dots correspond to the periodic solution for different values of the numbers of
period, $q$, that are used in making the $S^1$ compactification. The remaining discrete points correspond to 
$\mathcal{N}=1$ S-fold solutions with $\varphi$ an LPP function, for representative values of $q=1,2,3$. 
Starting from the left, for a given $q$, we have $n=3$ at the left and then rising to infinity as one 
approaches the bounce solution or the $\mathcal{N}=4$ S-fold solution at
$\mathcal{E}=1/2$, where the free energy diverges. Moving further to the right the value of $n$ decreases from infinity down to a bounded value
$[2\cosh q 3\pi]$, at the intersection with the $\mathcal{N}=1$ solutions on the blue line, which can be deduced from the perturbative analysis
\eqref{pertepsilon}. 

     \section{5-scalar model, $SU(2)$ invariant}\label{su2model}
This model is obtained from the 10-scalar model by setting $z^1=-z^3$, $z^2=-z^4$, or equivalently
$\alpha_1=\alpha_2=0$, $\phi_1=\phi_2=0$, $\beta_2=0$.
This model involves five scalar fields parametrised by 
\begin{align}
\beta_1,\,\, z^1=\tanh\big[\frac{1}{2}\big(\alpha_3+\varphi-i\phi_3+i\phi_4\big)\big],\, \,z^2=\tanh\big[\frac{1}{2}\big(\alpha_3-\varphi-i\phi_3-i\phi_4\big)\big]\,.
\end{align}
In addition to the symmetry \eqref{genz2},
this model is also invariant under the symmetry 
\begin{align}\label{5scalsym}
\phi_3\to-\phi_3\,,\qquad\alpha_3\to -\alpha_3\,,
\end{align}
with $\beta_1,\phi_4,\varphi$ unchanged, which is a remnant of 
the discrete transformations given in \eqref{gens33} for the 10-scalar truncation.
This additional symmetry will clearly manifest itself in the set of solutions we construct.
The integral of motion \eqref{eq:IntegOfMotion}
for this truncation is now given by
\begin{align}\label{eeforsu2}
\mathcal{E}=\frac{1}{L^3}e^{3A}[-\tan(\phi_3-\phi_4)+\tan(\phi_3+\phi_4)]\,.
\end{align}

If we further set $z^1=-z^2$, equivalently, $\alpha_3=\phi_3=0$, as well as $\beta_1=0$ then we 
obtain a two-scalar model depending $\varphi,\phi_4$ that  overlaps
with the truncation considered in the context of $\mathcal{N}=1$ S-folds in section 4 of \cite{Bobev:2020fon}, which we also 
discussed in the previous section.
In particular the $AdS_4\times\mathbb{R}$ solution associated with the $\mathcal{N}=1$ S-folds is given by
\begin{align}\label{nonesfold2}
\varphi&=\frac{\sqrt{5}}{2}r,\qquad
\phi_4=\cos^{-1}\sqrt{\frac{5}{6}}\,,\qquad
e^{A}=\frac{5L}{6}\,,\nn
\beta_1&=\alpha_3=\phi_3=0\,,
\end{align}
with $\mathcal{E}=\frac{25\sqrt{5}}{108}$.
There is another $\mathcal{N}=1$ S-fold solution that can be obtained from the symmetry
\eqref{genz2}, with opposite sign for $\mathcal{E}$.

On the other hand if we set $z^2=\bar z^2$ or equivalently 
$\phi_3=-\phi_4$ then we obtain a four-scalar model depending on $\phi_3,\alpha_3,\varphi,\beta_1$ that
overlaps\footnote{They consider a model with seven
scalars: $(\varphi,\chi,\alpha,\lambda,c,\omega,\psi)$. One should set $c=\omega=\psi=0$ and then identify
$\alpha=\beta_1$, $\lambda=\alpha_3$, $\sin2 \phi_3=-\tanh 2\chi$ as well as $g=2/L$.
} 
with the truncation considered
in the context of $\mathcal{N}=2$ S-folds in section 3 of \cite{Bobev:2020fon}. 
Also note that after utilising the symmetry \eqref{5scalsym} we can also truncate
to a 4-scalar model by taking $z^1=\bar z^1$, or equivalently $\phi_3=+\phi_4$.
The $\mathcal{N}=2$ S-fold solution, with $\phi_3=-\phi_4$, can be written
\begin{align}
\varphi&=r\,,\quad
\phi_3=-\phi_4=-\frac{\pi}{8}\,,\quad
\beta_1=-\frac{1}{12}\log 2\,,\quad
e^A=\frac{L}{2^{1/3}}\,,\qquad
\alpha_3=0\,,
\end{align}
with $\mathcal{E}=\frac{1}{2}$.
After using the symmetries \eqref{genz2} and \eqref{5scalsym} there are now
a total of four $\mathcal{N}=2$  $AdS_4\times\mathbb{R}$ S-fold solutions with $\varphi$ linear in $r$.
From  \eqref{freeensfold} the free energy of these solutions is given by
\begin{align}
\mathcal{F}_{S^3}=\frac{1}{2}\mathrm{arccosh}\frac{n}{2}{N^2}\,.
\end{align}
in agreement with \cite{Bobev:2020fon}.

Finally, if we set $z^1=z^2$ or equivalently $\phi_4=\varphi=0$ then we
obtain the $\mathcal{N}=1^*$ one-mass truncation used in \cite{Arav:2020obl}, which contains three scalars $\beta_1,\phi_3,\alpha_3$ and retains the symmetry \eqref{5scalsym}. This truncation also contains two LS $AdS_5$ fixed point solutions, LS$^\pm$, which are related by \eqref{5scalsym}
and given by
\begin{align}\label{strass}
\beta_1=-\frac{1}{6}\log 2,\quad
\phi_3=\pm\frac{\pi}{6},\quad\alpha_3=0,\quad \tilde L=\frac{3}{2^{5/3}} L\,,
\end{align}
where $\tilde L$ is the radius of the $AdS_5$.

\subsection{Periodic perturbation about the $\mathcal{N}=1$ S-fold}
Much as in the last section, within the 5-scalar truncation we can build a perturbative
solution about the $\mathcal{N}=1$ S-fold solution given in \eqref{nonesfold2}.
The key point is that there is now a periodic linearised perturbation of the form
\begin{align}
\delta\alpha_3&= \sin\frac{\sqrt{5} r}{3}\,,\qquad
\delta\phi_3=-\sqrt{5}  \cos\frac{\sqrt{5} r}{3}\,.
\end{align}

With some effort we can use this to construct a perturbative expansion in a parameter $\epsilon$, that takes the form
\begin{align}\label{pertsoltwo}
\alpha_3&= \sum_{m,p\in odd}^\infty
a_{m,p}^{(\alpha_3)}\epsilon^m \sin{}pKr\,,\qquad\qquad\quad
\phi_3= \sum_{m,p\in odd}^\infty
a_{m,p}^{(\phi_3)}\epsilon^m \cos{}pKr\,,\nn
\phi_4&=\phi_4^{zm}(\epsilon)+\sum_{m,p\in even}^\infty
a_{m,p}^{(\phi_4)}\epsilon^m \cos{}pKr\,,\quad
\varphi= k(\epsilon)r +\sum_{m,p\in even}^\infty
a_{m,p}^{(\varphi)}\epsilon^m \sin{}pKr
\,,\nn
\beta_1&=\beta_1^{zm}(\epsilon)+\sum_{m,p\in even}^\infty a_{m,p}^{(\beta_1)}\epsilon^m \cos{}pKr \,,\quad
{A}= A^{zm}(\epsilon)+\sum_{m,p\in even}^\infty
a_{m,p}^{({A})}\epsilon^m \cos{}pKr\,,
\end{align}
where the sums over odd integers start from 1 and the sums over even integers
start from 2. All functions, except $\varphi$ are periodic in the radial direction with period $\Delta r=\frac{2\pi}{K}$, with $\varphi$
an LPP function, exactly as in \eqref{sfoldansatz}-\eqref{varphidef}.
The wavenumber $K$ is itself given by the following series in $\epsilon$:
\begin{align}
K\equiv \frac{2\pi}{\Delta r}&=\frac{\sqrt{5}}{3}-\frac{292 \sqrt{5}}{117}\epsilon ^2-\frac{3316328 \sqrt{5}}{59319}\epsilon ^4-\frac{241179878834}{30074733 \sqrt{5}}\epsilon ^6+\cdots\,,
\end{align}
which we notice is decreasing as we move away from the $\mathcal{N}=1$ S-fold solution.

Notice that both $\alpha_3$ and $\phi_3$ have vanishing zero mode in this expansion.
The zero modes of the remaining periodic functions are explicitly given by
\begin{align}
\phi_4^{zm}&=\cos^{-1}\Bigg(\sqrt{\frac{5}{6}}\Bigg)-\frac{\sqrt{5}}{3}\epsilon^2-\frac{4861 \sqrt{5}}{6084}\epsilon ^4-\frac{185672641 \sqrt{5}}{9253764} \epsilon ^6+\cdots\,,\nn
\beta_1^{zm}&=-\frac{2}{3} \epsilon ^2-\frac{755}{78}  \epsilon ^4-\frac{5171099}{19773} \epsilon ^6
+\cdots\,,\nn
A^{zm}&= \log \left(\frac{5 L}{6}\right)- \epsilon ^2-\frac{10241}{3042}  \epsilon ^4-\frac{663866873}{4626882}\epsilon ^6+\cdots\,,
\end{align}
In addition the slope of $\varphi$ takes the form
\begin{align}
k&=\frac{\sqrt{5} }{2}-\frac{3\sqrt{5} }{2} \epsilon ^2-\frac{311\sqrt{5}}{1352} \epsilon ^4-\frac{19753429\sqrt{5}}{228488}\epsilon ^6+\cdots\,.
\end{align}
Furthermore, we also have $\Delta\varphi\equiv k\Delta r$ is given by
\begin{align}\label{pertepstwo}
\Delta\varphi=3 \pi+\frac{175 \pi}{13}  \epsilon ^2+\frac{5295375 \pi}{8788} \epsilon ^4+\frac{153607091549 \pi}{7425860}  \epsilon ^6+\cdots\,.
\end{align}
The integral of motion \eqref{eeforsu2} is given by
\begin{align}
&\mathcal{E} = \frac{25 \sqrt{5}}{108}\big(1-2 \epsilon ^2+\frac{4598}{507}\epsilon ^4+\frac{96057473}{771147}\epsilon ^6+\cdots\big)\,.
\end{align}

We now write the periodic functions collectively as
 $\Psi_1=\{A,\phi_4,\beta_1\}$  and
  $\Psi_2=\{\alpha_3,\phi_3\}$ 
so that the whole solution is specified by $\Psi_1(\epsilon,r)$, $\Psi_2(\epsilon,r)$ and $\varphi(\epsilon,r)$. 
We then find
\begin{align}
\Psi_1(\epsilon,r+\pi/K)&=\Psi_1(-\epsilon,r)=+\Psi_1(\epsilon,r),\nn
\Psi_2(\epsilon,r+\pi/K)&=\Psi_2(-\epsilon,r)=-\Psi_2(\epsilon,r),\nn
\varphi(\epsilon,r+\pi/K)&=\varphi(-\epsilon,r) + \text{constant}\,,
\end{align}
where the constant can be removed by \eqref{dilshift} and we note that the last equalities in the first
two lines are associated with the symmetry \eqref{5scalsym}.

After uplifting to type IIB and carrying out the S-fold procedure as described in section \ref{sfoldproc}, we obtain new S-folds of type IIB provided that we can solve
\eqref{keycond}.
This can be done as in the discussion following \eqref{freeone} and, in particular,  
the smallest value of $n$ that can be reached in
\eqref{keycond} is $n=12392$, which occurs for $q=1$ and $\epsilon\sim 0.0008$.
The free energy for the S-folded solutions can be obtained from \eqref{fenexp} and is given by
 \begin{align}
 \mathcal{F}_{S^3}
 &=\frac{25\sqrt{5}}{108}\left(1-\frac{175}{39}\epsilon^4 -\frac{13887100}{39^3}\epsilon^6+\dots\right)\mathrm{arccosh}\frac{n}{2}N^2\,.
 \end{align}

This truncation also contains the known $AdS_4\times\mathbb{R}$ $\mathcal{N}=2$ S-fold solutions, but there
is no longer a linearised periodic mode within this truncation in which to build an analogous solution. This is
similar to the known $AdS_4\times\mathbb{R}$ $\mathcal{N}=4$ S-fold solutions in the $SO(3)$ invariant truncation that we considered in
the previous section.

 \subsection{New S-fold solutions}  
   
The new $AdS_4\times \mathbb{R}$ solutions, with $\varphi$ a LPP function, can be constructed as limiting cases of Janus solutions, much as in the last section. We again start by constructing Janus solutions with turning point of $A$ at $r=r_{tp}$, with $r_{tp}=0$.
We can use the shift symmetry \eqref{dilshift} to
choose $\varphi(r_{tp})=0$. We then focus\footnote{As in the previous section, if we relax the condition that the initial data is invariant under the $\mathbb{Z}_2$ symmetry, then we only find limiting solutions that are in the  ``one-sided" Janus class
discussed in section \ref{onesided}. We also note that the perturbative solution \eqref{pertsoltwo} is invariant under this symmetry.
} on solutions that are invariant under the $\mathbb{Z}_2$ symmetry, obtained by
combining \eqref{genz2} and \eqref{z2twosymgm1}, 
\begin{align}\label{z2twosymgm122}
r\to-r,\qquad z^A\to -\bar z^A,\qquad\xi\to -\xi+\pi\,.
\end{align}
This implies that $\phi_3,\phi_4$ are even functions of $r$ and $\alpha_3,\varphi$ are 
odd functions.
Thus, we again take $\alpha_3(r_{tp})=0$ as part of our initial value data for the solutions. From 
\eqref{eq:RewrittenJanusBPSAeq}-\eqref{eq:RewrittenJanusBPSzeq}, and as explained in
section 5 of \cite{Arav:2020obl}, the solutions are now specified by the values of $\phi_3(r_{tp})$ and $\phi_4(r_{tp})$, with the value of
$\beta_1(r_{tp})$ fixed by this data. By suitably tuning the values of the scalar field at the turning points we are able to construct the limiting cases of solutions associated with the S-folds.

\begin{figure}[h!]
\centering
{\includegraphics[scale=0.75]{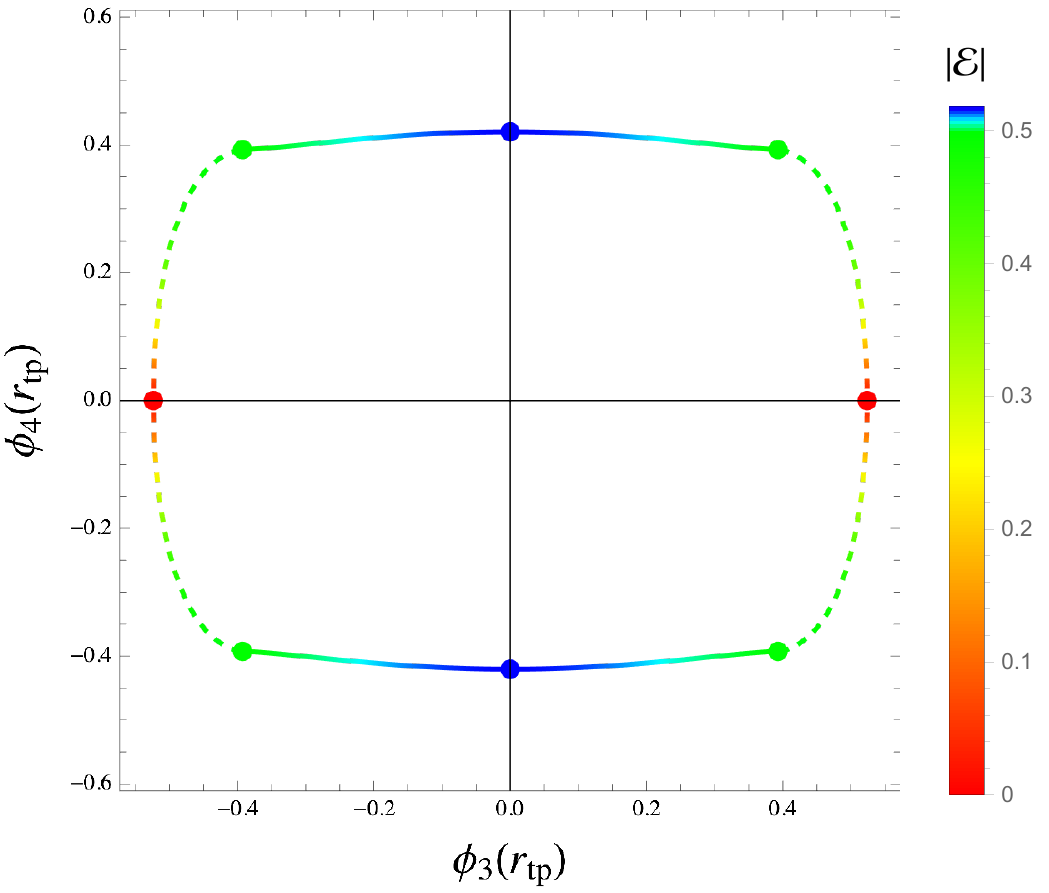}}
\caption{Turning point initial data for the $AdS_4\times\mathbb{R}$ solutions of the 5-scalar $SU(2)$ invariant model.
The blue dots correspond to the $\mathcal{N}=1$ linear dilaton solutions
while the green dots correspond to the 
$\mathcal{N}=2$ linear dilaton solutions, as well as associated soliton solutions. The red dots correspond to the two LS
$AdS_5$ solutions, LS$^\pm$. The remaining points on the solid lines correspond to $AdS_4\times\mathbb{R}$ solutions with $\varphi$ a LPP function of $r$, with the same colour representing the same physical solution. 
All points inside the curve correspond to Janus solutions of $\mathcal{N}=4$ SYM theory while points outside the curve have singularities.
The dashed lines correspond to an  LS$^\pm$ to LS$^\pm$ Janus solution.}\label{fig:su2_phi_tp}
\end{figure}

The space of solutions we have found in this way is summarised by the curve shown in figure \ref{fig:su2_phi_tp}. 
If one starts with turning point data that lies anywhere within the curve, one obtains a Janus solution of $\mathcal{N}=4$ SYM theory 
with fermion and boson masses and a coupling constant that varies as one crosses the interface. 
On the other hand if one starts outside the curve then one finds that the solution becomes singular on both sides of the interface.

Observe that the figure is symmetric under changing the signs of either $\phi_3(r_{tp})$ or $\phi_4(r_{tp})$. This is a result of
the symmetries \eqref{genz2} and \eqref{5scalsym}. The associated $AdS_4\times\mathbb{R}$ solutions obtained using these symmetries, 
which for the Janus solutions are a combination of a discrete $R$-symmetry and an $S$-duality transformation
(in the case of \eqref{genz2}), are physically equivalent.
The value of $\mathcal{E}$ is positive for the upper part of the curve and negative for the lower part.
We next point out that the blue dots correspond to the 
$\mathcal{N}=1$ $AdS_4\times\mathbb{R}$ S-fold solutions which have $\varphi$ a linear function of $r$. 
The green dots represent the $\mathcal{N}=2$ $AdS_4\times\mathbb{R}$ S-fold solutions as well as associated ``soliton" solutions
that we discuss further below.
The remaining points on the coloured, solid lines all correspond to $AdS_4\times\mathbb{R}$ solutions with $\varphi$ an LPP function of $r$. 
Also,
if one starts at the $\mathcal{N}=1$ S-fold solution at the top of the curve, then one can match on to the perturbative family of solutions that we constructed in the previous subsection.

Points on the solid curve with the same colour represent, essentially, the same LPP solution, up to dilaton shifts and possible discrete symmetries. 
Moving from the right of the blue dot at the top all the way to the green dot at the right one finds LPP solutions that are essentially the same as those as one moves to the left; although the turning point data at $r=r_{tp}$ is different, the data of one of the solutions at $r=r_{tp}$ 
agrees with the turning point data of the other solution at $r=r_{tp}+\Delta r/2$, after making a suitable shift of $\varphi$ using 
\eqref{dilshift}. Note that the two sets of turning point data are also related
by \eqref{5scalsym}. 
One can explicitly check these features analytically for the perturbative solution \eqref{pertsoltwo}.

\begin{figure}[h!]
\centering
{\includegraphics[scale=0.5]{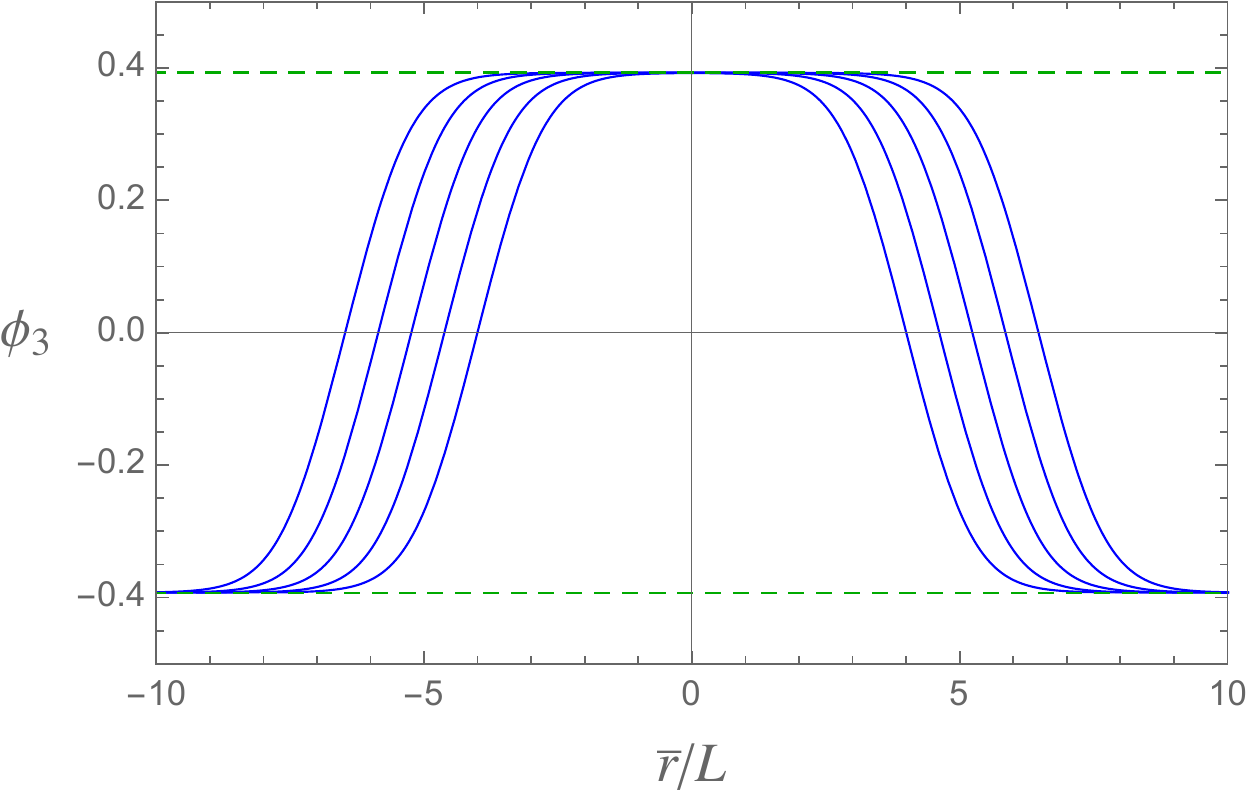}}\qquad
{\includegraphics[scale=0.48]{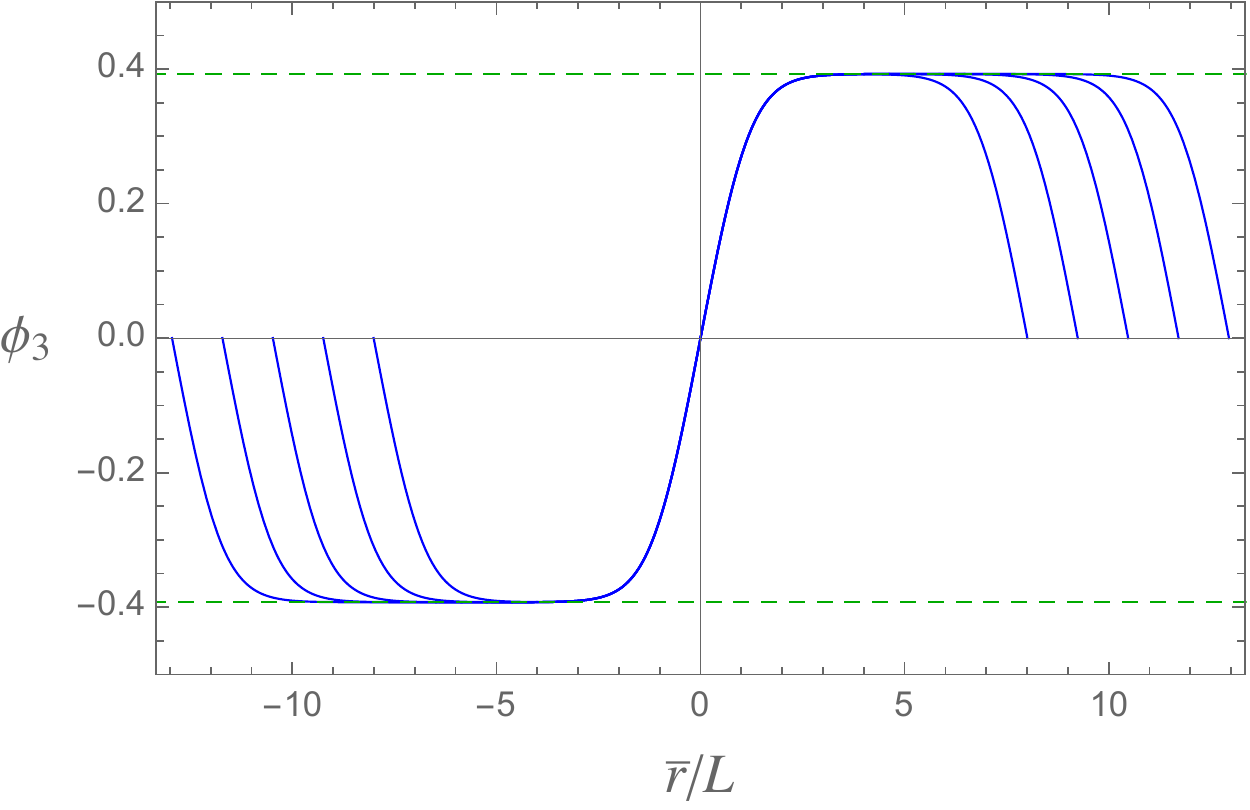}}
\caption{Limiting families of solutions for the 5-scalar $SU(2)$ invariant model, with 
just the periodic behaviour of $\phi_3$ displayed for clarity. The
left panel illustrates the approach to the green dots in figure \ref{fig:equalmass_phi_tp}, along the coloured curve; one finds that the solution will approach
the $\mathcal{N}=2$ linear dilaton solution associated with the upper green dashed line for all $\bar r$. In the right panel we
display a different limiting solution, obtained by fixing $\phi_3(0)=0$, which 
degenerates into a soliton solution that approaches one 
$\mathcal{N}=2$ linear dilaton solution, at $\bar r\to -\infty$ and another $\mathcal{N}=2$ linear dilaton solution at $\bar r\to\infty$ with opposite
sign of $\phi_3$ (related by \eqref{5scalsym}).}\label{fig:su2_phi_tpdeg}
\end{figure}

In the limit of approaching the green dots in figure \ref{fig:su2_phi_tp} along the solid
curve, the LPP solutions degenerate into 
the $AdS_4\times\mathbb{R}$ $\mathcal{N}=2$ S-fold solutions as illustrated in the left panel in figure
\ref{fig:su2_phi_tpdeg} for one of the periodic functions, $\phi_3(r)$. As one approaches
the critical initial data associated with the green dot which has 
$\phi_3=\frac{\pi}{8}\sim 0.39$, the solution degenerates into the $\mathcal{N}=2$ S-fold solution, with the region around $\bar r=0$ 
extending out all the way to infinity. Interestingly, essentially using the same family of solutions,
one can construct another limiting solution which is a kind of ``soliton" solution
that approaches one of the $AdS_4\times\mathbb{R}$ $\mathcal{N}=2$ S-fold solutions as $\bar r \to-\infty$ and a different 
$AdS_4\times\mathbb{R}$ $\mathcal{N}=2$ S-fold solution, related by flipping the sign of $\phi_3$, as $\bar r \to\infty$. This limiting
solution is
illustrated in the right panel of figure \ref{fig:su2_phi_tpdeg}.

We next turn to the remaining points in figure \ref{fig:su2_phi_tp}. The red dots are the two LS $AdS_5$ fixed points given in 
\eqref{strass}, which we refer to as LS$^\pm$. Moving along the class of Janus solutions on the horizontal axis towards the red dots at the right, say, one finds that the Janus 
solutions degenerate into three components; 
a Poincar\'e invariant RG flow solution that starts off at the $AdS_5$ vacuum and then approaches the
LS$^+$ $AdS_5$ fixed point, the LS$^+$ fixed point solution itself and then another 
Poincar\'e invariant RG flow solution going between LS$^+$ and the $AdS_5$ vacuum.
The dashed curves correspond to another interesting degeneration of the Janus solutions. 
As one approaches the dashed curve on the right side of the figure one again finds
three components: there is the same two Poincar\'e invariant components on the outside and the middle component is now
an LS Janus solution that moves between LS$^+$ and LS$^+$ on either side of the interface, with $\varphi$ linear in $\bar r$. 
There is similar behaviour as one approaches the red dot or the dashed line on the left 
side of the figure with LS$^-$ replacing LS$^+$.

\begin{figure}[h!]
\centering
{
\includegraphics[scale=0.65]{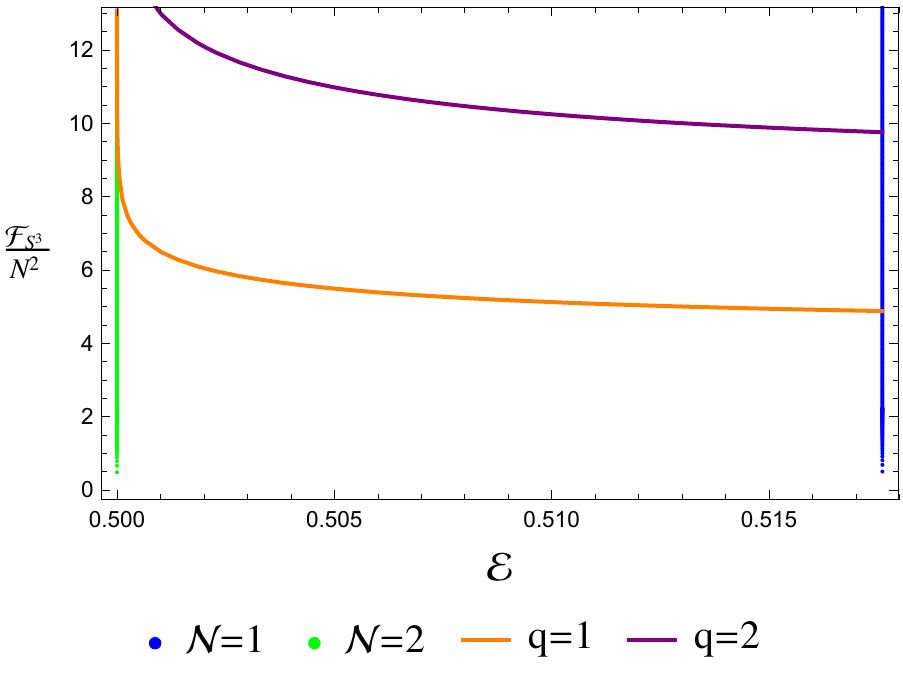}
}
\caption{Plot of the discrete S-folded solutions and the associated free energy of the dual field theory, $\mathcal{F}_{S^3}$, for the 
5-scalar $SU(2)$ invariant model as in figure \ref{fig:su2_phi_tp}.
The discrete points rapidly become indistinguishable from a continuous line.}\label{fig:equalmass_F_T_varphisu2}
\end{figure}

To obtain S-fold solutions of type IIB string theory we also need to impose the quantisation condition \eqref{keycond}. 
In figure \ref{fig:equalmass_F_T_varphisu2} we have plotted some of these discrete solutions as well as $\mathcal{F}_{S^3}$ given in
\eqref{fenexp}. 
The discrete set of vertical points coloured blue and green correspond to the $\mathcal{N}=1$ and 
$\mathcal{N}=2$ S-fold solutions with linear dilatons, respectively, and $n$ increasing from 3 to infinity as one goes up.
The remaining discrete points correspond to 
$\mathcal{N}=1$ S-fold solutions with $\varphi$ an LPP function, for representative values of $q=1,2$. 
Starting from the right at the blue dots, for a given $q$, we have $n$ starting from $[2\cosh  q 3\pi]$,
 which can be deduced from the perturbative analysis \eqref{pertepstwo}, and then 
rising to infinity as one approaches the $\mathcal{N}=2$ S-fold solution at
$\mathcal{E}=1/2$, where the free energy diverges. 

   \section{One-sided Janus solutions}\label{onesided}
   In this section we discuss a novel class of $D=5$ solutions within the ansatz 
   \eqref{metjanus},
   that at one end of $\mathbb{R}$ approach the $AdS_5$ vacuum, while at the other end approach an $AdS_4\times\mathbb{R}$ solution 
   with the $D=5$ dilaton, $\varphi$, either a linear function or an LPP function of $r$. We can also construct solutions that approach the periodic 
   $AdS_4\times\mathbb{R}$ solution at the other end. 
We refer to these solutions as ``one-sided Janus" solutions. In contrast to other one sided Janus solutions that have been 
   previously constructed, for example in \cite{Gutperle:2012hy,Bobev:2013yra,Arav:2020asu,Arav:2020obl},  
   remarkably these new solutions are free from singularities.

\subsection{An analytic solution preserving $\mathcal{N}=4$ supersymmetry}\label{nfourstuff}
We first consider an analytic solution that lies within the $SO(3)\times SO(3)$ invariant truncation that
involves 3 scalar fields, $\phi_1=\phi_2=\phi_3=-\phi_4$, $\alpha_1=\alpha_2=\alpha_3$ and $\varphi$.
   
Using the proper distance gauge with radial coordinate $\bar r$, we find the following solution
\begin{align}
\label{eq:one-sided Janus solutions proper distance}
\tan 4\phi_1&=-\frac{2 \sqrt{2} e^{-3{\bar r}/L}}{\left(1+e^{-2{\bar r}/L}\right)^{3/2}}\,,\qquad\qquad
\cosh 4\alpha_1=\frac{1+\frac{2 e^{-4{\bar r}/L}}{\left(1+e^{-2{\bar r}/L}\right)^2}}{\sqrt{1+\frac{8 e^{-6{\bar r}/L}}{\left(1+e^{-2{\bar r}/L}\right)^3}}}\,,\nn
e^{8\varphi-8\varphi_{(s)}}&=\frac{1+3 e^{-2{\bar r}/L}}{\left(1+e^{-2{\bar r}/L}\right)^3 \left(1+3 e^{-4{\bar r}/L}\right)}\,,\qquad
e^A=\frac{L e^{{\bar r}/L} \sqrt{1+e^{-2{\bar r}/L}}}{\sqrt{2} }\,.
\end{align}
For these solutions, in which the warp factor $A$ does not have a turning point, we find that the integral of motion is given by
$\mathcal{E}=\frac{1}{2}$. Recall that the $\mathcal{N}=4$ S-fold solution with a linear dilaton given in 
\eqref{nfoursfold} also had $\mathcal{E}=\frac{1}{2}$. In other words, taking the limit $\mathcal{E}\to\frac{1}{2}$ in the family of
Janus solutions in this truncation can either give the $\mathcal{N}=4$ S-fold solution or this new solution, which describes a one-sided Janus solution.

At the $\bar r\to +\infty$ end these solutions approach the $AdS_5$ vacuum solution, dual to $\mathcal{N}=4$ SYM theory.
After shifting the radial coordinate ${\bar r}\to {\bar r}-L\log \frac{L}{\sqrt{2}}$, so we can easily compare with
\cite{Arav:2020obl}, we find that as $\bar r\to\infty$
we have the asymptotic expansion
\begin{align}
\label{eq:SYM_AdS5_expansion}
\phi_1&=-\frac{L^3}{4}e^{-3{\bar r}/L}+\cdots\,,\qquad\qquad
\alpha_1=\frac{L^2}{4}e^{-2{\bar r}/L}-\frac{L^4}{4}e^{-4{\bar r}/L}+\cdots\,,\nn
\varphi&=\varphi_{(s)}-\frac{3L^4}{16}e^{-4{\bar r}/L}+\cdots\,,\qquad\qquad
A=\frac{{\bar r}}{L}+\frac{L^2}{4}e^{-2{\bar r}/L}-\frac{L^4}{16}e^{-4{\bar r}/L}+\cdots\,.
\end{align}
From the results given in \cite{Arav:2020obl} we can immediately deduce that all sources for the operators dual to the scalar fields vanish.
Furthermore, we can also determine the one point functions. As explained in detail \cite{Arav:2020obl}, and refined in appendix \ref{appren},
we can determine the one-point functions
that are associated with $\mathcal{N}=4$ SYM theory on flat spacetime\footnote{As opposed to $AdS_4$ to which it is related by a Weyl transformation. We also
note that the analysis in \cite{Arav:2020obl} assumed $\varphi_{(s)}=0$ which can be achieved by a dilaton shift.} with coordinates $(t,y_i)$; we find that the one-point functions having spatial dependence on one of the spatial directions, say $y_3$,
with\footnote{Note that the operators have not been canonically normalised, which explains the factors of $L$ appearing on the left hand side.}
\begin{align}\label{vevsonesided}
L^2\langle\mathcal{O}_{\alpha_1}\rangle=\frac{N^2}{8 \pi^2}\frac{1}{y_3^2}\,,\qquad
L\langle\mathcal{O}_{\phi_1}\rangle=-\frac{N^2}{4 \pi^2}\frac{1}{y_3^3}\,,\qquad 
\langle\mathcal{O}_{\varphi}\rangle=-\frac{3N^2}{8 \pi^2}\frac{1}{y_3^4}\,,
\end{align}
where we used \eqref{stdresult}.
These expressions display the appropriate dependence on $y_3$ that is consistent with $d=3$ conformal invariance with respect to the $(t,y_1,y_2)$
for dual operators of scaling dimension
$\Delta=2,3$ and 4, respectively.

At the other end, as ${\bar r}\rightarrow -\infty$, again after shifting ${\bar r}\to {\bar r}-L\log \frac{L}{\sqrt{2}}$,
the asymptotic expansion is given by
\begin{align}
\label{eq:another_end_expansion}
\phi_1&=-\frac{1}{4} \tan ^{-1}\left(2 \sqrt{2}\right)+\frac{1}{3 \sqrt{2} L^2}e^{2{\bar r}/L}+\frac{1}{18 \sqrt{2} L^4}e^{4{\bar r}/L}+\cdots\,,\nn
\alpha_1&=\frac{1}{3 L^2}e^{2{\bar r}/L}-\frac{4 }{9 L^4}e^{4{\bar r}/L}+\cdots\,,\nn
\varphi&=\frac{{\bar r}}{L}+\varphi_{(s)}-\log\frac{L}{\sqrt{2}}-\frac{2 }{3 L^2}e^{2{\bar r}/L}+\frac{5 }{9 L^4}e^{4{\bar r}/L}+\cdots\,,\nn
A&=\log\frac{L}{\sqrt{2}}+\frac{1}{L^2}e^{2{\bar r}/L}-\frac{1}{L^4}e^{4{\bar r}/L}+\cdots\,.
\end{align}
This shows that the solution at this end is precisely approaching the $\mathcal{N}=4$ $AdS_4\times\mathbb{R}$ S-fold solution with $\varphi$
a linear function of $r$, which was   
given in \eqref{nfoursfold}.

The solution solves the BPS equations \eqref{eq:RewrittenJanusBPSAeq}-\eqref{eq:RewrittenJanusBPSzeq} and hence it preserves at least $\mathcal{N}=1$ supersymmetry. However, since it is a solution that lies within the $SO(3)\times SO(3)$ invariant truncation it
actually preserves $\mathcal{N}=4$ supersymmetry. Furthermore, after uplifting the solution to type IIB, using the formulae in appendix \ref{so3so3model},
we obtain a $D=10$ metric of the form
\begin{align}
ds^2=f^2_4 ds^2_{AdS_4}+f^2_1 d\Omega_2^2+f^2_2 d\tilde{\Omega}_2^2+ds^2({\Sigma})\,,
\end{align} 
where $d\Omega_2^2$ and $d\tilde{\Omega}_2^2$ are metrics on round two-spheres and 
$f_1,f_2$ and $f_4$ are functions of the coordinates on $\Sigma$. A full classification of such solutions which preserve
$\mathcal{N}=4$ supersymmetry can be found in \cite{DHoker:2007zhm,DHoker:2007hhe}. In appendix 
\ref{connection} we explicitly show that our uplifted solution lies within this framework. In particular, the
Riemann surface is taken to be an infinite strip with complex coordinate $w$ with
\begin{align}\label{eq:coordinate_transformation_Gutperletext}
w=x+i\psi\,, \end{align}
where $-\infty<x <\infty$ and $\psi\in [0,\pi/2]$.
The solution is completely specified by two harmonic functions on the strip which are given by
\begin{align}
h_1&=-i\frac{ L^2}{2 \sqrt{2}e^{\varphi_{(s)}}} \left(\sinh w-\sinh\bar{w}\right)=\frac{L^2}{\sqrt{2}e^{\varphi_{(s)}} }\cosh x \sin \psi\,,\nn
h_2&=\frac{ e^{\varphi_{(s)}}L^2}{4 \sqrt{2}} \left(e^{w}+e^{\bar{w}}\right)=\frac{e^{\varphi_{(s)}} L^2}{2 \sqrt{2}} e^{x}\cos \psi\,,
\label{eq:onesided_harmonic}
\end{align} 
and in comparing with \eqref{eq:one-sided Janus solutions proper distance} we should identify $x=\bar r/L$.

It is interesting to compare this solution with the  
supergravity solutions associated with the near horizon limit of a collection of $N_3$ D3-branes ending on $N_5$ coincident D5-branes. More specifically, we want $N_3=K N_5$ where $K \in \mathbb{Z}$, the linking number, is the same for all D5-branes.
From the results of \cite{DHoker:2007zhm,DHoker:2007hhe,Aharony:2011yc,Assel:2011xz,Assel:2012cj} we can write the harmonic functions
for such solutions as
\begin{align}\label{onesdiedjus}
h_1 &=\frac{\sqrt{\pi N_3} \ell_s^2 }{\sqrt{2 g_s}} \left( e^{x} \sin\psi +\frac{\sqrt{g_s N_3}}{2^{3/2} \sqrt{\pi} K} \log \left[
\frac{
2 g_sN_3^2 e^{2x}  +\pi K^2+ 2^{3/2}\sqrt{\pi g_s N_3}K \sin\psi e^{x}
}{
2 g_sN_3^2 e^{2x}  +\pi K^2- 2^{3/2}\sqrt{\pi g_s N_3}K \sin\psi e^{x}
}
 \right] \right), \nn
h_2 &=\frac{\sqrt{\pi g_s N_3}  \ell_s^2 }{\sqrt{2}}e^{x} \cos\psi\,,
\end{align}
where $g_s=e^{2\varphi_{(s)}}$ is the string coupling constant and $\ell_s$ is the string length.
In the large $x$ limit, as we approach the $\mathcal{N}=4$ SYM end, this solution
behaves as 
\begin{align}\label{exps}
h_1 &= \frac{\sqrt{2\pi N_3} \ell_s^2 }{\sqrt{g_s}}  \left( \cosh x \sin\psi 
 +
\frac{\pi K^2}{12 g_s N_3}e^{-3x}\sin 3\psi 
 +\cO(e^{-5x}) 
 \right)
 \,,\nn
 h_2 &=\frac{\sqrt{\pi g_s N_3}  \ell_s^2 }{\sqrt{2}}e^{x} \cos\psi\,.
\end{align}
Thus, after identifying the Einstein frame $AdS_5$ curvature $\sqrt{4\pi N_3}\ell_s^2=L^2$, as $x\to \infty$ we see that this solution
has the same asymptotic form as \eqref{eq:onesided_harmonic}, with sub-leading corrections.
Moreover, note that we also obtain the expansion \eqref{exps} by taking the limit $N_3\to\infty$ while
holding the linking number $K$ fixed.

\subsection{Other constructions}
It is straightforward to construct additional one-sided Janus solutions
numerically. In fact we have found no obstruction to constructing solutions that approach the $AdS_5$ vacuum at one end and
any of the $AdS_4\times\mathbb{R}$ solutions that we have discussed in the previous sections at the other end; namely the $\mathcal{N}=1,2$ S-fold solutions with
$\varphi$ a linear function, the more general S-fold solutions with $\varphi$ an LPP function or the periodic solution.
The one-sided Janus solutions approaching the S-folds with linear dilaton do not have any turning points. The solutions approaching the S-folds with either $\varphi$ an LPP function or the
periodic solution do have turning points, but the turning point data is not symmetric under the $\mathbb{Z}_2$ 
symmetry as we imposed for the solutions summarised in figures \ref{fig:equalmass_phi_tp} and \ref{fig:su2_phi_tp}.
All of these one-sided Janus solutions are regular.

\begin{figure}[h!]
\centering
{
\includegraphics[scale=0.28]{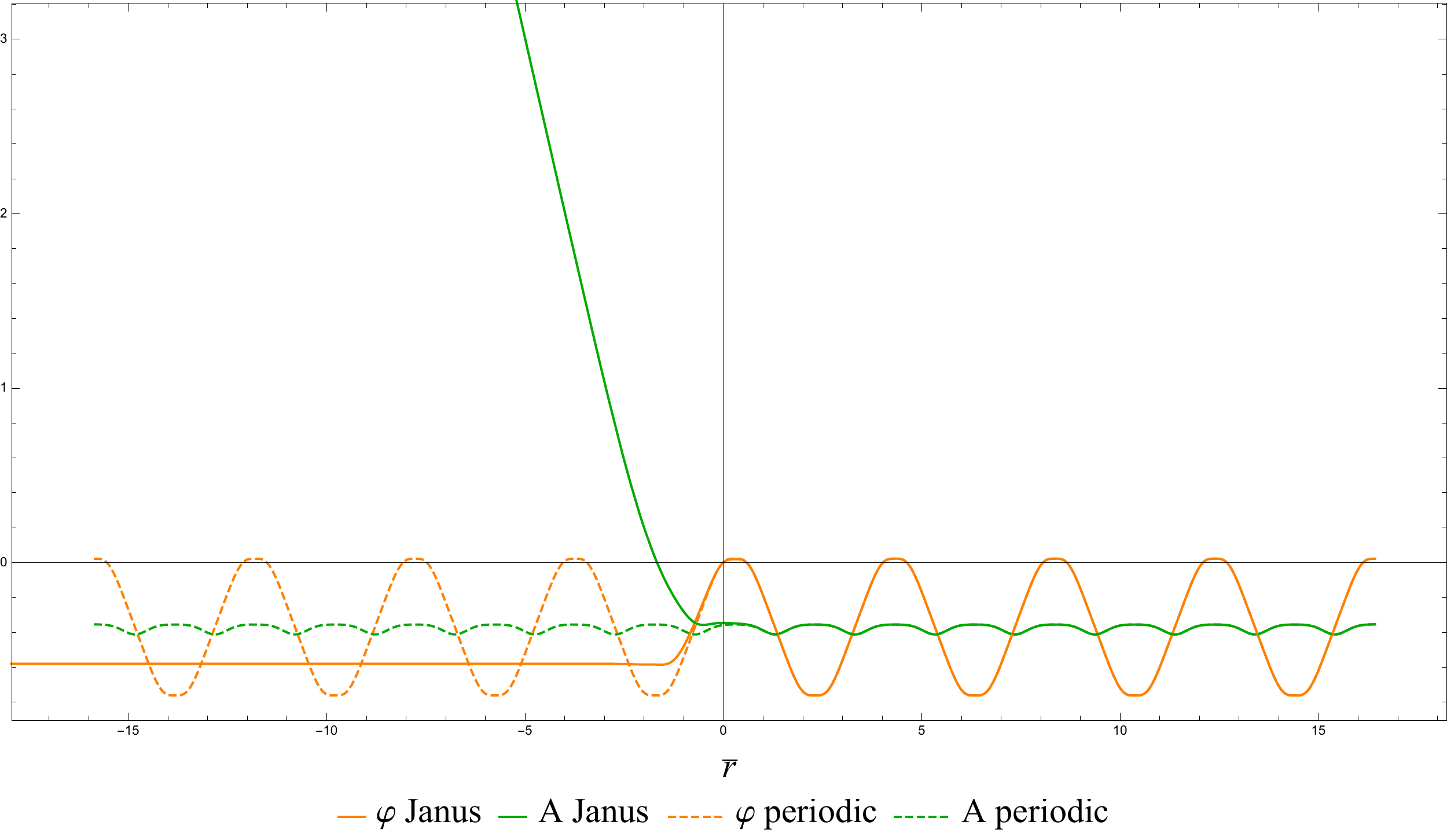}\qquad
\includegraphics[scale=0.28]{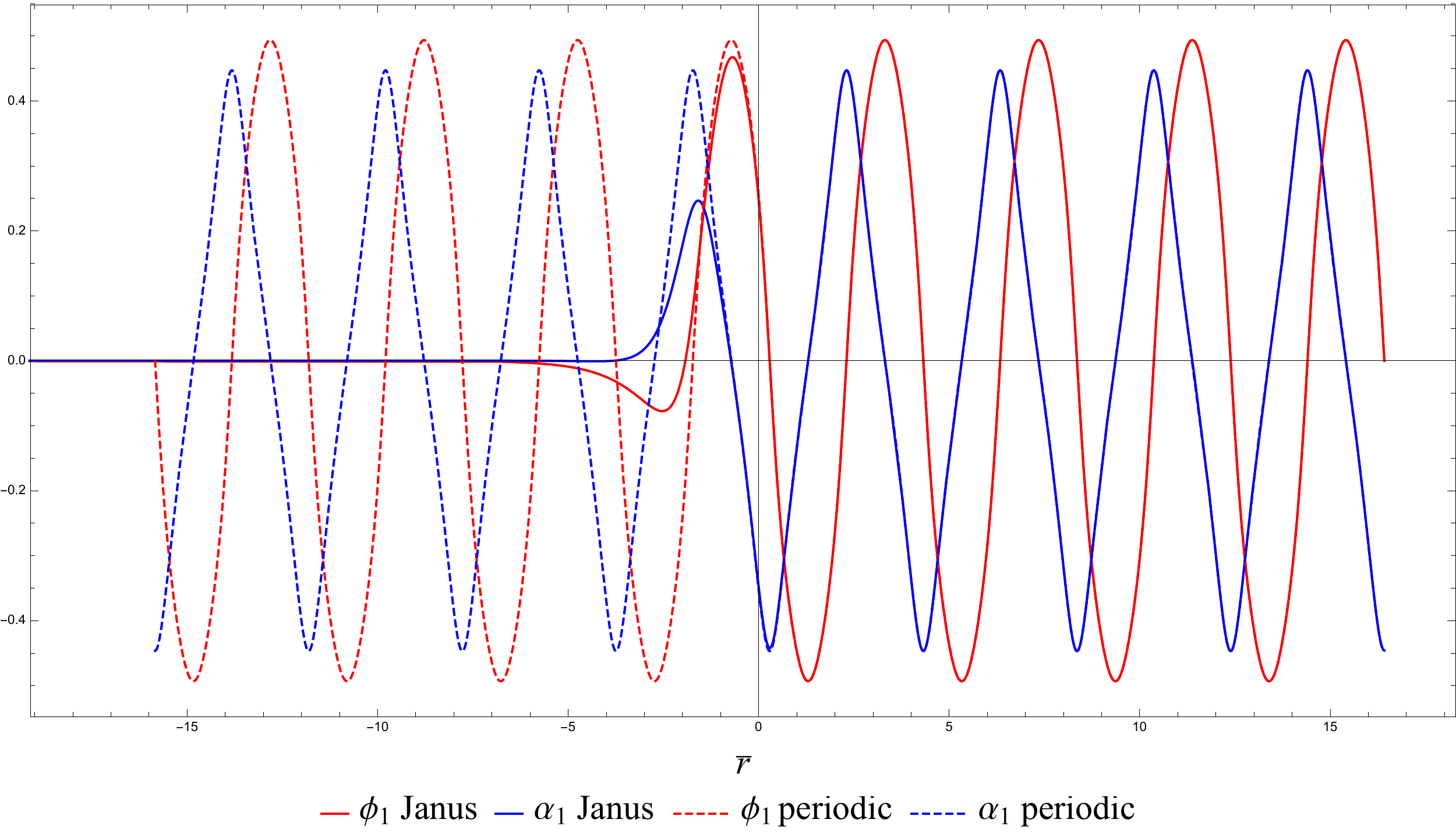}
}
\caption{A one sided Janus solution (solid lines) for the $\mathcal{N}=1^*$ equal mass $SO(3)$ invariant model
that approaches $AdS_5$ as $\bar r\to-\infty$ and approaches, very rapidly, the exactly periodic $AdS_4\times \mathbb{R}$ solution (dashed lines)
as $\bar r\to +\infty$. The left panel plots the behaviour of the warp factor $A$ and the $D=5$ dilaton
$\varphi$ and the right panel plots the scalar fields $\alpha_1,\phi_1$ (we have not plotted $\phi_4$ for clarity).}\label{onesidedj}
\end{figure}

To illustrate we have displayed in figure \ref{onesidedj} a solution constructed in
the $\mathcal{N}=1^*$ equal mass $SO(3)$ invariant model of section \ref{so3model} that approaches the
$AdS_5$ vacuum at $\bar r\to -\infty$ and the periodic $AdS_4\times\mathbb{R}$ solution at $\bar r\to+\infty$. 
Notice that this particular Janus solution has the feature that the dilaton $\varphi$ is bounded.

   \section{Discussion}
We have constructed a rich set of new S-fold solutions of type IIB string theory of the form $AdS_4\times S^1\times S^5$ which are dual to 
$\mathcal{N}=1$ SCFTs in $d=3$.
The solutions are patched together along the $S^1$ direction using a non-trivial $SL(2,\mathbb{Z})$ transformation in the hyperbolic conjugacy class.
The solutions are first constructed in $D=5$ gauged supergravity and then uplifted to $D=10$. In the previously known 
$AdS_4\times \mathbb{R}$ solutions associated with S-folds preserving $\mathcal{N}=1,2,4$ supersymmetry, the $D=5$ dilaton is a linear function of a coordinate on the
$\mathbb{R}$ direction. Crucially, in the new solutions the $D=5$ dilaton is now a linear plus periodic (LPP) function. 
We also showed that some of the new families of LPP $AdS_4\times \mathbb{R}$ solutions can be seen in a perturbative expansion about
the $\mathcal{N}=1$ S-fold solution with a linear dilaton. In addition, 
for the $SO(3)$ invariant model the numerical construction
of such solutions revealed additional branches of LPP $AdS_4\times \mathbb{R}$ solutions, not perturbatively connected with any known S-fold solutions.

An interesting feature of the new $AdS_4\times {S^1}\times S^5$ solutions is that we can make the
size of the $S^1$ parametrically larger than the size of the $S^5$, by carrying out the S-folding procedure after multiple periods with respect to the underlying periodic structure. This will gives rise to an interesting hierarchy of scaling dimensions in the $\mathcal{N}=1$ $d=3$ SCFT.
   
A proposal for the $\mathcal{N}=4$ SCFT in $d=3$ dual to the $\mathcal{N}=4$ S-folds of \cite{Inverso:2016eet} 
was given in \cite{Assel:2018vtq}. One takes the strongly coupled $[TU(N)]$ theory of \cite{Gaiotto:2008sd} 
and then gauges the global $U(N)\times U(N)$ global symmetry using an $\mathcal{N}=4$ vector multiplet. In addition one adds 
a Chern-Simons term at level $n$, where $n$ is the integer that is used to make the S-folding identifications
(see \eqref{keycond}). Proposals for the $\mathcal{N}=4$ SCFT in $d=3$ dual to the $\mathcal{N}=2$ S-folds of \cite{Guarino:2020gfe} 
were also discussed in \cite{Bobev:2020fon}. It would be very interesting to identify the 
$\mathcal{N}=1$ SCFTs in $d=3$ that are dual to the S-fold solutions of \cite{ Guarino:2019oct}, the new constructions in this paper,
as well as the periodic $AdS_4\times S^1\times S^5$ solution of \cite{Arav:2020obl}.
The small amount of supersymmetry makes this challenging, but one can hope that the connection with Janus solutions
which we have highlighted in this paper, as well as in \cite{Arav:2020obl}, will allow progress to be made.
   
We have seen that the periodic $AdS_4\times \mathbb{R}$ solution found in \cite{Arav:2020obl}, which uplifts to smooth 
$AdS_4\times S^1\times S^5$ of type IIB supergravity, is a rather exceptional solution in the general constructions of this paper. It would be very interesting to know whether or not there are additional such solutions of the form $AdS_d\times T^n\times M_k$ either in $D=10$ or $D=11$ supergravity.

We have focussed on constructing supersymmetric S-fold solutions, but one can also investigate non-supersymmetric possibilities. 
In fact a non-supersymmetric $AdS_4\times {\mathbb{R}}\times M_5$ solution of type IIB supergravity was discussed long ago in \cite{Robb:1984uj} and\cite{Quevedo:1985zx}. 
These solutions are associated with the $D=10$ dilaton linear in the $\mathbb{R}$ direction, and have been subsequently rediscovered
several times \cite{Freedman:2003ax,DallAgata:2011aa,Jain:2014vka,Guarino:2019oct}. However, 
in \cite{DallAgata:2011aa,Jain:2014vka,Guarino:2019oct}
it was argued that these solutions are unstable (in contrast to the claim in \cite{Robb:1984uj}) and hence are not of interest for S-folds with CFT duals.

Our constructions have also revealed a novel class of non-singular ``one-sided Janus" solutions preserving $\mathcal{N}$=1,2 or 4 supersymmetry.
These regular solutions approach the $AdS_5$ vacuum on one side and an $AdS_4\times\mathbb{R}$ solution with 
the $D=5$ dilaton a linear function of the radial coordinate or an LPP function. We also constructed a solution
that approaches the periodic $AdS_4\times \mathbb{R}$ solution of \cite{Arav:2020obl} on the other side, which is both regular and has
bounded dilaton.
For the solution that approaches the $\mathcal{N}= 4$ S-fold solution with linear dilaton we were able to construct an analytic solution. 
Using the results of \cite{DHoker:2007zhm,DHoker:2007hhe,Aharony:2011yc,Assel:2011xz,Assel:2012cj} we interpreted this solution as arising from D3-branes ending on D5-branes and it will be worthwhile to
investigate this in more detail.

It seems likely that it will be possible to construct additional LPP $AdS_4\times \mathbb{R}$ and one-sided Janus solutions 
within the 10 scalar truncation and more generally within the full $SO(6)$ gauged supergravity with 42 scalars. 
It may also be possible to construct new
type IIB solutions of the form $AdS_4\times S^1\times SE_5$, where $SE_5$ is a Sasaki-Einstein manifold, generalising
the work of \cite{Bobev:2019jbi}. More generally, one can try to construct non-geometric solutions of the form $AdS_d\times T^n\times M_k$, where $T^n$ is an $n$-dimensional torus and the solutions are patched together in the $T^n$ directions using U-duality transformations \cite{Hull:1994ys}.

\subsection*{Acknowledgments}
We thank Alessandro Tomasiello for helpful discussions.
This work was supported by STFC grant ST/T000791/1. KCMC is supported by an Imperial College President's PhD Scholarship. 
JPG is supported as a KIAS Scholar and as a Visiting Fellow at the Perimeter Institute. 
The work of CR is funded by a Beatriu de Pin\'os
Fellowship.

\appendix
\section{Uplifting to type IIB supergravity}\label{upliftform}
\subsection{The 10-scalar model in maximal gauged supergravity}
We first discuss how the 10-scalar model is obtained from maximal $SO(6)$ gauged supergravity in $D=5$. 
The 42 scalars of $SO(6)$ gauged supergravity parametrise the coset $E_{6(6)}/USp(8)$, with $USp(8)$ the maximal compact subgroup of $E_{6(6)}$. To describe this coset space, it is convenient to work in a basis for $E_{6(6)}$ that is adapted to its maximal subgroup $SL(6)\times SL(2,\mathbb{R})$, recalling that the gauge group $SO(6)\subset SL(6)$. Following \cite{Gunaydin:1984qu}, we write the generators of $E_{6(6)}$ in the fundamental ${\bf 27}$ representation in this basis as
\begin{equation}\label{eq:E6gen}
\mathbb{X} = 
\begin{pmatrix}
-4\Lambda_{[I}\,^{[M}\delta_{J]}\,^{N]} & \sqrt{2}\Sigma_{IJP\beta}\\
&\\
\sqrt{2}\Sigma^{MNK\alpha} & \Lambda_P\,^K\delta_\beta\,^\alpha + \Lambda_\beta\,^\alpha\delta_P\,^K
\end{pmatrix}\,,
\end{equation}
where the indices $I,J,... = 1,2,\ldots,6$, raised and lowered with $\delta_{IJ}$, label the fundamental of $SL(6)$, while the indices $\alpha,\beta,...=1,2$, raised and lowered with $\epsilon_{\alpha\beta}$, are $SL(2,\mathbb{R})$ indices. It is often convenient to consider $\mathbb{X}$ as a 27$\times27$ matrix associated with the branching of the fundamental of $E_{6(6)}$ under $SL(6)\times SL(2,\mathbb{R})$, like $\bf{27}\to (15,1) + (6,2)$. From this perspective, a fundamental index of $E_{6(6)}$, $A=1,2,\ldots 27$ splits according to $\{A\} = \{[IJ],I\alpha\}$, where $[IJ]$ are the 15 antisymmetric pairs of $SL(6)$ indices.

The non-compact part of this algebra is generated by the 20 symmetric, traceless $\Lambda_I\,^J \in SL(6)$, the 2 symmetric, traceless  $\Lambda_\alpha\,^\beta \in SL(2,\mathbb{R})$ and the 20 $\Sigma_{IJK\alpha}$ antisymmetric in $IJK$ and satisfying  $\Sigma_{IJK\alpha} = \frac{1}{6}\epsilon_{IJKLMN}\epsilon_{\alpha\beta}\Sigma^{LMN\beta}$. It is possible to choose a gauge for the coset element such that these 42 non-compact generators are in one-to-one correspondence with the scalar fields of the gauged supergravity. 

In this gauge, the truncation to the 10-scalar model discussed \cite{Bobev:2016nua}, retains the metric and the ten scalar fields 
$\{\beta_1,\beta_2,\bar\alpha_1,\bar\alpha_2,\bar\alpha_3,\bar\phi_1,\bar\phi_2,\bar\phi_3,\bar\phi_4,\bar\varphi\}$ defined by
\begin{align}\label{base1}
\Lambda_I\,^J &= \mathrm{diag}\left(\bar\alpha_1+\beta_1+\beta_2,-\bar\alpha_1+\beta_1+\beta_2, \bar\alpha_2+\beta_1-\beta_2, -\bar\alpha_2 +\beta_1-\beta_2,\bar\alpha_3 -2\beta_1,-\bar\alpha_3-2\beta_1 \right)\,,\nonumber\\
\Lambda_\alpha\,^\beta & =\mathrm{diag}\left(\bar\varphi,-\bar\varphi \right)\,,
\end{align}
and
\begin{align}\label{base2}
\Sigma_{1351}  = -\Sigma_{2462}  = &\frac{1}{2\sqrt{2}}\left(\bar\phi_1+\bar\phi_2+\bar\phi_3-\bar\phi_4\right)\,,\nonumber\\
\Sigma_{1461}  = -\Sigma_{2352}  = & \frac{1}{2\sqrt{2}}\left(-\bar\phi_1+\bar\phi_2+\bar\phi_3+\bar\phi_4\right)\,,\nonumber\\
\Sigma_{2361}  = -\Sigma_{1452} = & \frac{1}{2\sqrt{2}}\left(\bar\phi_1-\bar\phi_2+\bar\phi_3+\bar\phi_4\right)\,,\nonumber\\
\Sigma_{2451}  = -\Sigma_{1362} = &\frac{1}{2\sqrt{2}}\left(\bar\phi_1+\bar\phi_2-\bar\phi_3+\bar\phi_4\right)\,.
\end{align}
These barred scalar fields are non-linearly related to the unbarred scalar fields that we use in \eqref{zedintermsofscs}, however they do agree at linear order.
It is straightforward to demonstrate that the generators associated with this truncation generate $SO(1,1)^2\times SU(1,1)^4\subset E_{6(6)}$. 
Specifically, if we let $B^{(1)}$, $B^{(2)}$ each generate an $SO(1,1)$, and $S_{1,2,3}^{(A)}$ for $A = 1,2,3,4$ 
generate four commuting copies of $SU(1,1)$ satisfying
\begin{equation}
\big[S_1^{(A)},S_2^{(A)} \big]=2 S_3^{(A)}, \qquad \big[S_1^{(A)},S_3^{(A)} \big]=2 S_2^{(A)}, \qquad \big[S_2^{(A)},S_3^{(A)} \big]=-2 S_1^{(A)} \,,
\end{equation}
then we can explicitly identify the generators using table \ref{tab:su11x4}.

\begin{table}
\begin{center}
\begin{tabular}{c c c c c c c c c c c}
 & $B^{(1)}$ & $B^{(2)}$ & $S^{(1)}_1$ & $S^{(1)}_2$ & $S^{(2)}_{1}$ & $S^{(2)}_2$ & $S^{(3)}_1$ & $S^{(3)}_2$ & $S^{(4)}_1$ & $S^{(4)}_2$\\
\hline
\hline
$\bar\alpha_1$ & 0 & 0 &  $\frac{1}{2}$& 0 & $\frac{1}{2}$ &0 &$\frac{1}{2}$ &0 & $\frac{1}{2}$ & 0\\ 
$\bar\alpha_2$ & 0 & 0 & $\frac{1}{2}$ & 0 & $-\frac{1}{2}$ &0 &$\frac{1}{2}$ & 0 & $-\frac{1}{2}$ & 0\\
$\bar\alpha_3$ & 0 & 0 & $ \frac{1}{2}$ & 0 &$\frac{1}{2}$ &0&$-\frac{1}{2}$ & 0 & $-\frac{1}{2}$ & 0\\
$\bar\varphi$ & 0 & 0 & $\frac{1}{2}$ & 0 &$ -\frac{1}{2}$ & 0 &$-\frac{1}{2}$ &0 & $\frac{1}{2}$ & 0\\
$\bar\phi_1$ & 0 & 0 & 0 & $\frac{1}{2}$ & 0 &$\frac{1}{2}$& 0 & $\frac{1}{2}$ & 0 &$\frac{1}{2}$\\
$\bar\phi_2$ & 0 & 0 & 0 & $\frac{1}{2}$ & 0 & $-\frac{1}{2}$ & 0 & $\frac{1}{2}$ & 0 &$-\frac{1}{2}$\\
$\bar\phi_3$ & 0 & 0 & 0 & $\frac{1}{2}$  & 0 &$\frac{1}{2}$ & 0 &$-\frac{1}{2}$ & 0 & $-\frac{1}{2}$\\
$\bar\phi_4$ & 0 & 0 & $0$ & $-\frac{1}{2}$  & 0 & $\frac{1}{2}$ & 0 &$\frac{1}{2}$ &0& $-\frac{1}{2}$\\
$\beta_1$ & 1 & 0 & $0$ & $0$ & 0 &$0$ & 0 & 0 & 0 & 0\\
$\beta_2$ & 0 & 1 & 0 & 0 & $0$ & 0 & $0$ & $0$ & 0 & $0$
\end{tabular}
\end{center}
\caption{The non-compact generators of the $SO(1,1)^2\times SU(1,1)^4\subset E_{6(6)}$ algebra in the ${\bf 27}$ that are associated with the ten scalar truncation can be obtained from this table and \eqref{base1},\eqref{base2}.}\label{tab:su11x4}
\end{table}

The ten scalar fields which are retained in the truncated theory parametrise the coset $SO(1,1)^2\times \left[SU(1,1)/U(1)\right]^4$. It is
convenient to parametrise this coset in terms of two real scalars $\beta_{1,2}$ and four complex scalars $z^A$, which are functions
of the remaining scalars  
$\{\bar\alpha_1,\bar\alpha_2,\bar\alpha_3,\bar\phi_1,\bar\phi_2,\bar\phi_3,\bar\phi_4,\bar\varphi\}$, 
with the $z^A$ transforming linearly under the $U(1)\subset SU(1,1)$. To do this we first move to a basis 
for each of the $SU(1,1)$ algebras with definite $U(1)$ charge, by defining the generators
\begin{equation}
E^{(A)} = \frac{1}{2}\big( S^{(A)}_1+i S^{(A)}_2\big)\,,\qquad \mathrm{and} \qquad F^{(A)} = \frac{1}{2}\big(S^{(A)}_1-iS^{(A)}_2 \big).
\end{equation}
The desired parametrisation of the coset is then given by
\begin{equation}\label{veeexplicit}
\mathcal{V} = e^{\beta_1 B^{(1)}+\beta_2  B^{(2)}}\cdot \prod_{a}e^{s(|z^A|)\left(z^A E^{(A)}+\bar{z}^A F^{(A)} \right)}\,,
\end{equation}
where 
\begin{equation}
s(|z^A|) =\frac{1}{|z^A|}\mathrm{arcsech}\sqrt{1-|z^A|^2}.
\end{equation}
We will work with right cosets, in which $\mathcal{V}$ transforms from the left under global elements of $SO(1,1)^2\times SU(1,1)^4$ and from the right under local $U(1)^4$ rotations. The $U(1)^4$ invariant tensor defined by
\begin{equation}\label{eq:M10}
\mathcal{M} = \mathcal{V}\cdot\mathcal{V}^\dagger\,,
\end{equation}
can then be used to construct the kinetic terms for the scalar fields of the $D=5$ 10-scalar model via
\begin{equation}
\mathcal{L}_{10}^{(k)} = \frac{1}{96}\mathrm{tr}\left(\partial_\mu \mathcal{M}\partial^\mu\mathcal{M}^{-1} \right),
\end{equation}
as given in \eqref{bulkaction}.
It will also play a distinguished role in the uplift of this model to ten dimensions as we discuss below.

The scalar potential $\mathcal{P}$ of the 10-scalar model appearing in \eqref{bulkaction}
can be obtained from this coset representative using the general results 
for the form of the
scalar potential in the $SO(6)$ gauged supergravity given in \cite{Gunaydin:1984qu}. To do this, and following \cite{Gunaydin:1984qu}, 
it is helpful to change to a basis adapted to $USp(8)\subset E_{6(6)}$ using the antisymmetric hermitian gamma matrices of $\mathrm{Cliff}(7)$. An explicit representation is provided by the set of $8\times 8$ matrices $(\Gamma_0, \Gamma_I)$ given by
\begin{align}
\Gamma_0 & = -\sigma_2\otimes\sigma_3\otimes\sigma_3\,, \qquad
\Gamma_1  = \sigma_1\otimes\sigma_1\otimes\sigma_2\,, \nonumber\\
\Gamma_2 & = \sigma_3\otimes\sigma_1\otimes\sigma_2\,, \qquad\,\,\,\,
\Gamma_3  = -\sigma_2\otimes\sigma_1\otimes1\,, \nonumber\\
\Gamma_4 & = 1\otimes\sigma_2\otimes 1\,, \qquad\quad\,\,\,
\Gamma_5  = \sigma_2 \otimes \sigma_3\otimes\sigma_1\,, \nonumber\\
\Gamma_6 & = -1\otimes\sigma_3\otimes\sigma_2\,, 
\end{align}
where the $\sigma_{1,2,3}$ are Pauli matrices. From these one constructs
\begin{equation}\label{eq:USp8basis}
\Gamma_{IJ} =\frac{1}{2}\left[\Gamma_I,\Gamma_J\right] \qquad \mathrm{and} \qquad \Gamma^{I\alpha} = (\Gamma_I, i\Gamma_I\Gamma_0),
\end{equation}
whose ``spinor" indices $a, b$ are $USp(8)$ indices. In particular $(\Gamma_{IJ})^{ab}$ transforms in the $\bf{27}$ of $USp(8)$, indexed by the symplectic traceless index pairs $[ab]$. The symplectic trace is taken with respect to the invariant tensor
\begin{equation}\label{eq:symInv}
\Omega^{ab} = -\Omega_{ab} = -i\left(\Gamma_0\right)^{ab}.
\end{equation}
Introducing the notation 
\begin{equation}
\mathcal{V}_{A}\,^{ab} = \left(V_{IJ}\,^{ab},V^{I\alpha ab} \right)\,, \qquad \mathrm{and} \qquad 
\mathcal{V} = 
\begin{pmatrix}
U_{IJ}\,^{PQ} & U_{IJ, R\beta} \\
U^{K\alpha,PQ} & U^{K\alpha}\,_{R\beta}
\end{pmatrix}\,,
\end{equation}
for the coset representative in the $USp(8)$  and $SL(6)\times SL(2,\mathbb{R})$ bases, respectively,  one can use (\ref{eq:USp8basis}) to relate the two:
\begin{align}
V_{PQ}\,^{ab} & = \frac{1}{8}\big[\left(\Gamma_{IJ}\right)^{ab}U_{PQ}\,^{IJ}+2 \left(\Gamma^{I\alpha}\right)^{ab} U_{PQ,I\alpha} \big]\,, \nonumber\\
V^{K\alpha ab} & = \frac{1}{4\sqrt{2}}\big[\left(\Gamma_{IJ}\right)^{ab} U^{K\alpha, IJ}+2 \left(\Gamma^{I\beta}\right)^{ab} U^{K\alpha}\,_{I\beta} \big].
\end{align}
The $W$ tensors in \cite{Gunaydin:1984qu} are then given by
\begin{equation}\label{eq:Wtens}
W_{abcd} = \delta_{IJ}\epsilon_{\alpha\beta}V^{I\alpha a'b'} V^{J\beta c'd'}\Omega_{aa'}\Omega_{bb'}\Omega_{cc'}\Omega_{dd'}, \qquad W_{ab} = \Omega^{dc}W_{cadb}\,,
\end{equation}
and the scalar potential of the $SO(6)$ gauged supergravity is
\begin{equation}
\mathcal{P} = -\frac{g^2}{32}\left(2 W_{ab} W^{ab}-W_{abcd}W^{abcd} \right)\,,
\end{equation}
where $USp(8)$ indices are raised and lowered with the symplectic invariant (\ref{eq:symInv}) according to the rules implicit in (\ref{eq:Wtens}).
After substituting \eqref{veeexplicit}, using
\begin{align}\label{gell}
g=\frac{2}{L}\,,
\end{align}
and some calculation we obtain \eqref{peeform} for the 10-scalar truncation.

\subsection{The uplift to type IIB supergravity}
The uplift of the bosonic sector of the maximal gauged supergravity to type IIB supergravity is given in \cite{Baguet:2015sma}. 
The $D=10$ Einstein metric can be written in the form 
\begin{equation}
\dd s^2_{10} = \Delta^{-2/3}\left(\dd s^2_5 + G_{mn}\dd \theta^m\dd \theta^n \right),
\end{equation}
where $ds^2_5$ is the $D=5$ metric, $\theta^m$, $m=1,2,...,5$, parametrise $S^5$ and the metric $G_{mn}$ and the warp factor $\Delta$ are defined below.
The type IIB dilaton, $\Phi$, and axion, $C_0$, parametrise the coset $SL(2,\mathbb{R})/SO(2)$ and can be packaged in terms of a two-dimensional matrix via
\begin{equation}
m_{\alpha\beta} = 
\begin{pmatrix}
e^{\Phi}C_0\,^2+e^{-\Phi} & -e^{\Phi}C_0\\
 -e^{\Phi}C_0 & e^{\Phi}
\end{pmatrix},
\end{equation}
with $\det m=1$.
The remaining type IIB fields consist of 
two-form potentials $(A_{(2)}^1,A_{(2)}^2)$, which transform as 
an $SL(2,\mathbb{R})$ doublet and from which we identify the NS-NS two-form $B_{(2)}$ and the RR
two-form $C_{(2)}$ via
\begin{align}
B_{(2)}=A_{(2)}^1\,,\qquad 
C_{(2)}=A_{(2)}^2\,,
\end{align}
as well as the four-form potential $C_{(4)}$ that is associated with the
self-dual five-form flux as in \cite{Baguet:2015sma}.

We focus on uplifting the gravity-scalar sector of the $D=5$ theory for which the scalar matrix $\mathcal{M}$ introduced in (\ref{eq:M10}) 
plays a key role. In the $SL(6)\times SL(2,\mathbb{R})$ basis we can write the components of $\mathcal{M}$ and its inverse $\mathcal{M}^{-1}$ as
\begin{equation}
\mathcal{M} = 
\begin{pmatrix}
M_{IJ,PQ} & M_{IJ}\,^{R\beta}\\
M^{K\alpha}\,_{PQ} & M^{K\alpha,R\beta}
\end{pmatrix}\,,
\qquad 
\mathcal{M}^{-1} =
\begin{pmatrix}
M^{IJ,PQ} & M^{IJ}\,_{R\beta}\\
M_{K\alpha}\,^{PQ} & M_{K\alpha,R\beta}
\end{pmatrix}\,.
\end{equation}
We also introduce the round metric on the five-sphere, $\mathring{G}_{mn}$, with inverse $\mathring{G}^{mn}$.
We can write the Killing vectors of the round metric in terms of constrained coordinates $Y^I$ on $S^5$, satisfying $Y^I Y^I =1$, via
\begin{equation}
\mathcal{K}_{IJ}\,^m = -\frac{1}{L}\mathring{G}^{mn}Y_{[I}\partial_n Y_{J]}\,.
\end{equation}

In term of these quantities, the ten-dimensional fields of the uplifted $D=5$ gravity-scalar sector are given by
\begin{align} \label{eq:10daxiodil}
G^{mn} & = \mathcal{K}_{IJ}\,^m\mathcal{K}_{PQ}\,^n M^{IJ,PQ}\,,\nn
m^{\alpha\beta} & = \left(m_{\alpha\beta} \right)^{-1} = \Delta^{4/3}Y_I Y_J M^{I\alpha,J\beta}\,,\nn
A^{\alpha}_{mn} & = -L\epsilon^{\alpha\beta}G_{nk}\mathcal{K}_{IJ}^kM^{IJ}\,_{P\beta}\partial_m Y^P\,,\nn
C_{mnkl} & = \frac{L^4}{4}\left(\sqrt{\mathring{G}}\epsilon_{mnklp}\mathring{G}^{pq}\Delta^{4/3}m_{\alpha\beta}\partial_q\left(\Delta^{-4/3}m^{\alpha\beta} \right) +\mathring{\omega}_{mnkl} \right)\,,
\end{align}
where $\dd\mathring{\omega} = 16\mathrm{vol}_{S^5}$. Note that the $D=10$ warp factor $\Delta$ is defined implicitly using the 
fact that the axio-dilaton matrix (\ref{eq:10daxiodil}) satisfies $\det m=1$.

Restricting now to the 10-scalar model, we can illustrate the above formulae by writing down the components of the
axion and dliaton matrix:
\begin{align}
&\Delta^{-4/3}m^{11}=\nn
 & e^{2\beta_1+2\beta_2}\Big(\frac{(1+z^1)(1+\bar{z}^1)(1+z^4)(1+\bar{z}^4)}{(1-z^1\bar{z}^1)(1-z^4\bar{z}^4)} (Y_1)^2
+\frac{(1-z^2)(1-\bar{z}^2)(1-z^3)(1-\bar{z}^3)}{(1-z^2\bar{z}^2)(1-z^3\bar{z}^3)} (Y_2)^2\Big)\nonumber\\
&+e^{2\beta_1-2\beta_2}\Big(\frac{(1+z^1)(1+\bar{z}^1)(1-z^2)(1-\bar{z}^2)}{(1-z^1\bar{z}^1)(1-z^2\bar{z}^2)}(Y_3)^2
+\frac{(1-z^3)(1-\bar{z}^3)(1+z^4)(1+\bar{z}^4)}{(1-z^3\bar{z}^3)(1-z^4\bar{z}^4)}(Y_4)^2\Big)\nonumber\\
&+e^{-4\beta_1}\Big(\frac{(1+z^1)(1+\bar{z}^1)(1-z^3)(1-\bar{z}^3)}{(1-z^1\bar{z}^1)(1-z^3\bar{z}^3)}(Y_5)^2
+\frac{(1-z^2)(1-\bar{z}^2)(1+z^4)(1+\bar{z}^4)}{(1-z^2\bar{z}^2)(1-z^4\bar{z}^4)}(Y_6)^2 \Big)
\end{align}
\begin{align}
\Delta^{-4/3}m^{12} =
& e^{2\beta_1+2\beta_2}\Big(\frac{(z^2-\bar{z}^2)(z^3-\bar{z}^3)}{(1-z^2\bar{z}^2)(1-z^3\bar{z}^3)}-\frac{(z^1-\bar{z}^1)(z^4-\bar{z}^4)}{(1-z^1\bar{z}^1)(1-z^4\bar{z}^4)} \Big)Y_1 Y_2\nn
&+ e^{2\beta_1-2\beta_2}\Big(\frac{(z^1-\bar{z}^1)(z^2-\bar{z}^2)}{(1-z^1\bar{z}^1)(1-z^2\bar{z}^2)}-\frac{(z^3-\bar{z}^3)(z^4-\bar{z}^4)}{(1-z^3\bar{z}^3)(1-z^4\bar{z}^4)} \Big)Y_3Y_4\nonumber\\
& + e^{-4\beta_1}\Big(\frac{(z^1-\bar{z}^1)(z^3-\bar{z}^3)}{(1-z^1\bar{z}^1)(1-z^3\bar{z}^3)}-\frac{(z^2-\bar{z}^2)(z^4-\bar{z}^4)}{(1-z^2\bar{z}^2)(1-z^4\bar{z}^4)} \Big)Y_5 Y_6
\end{align}
\begin{align}
&\Delta^{-4/3}m^{22}  = \nn
&e^{2\beta_1+2\beta_2}\Big(\frac{(1+z^2)(1+\bar{z}^2)(1+z^3)(1+\bar{z}^3)}{(1-z^2\bar{z}^2)(1-z^3\bar{z}^3)} (Y_1)^2
 +\frac{(1-z^1)(1-\bar{z}^1)(1-z^4)(1-\bar{z}^4)}{(1-z^1\bar{z}^1)(1-z^4\bar{z}^4)} (Y_2)^2\Big)\nonumber\\
&+e^{2\beta_1-2\beta_2}\Big(\frac{(1+z^3)(1+\bar{z}^3)(1-z^4)(1-\bar{z}^4)}{(1-z^3\bar{z}^3)(1-z^4\bar{z}^4)}(Y_3)^2
+\frac{(1-z^1)(1-\bar{z}^1)(1+z^2)(1+\bar{z}^2)}{(1-z^1\bar{z}^1)(1-z^2\bar{z}^2)}(Y_4)^2\Big)\nonumber\\
&+e^{-4\beta_1}\Big(\frac{(1+z^2)(1+\bar{z}^2)(1-z^4)(1-\bar{z}^4)}{(1-z^2\bar{z}^2)(1-z^4\bar{z}^4)}(Y_5)^2
+\frac{(1-z^1)(1-\bar{z}^1)(1+z^3)(1+\bar{z}^3)}{(1-z^1\bar{z}^1)(1-z^3\bar{z}^3)}(Y_6)^2 \Big)
\end{align}

There are a number of additional sub-truncations of the 10-scalar model as summarised in figure \ref{truncdiag}.
In this paper we are particularly interested in the $SO(3)$ invariant 4-scalar model as well as the 
$SU(2)$ invariant 5-scalar model and their sub-truncations.

\subsubsection{The $SO(3)$ invariant 4-scalar model}
This truncation is obtained from the 10-scalar model by taking $\beta_1=\beta_2 =0$ and $z^4=-z^3=-z^2$. The truncation is invariant under $SO(3)\subset SU(3)\subset SO(6)$.
Similar to \cite{Bobev:2018eer} a useful parametrisation of the five-sphere adapted to this isometry is given by
\begin{equation}
\begin{pmatrix}
Y^1+i Y^2\\
Y^3+iY^4\\
 Y^5+iY^6
\end{pmatrix} =
e^{i\alpha}\cos\chi\mathcal{R}
\begin{pmatrix}
1\\
0\\
0\\
\end{pmatrix} + ie^{i\alpha} \sin\chi\mathcal{R}
\begin{pmatrix}
0\\
1\\
0
\end{pmatrix}.
\end{equation}
Here $0\le\alpha\le 2\pi$, $0\le\chi\le\pi/4$,
$\mathcal{R}=e^{\xi_1 g_1}e^{\omega g_2}e^{\xi_2 g_1}$ is an $SO(3)$ rotation matrix parametrised by three Euler angles
$\omega, \xi_1,\xi_2$
where $g_1,g_2$ are
the $3\times 3$ matrices
\begin{equation}
g_1 = e_{21}-e_{12}\,,\qquad \mathrm{and} \qquad g_2 = e_{31}-e_{13}\,,
\end{equation}
with $e_{ij}$ having a unit in the $i,j$ position and zeroes elsewhere.
In this parametrisation, the round metric on the five-sphere is written as a $U(1)$ fibration over $\mathbb{C}P^2$ as
\begin{equation}
\dd\mathring{\Omega}^2_5 = \dd s^2_{\mathbb{C}P^2}+\left(\dd\alpha-\sin2\chi \tau_3\right)^2\,,
\end{equation}
where
\begin{equation}\label{so3cp2}
\dd s^2_{\mathbb{C}P^2} = \dd\chi^2 +\sin^2\chi\,\tau_1^2+\cos^2\chi\,\tau_2^2+\cos^2 2\chi\,\tau_3^2\,,
\end{equation}
and the $\tau_{1,2,3}$ are locally left-invariant one-forms for $SO(3)$ given by
\begin{align}\label{lioneforms}
\tau_1 & = -\sin\xi_2\dd\omega+\cos\xi_2\sin\omega\dd\xi_1\,,\nonumber\\
\tau_2 & = \cos\xi_2\dd\omega+\sin\xi_2\sin\omega\dd\xi_1\,, \nonumber\\
\tau_3 & = \dd\xi_2+\cos\omega\dd\xi_1\,.
\end{align}
This parametrisation of $\mathbb{C}P^2$ is cohomogeneity-one with principle orbits actually given by $SO(3)/\mathbb{Z}_2\subset SU(3)$ (rather than $SO(3)$).
The singular orbits are an $\mathbb{RP}^2$ at $\chi=0$ and an $S^2$ at $\chi=\pi/4$ 
(see e.g. \cite{ziller2007geometry}).

After uplifting solutions in the $SO(3)$ invariant model, the ten dimensional metric will, in general have non-trivial dependence on
$\alpha$ and more general dependence on $\chi$ than that given in \eqref{lioneforms}
and the symmetry will be the $SO(3)/\mathbb{Z}_2$ associated with the $\tau_i$. For the further truncation to the $SU(3)$ invariant model in figure \ref{truncdiag}, the $\chi$ dependence will be as in \eqref{so3cp2}, giving rise to $SU(3)$
symmetry associated with $\mathbb{C}P^2$, but there will be non-trivial dependence on $\alpha$.

\subsubsection{The $SO(3)\times SO(3)$ invariant 3-scalar model}\label{so3so3model}
The $SO(3)\times SO(3)$ invariant sector has three scalars, and can be obtained from the $SO(3)$ invariant model
just discussed by setting $z_2=\bar z_2$. Specifically, we have
\begin{align}
z^1=\tanh\Big[\frac{1}{2}\big(3\alpha_1+\varphi-4i\phi_1\big)\Big]\,,\quad z^2=\tanh\Big[\frac{1}{2}\big(\alpha_1-\varphi\big)\Big]\,,
\end{align}
with $\beta_1=\beta_2=0$.
For this case we can parametrise the five-sphere
using the coordinates 
\begin{align}\label{coordsforn4}
Y_1&=\cos\psi \sin\theta\cos\xi,\quad Y_3=\cos\psi \sin\theta\sin\xi,\quad Y_5=\cos\psi\cos\theta\,,\nn
Y_2&=\sin\psi \sin\tilde \theta\cos\tilde\xi,\quad Y_4=\sin\psi \sin\tilde\theta\sin\tilde\xi,\quad Y_6=\sin\psi\cos\tilde\theta\,,
\end{align}
with $0\le \theta,\tilde{\theta}\le \pi$, $0\le\xi,\tilde{\xi}\le 2\pi$ and $0\le \psi\le \pi/2$.
In these coordinates the round metric on the five-sphere is given by
\begin{equation}\label{roundsph}
\dd \mathring{\Omega}_5^2 = \dd \psi^2 + \cos^2\psi\, \dd\Omega_2^2 +\sin^2\psi\,\dd\tilde{\Omega}_2^2\,,
\end{equation}
with $\dd\Omega_2^2 = \dd\theta^2+\sin^2\theta\dd\xi^2$ and  $\dd\tilde{\Omega}_2^2 = \dd\tilde{\theta}^2+\sin^2\tilde{\theta}\dd\tilde{\xi}^2$.
The $SO(3)\times SO(3)$ symmetry of the gauged supergravity model is generated by the Killing vectors for each of the round two-spheres.

For this model it will be useful to write down some additional uplifting formulae.
The $D=10$ metric takes the form
\begin{align}
\label{eq:10d_metric_SO3SO3}
ds_{10}^2&=
\Delta^{-2/3}\Big[ds^2_5+L^2\big(d\psi^2+\frac{d\Omega_2^2}{e^{4 \alpha_1} \sec 4 \phi_1+\tan ^2\psi }
+\frac{d\tilde{\Omega}_2^2}{e^{-4 \alpha_1} \sec 4 \phi_1+\cot ^2\psi }\big)\Big]\,,
\end{align}
with the $D=10$ warp factor given below.
The axion-dilaton matrix is diagonal with
\begin{align}
\label{eq:axion_dilaton_SO3SO3}
m^{11}=&\Delta^{4/3}\Big[\cos^2\psi\frac{(1+z^1)(1+\bar{z}^1)(1-z^2)^2}{(1-\left|z^1\right|^2)(1-(z^2)^2)}+\sin^2\psi\frac{(1-z^2)^4}{(1-(z^2)^2)^2}\Big]\,,\nn
=&\Delta^{4/3}\left[e^{2 \varphi-2 \alpha_1} \sin ^2\psi  + e^{2 \alpha_1+2\varphi} \sec 4 \phi_1\cos ^2\psi \right]\,,
\nn
m^{22}=&\Delta^{4/3}\Big[\sin^2\psi\frac{(1-z^1)(1-\bar{z}^1)(1+z^2)^2}{(1-\left|z^1\right|^2)(1-(z^2)^2)}+\cos^2\psi\frac{(1+z^2)^4}{(1-(z^2)^2)^2}\Big]\,,\nn
=&\Delta^{4/3}\left[e^{-2 \varphi-2 \alpha_1}\sec 4 \phi_1 \sin ^2\psi  + e^{2 \alpha_1-2\varphi} \cos ^2\psi \right]\,,
\end{align}
and $m^{12}=m^{21}=0$,
where the $D=10$ warp factor is given by
\begin{align}
\Delta^{4/3}=\frac{e^{2\alpha_1}\sec^2\psi}{\sqrt{\left(e^{4 \alpha_1} \sec 4 \phi_1+\tan ^2\psi \right) \left(e^{4 \alpha_1}+\tan ^2\psi  \sec 4 \phi_1\right)}}\,.
\end{align}
Thus, we have vanishing axion, $C_0=0$, and $e^{\Phi}=m^{11}$.

The NS-NS and R-R two-forms are found to be
\begin{align}
B_{(2)}&=L^2\frac{ i \sin ^3\psi \,(z^2-1) (z^1-\bar{z}^1)}{ \Pi_1}\,\text{vol}_{\tilde{S}^2}\,,\nn
C_{(2)}&=-L^2\frac{ i  \cos ^3\psi  \,(z^2+1) (z^1-\bar{z}^1)}{ \Pi_2}\,\text{vol}_{S^2}\,,
\end{align}
where 
\begin{align}
\Pi_1&=z^1 \left[(z^2-1) \sin ^2\psi-\bar{z}^1 (z^2+\cos 2 \psi )\right]+z^2 \cos 2 \psi +\bar{z}^1 (z^2-1) \sin ^2\psi+1\,,\nn
\Pi_2&=z^1 \left[(z^2+1) \cos ^2\psi+\bar{z}^1 (z^2+\cos 2 \psi )\right]+z^2 \cos 2 \psi +\bar{z}^1 (z^2+1) \cos ^2\psi+1\,,
\end{align}
and
$\text{vol}_{S^2}=\sin\theta \,d\theta\wedge d\xi$, $\text{vol}_{\tilde{S}^2}=\sin\tilde\theta\, d\tilde\theta\wedge d\tilde\xi$.
Finally, the four-form potential is given by
\begin{align}
C_{(4)}=&\frac{L^4}{4}\hat{\omega}-\frac{L^4}{8} \sin^3 2 \psi\Big(
\frac{z^1 (z^2+2 \bar{z}^1-2)+z^2 (\bar{z}^1-4)-2 \bar{z}^1}{-3\Pi_1-\Pi_2+z^1 (1+z^2-\bar{z}^1z^2)+(z^2+1) \bar{z}^1-4}\nn
&\qquad
+\frac{z^1 (z^2+2 \bar{z}^1+2)+z^2 (\bar{z}^1+4)+2 \bar{z}^1}{\Pi_1+3\Pi_2 -z^1z^2 (\bar{z}^1+1)+z^1-z^2 \bar{z}^1+\bar{z}^1+4}\Big) 
\text{vol}_{S^2}\wedge \text{vol}_{\tilde{S}^2}\,,
\end{align}
where the four-form $\hat{\omega}$ is given by
\begin{align}
\hat{\omega}=\big(2\psi-\frac{1}{2}\sin4\psi\big)\text{vol}_{S^2}\wedge \text{vol}_{\tilde{S}^2},
\end{align}
and satisfies $d\hat{\omega}=16\text{vol}_{S^5}$, where the volume form is with respect to the round metric \eqref{roundsph}.

\subsubsection{The $SU(2)$ invariant 5-scalar model}
This truncation is obtained from the 10-scalar model by taking $\beta_2 =0$, $z^4=-z^2$ and $z^3 = -z^1$. The resulting truncation is invariant under $SU(2)\subset SU(3) \subset SO(6)$. To parametrise the five-sphere so that this symmetry is manifest,
similar to  \cite{Bobev:2020fon} one can define
\begin{align}
Y^1+i Y^2 & = e^{\frac{i}{2}(\xi_1+\xi_2)}\sin\rho\cos(\omega/2)\,,\nonumber\\
Y^3+iY^4 & = e^{\frac{i}{2}(-\xi_1+\xi_2)}\sin\rho\sin(\omega/2)\,,\nonumber\\
Y^5+iY^6& = e^{i\alpha}\cos\rho\,,
\end{align}
with $\omega, \xi_1,\xi_2$ Euler angles of $SU(2)$ with
\begin{align}\label{eulerangs2}
0\le\omega\le\pi,\quad 0\le\xi_1\le 2\pi,\quad 0\le\xi_2<4\pi\,,
\end{align}
and $0\le\rho\le\pi/2$, $0\le\alpha\le2\pi$.
In these coordinates the metric on the round sphere takes the form
\begin{equation}\label{su25}
\dd\mathring{\Omega}^2_5 = \dd\rho^2+\cos^2\rho\dd\alpha^2+\frac{1}{4}\sin^2\rho\left(\tau_1^2+\tau^2_2 +\tau^2_3 \right)\,,
\end{equation}
where the $\tau_i$ are $SU(2)$ left-invariant forms given in \eqref{lioneforms}.
The $SU(2)$ symmetry then corresponds to the Killing vector fields associated with the $SU(2)$ action. In general
$\partial_\alpha$ will not be a Killing vector of the uplifted solutions of the $SU(2)$ invariant 5-scalar model and furthermore, the
coefficients of the $\tau_i$ will differ from that of \eqref{su25}.

We can also write $\xi_2=2\alpha+\gamma$ so that
\begin{align}
Y^1+i Y^2 & = e^{i\alpha+\frac{i}{2}\xi_1+\frac{i}{2}\gamma}\sin\rho\cos(\omega/2)\,,\nonumber\\
Y^3+iY^4 & = e^{i\alpha-\frac{i}{2}\xi_1+\frac{i}{2}\gamma}\sin\rho\sin(\omega/2)\,,\nonumber\\
Y^5+iY^6& = e^{i\alpha}\cos\rho\,.
\end{align}
We then have
\begin{equation}
\dd\mathring{\Omega}^2_5 = \dd s^2_{\mathbb{C}P^2}+\big(\dd\alpha+\frac{1}{2}\sin^2\rho \tau_3 \big)^2\,,
\end{equation}
where
\begin{equation}\label{su25alt}
\dd s^2_{\mathbb{C}P^2} = \dd\rho^2 +\frac{1}{4}\sin^2\rho(\tilde\tau_1^2+\tilde\tau_2^2)+\frac{1}{16}\sin^2 2\rho\,\tilde\tau_3^2\,,
\end{equation}
and the $\tilde\tau_{1,2,3}$ are left-invariant one-forms for $SU(2)$
\begin{align}\label{lioneformsold}
\tilde\tau_1 & = -\sin\gamma\dd\omega+\cos\gamma\sin\omega\dd\xi_1\,,\nonumber\\
\tilde\tau_2 & = \cos\gamma\dd\omega+\sin\gamma\sin\omega\dd\xi_1\,, \nonumber\\
\tilde\tau_3 & = \dd\gamma+\cos\omega\dd\xi_1\,.
\end{align}
For the uplift of the $SU(2)$ invariant 5-scalar model, the metric will in general depend 
on $\alpha$ and moreover the extra $U(1)$ associated with rotating 
$\tilde\tau_1$ into $\tilde\tau_2$ that is manifest in \eqref{su25alt} will no longer be present.
Moving to the $SU(3)$ truncation in figure \ref{truncdiag} the uplifted metric will have a 
$\mathbb{C}P^2$ factor, as in \eqref{su25alt}, giving rise to the $SU(3)$ symmetry but there
will be dependence on $\alpha$, in general. Moving instead to the $SU(2)\times U(1)$
invariant truncation in figure \ref{truncdiag} the uplifted metric will in general have dependence
on $\alpha$, and the $U(1)$ associated with rotating 
$\tilde\tau_1$ into $\tilde\tau_2$ that is manifest in \eqref{su25alt} will be present.

\subsection{The $SL(2,\mathbb{R})$ action in five and ten dimensions}
Both the $D=5$ maximal gauged supergravity and the type IIB supergravity are invariant under global $SL(2,\mathbb{R})$ transformations. 
Focussing on the gravity and scalar sector of the $D=5$ theory the relationship between the two $SL(2,\mathbb{R})$ transformations can be made explicit
using uplift formulae in (\ref{eq:10daxiodil}).

Consider first the $D=5$ theory in which the $SL(2,\mathbb{R})\subset E_{6(6)}$ can be generated by the $\mathbb{X}$ of (\ref{eq:E6gen}) with
$\Lambda_\alpha\,^\beta$ a linear combination of the three matrices $\left(\Lambda^i\right)_\alpha{}^\beta$ given by
\begin{equation}
\left(\Lambda^1\right)_\alpha{}^\beta  = (\sigma^1)_\alpha{}^\beta,\quad
\left(\Lambda^2\right)_\alpha{}^\beta  = (\sigma^3)_\alpha{}^\beta,\quad
\left(\Lambda^3\right)_\alpha{}^\beta  =(-i\sigma^2)_\alpha{}^\beta,\quad 
\end{equation}
Explicitly, in terms of the 27 dimensional representation the $SL(2,\mathbb{R})$ generators are thus
\begin{align}
\mathbb{X}^i\big |_{SL(2,\mathbb{R})} = 
\begin{pmatrix}
0_{15\times 15} & & & & & &\\
& \left(\Lambda^i\right)_\alpha{}^\beta & & & & &\\
& & \left(\Lambda^i \right)_\alpha{}^\beta & & & &\\
& & & \left(\Lambda^i \right)_\alpha{}^\beta & & &\\
& & & & \left(\Lambda^i \right)_\alpha{}^\beta  & &\\
& & & & & \left(\Lambda^i \right)_\alpha{}^\beta &\\
& & & & & & \left(\Lambda^i \right)_\alpha{}^\beta
\end{pmatrix}\,.
\end{align}
A finite $SL(2,\mathbb{R})$ transformation in the $D=5$ theory, using the $i$th generator, can then be written $\mathcal{S}^i_{(5)} = e^{c\mathbb{X}^i |_{SL(2,\mathbb{R})} }$ where $c$ is constant. This transformation acts on the scalar matrix $\mathcal{M}$ given in (\ref{eq:M10}) via
\begin{equation}
\mathcal{M}\to\mathcal{M}' = \mathcal{S}_{(5)}^i \cdot \mathcal{M}\cdot \mathcal{S}_{(5)}^i\,^T.
\end{equation}
From this one can infer the corresponding transformation of the scalars parametrising the coset which, in general, is non-linear. 
For the specific case of the transformation associated with the $i=3$ generator, one finds the following action on the ten-scalar model:
\begin{align}
\beta_1\to &  \beta_1\,,\qquad \beta_2\to   \beta_2\,,\nn
z^1\to&  \frac{z^1 + \tanh\frac{c}{2}}{1+\tanh\frac{c}{2}z^1}\,,\qquad 
z^2\to  \frac{z^2 - \tanh\frac{c}{2}}{1-\tanh\frac{c}{2}z^2}\,,\nonumber\\
z^3\to&  \frac{z^3 - \tanh\frac{c}{2}}{1-\tanh\frac{c}{2}z^3}\,,\qquad
z^4\to   \frac{z^4 + \tanh\frac{c}{2}}{1+\tanh\frac{c}{2}z^4}.
\end{align}
From \eqref{zedintermsofscs} one can conclude that this transformation is equivalent to a simple 
shift in the five dimensional field $\varphi\to  \varphi +c$. Also note that the $SL(2,\mathbb{R})$ transformations
associated with the $i=1,3$ generators take us outside the 10-scalar truncation and will not play a role in this paper.

We now turn to the $SL(2,\mathbb{R})$ action in $D=10$. From (\ref{eq:10daxiodil}) we can conclude that the $D=5$ transformation by the element
$\mathcal{S}_{(5)}^i$ is equivalent to a transformation by 
\begin{equation}
\mathcal{S}_{(10)}^i = e^{c  \left(\Lambda^i\right)_\alpha\,^\beta }\,,
\end{equation}
in the $D=10$ theory. For example, and of most interest, the transformation associated with the $i=2$ generator
gives rise to
 \begin{equation}
m^{-1}\to m'\,^{-1} = \mathcal{S}_{(10)}^2\cdot m^{-1}\cdot \mathcal{S}_{(10)}^2\,^T \,,
\end{equation}
This transformation is equivalent to 
\begin{equation}
m_{\alpha\beta}\to m'_{\alpha\beta}=
\begin{pmatrix}
e^{-2c}m_{11} & m_{12} \\
m_{12} & e^{2c} m_{22}
\end{pmatrix}\,,
\end{equation}
and translates, in turn, into the following simple transformation of the $D=10$ dilaton and axion:
\begin{equation}
\Phi \to \Phi+2c \qquad \mathrm{and} \qquad C_0 \to e^{-2c}C_0\,.
\end{equation}
The transformation by $\mathcal{S}^2_{(10)}$ plays a key role for our solutions, as it allows one to S-fold
the $D=5$ solutions, as we discuss in the text (note that we call this transformation simply $\mathcal{S}$ in \eqref{lambdaexp}).

In checking that the S-fold procedure we employ does not break supersymmetry it is also useful
to see how an $\mathcal{S}_{(5)}^2\in SL(2,\mathbb{R})$ transformation acts on the $D=5$ supersymmetry parameters.
A transformation by any element of the $E_{6(6)}$ global symmetry group is associated with a local compensating
$USp(8)$ transformation, $\mathcal{H}$, which acts on the fermions. For the action of $\mathcal{S}^2_{(5)}$ 
we find that $\mathcal{H}\in U(1)^4\subset USp(8)$, in the fundamental representation, is explicitly given by
\begin{align}
\tiny
\mathcal{H}=\left(
\begin{array}{cccccccc}
\frac{k_1+\bar{k}_1}{2} & 0 & 0 & 0 & \frac{\bar{k}_1-k_1}{2} & 0 & 0 & 0 \\
 0 &\frac{k_2+\bar{k}_2}{2}& 0 & 0 & 0 & \frac{\bar{k}_2-k_2}{2} & 0 & 0 \\
 0 & 0 &\frac{k_3+\bar{k}_3}{2}  & 0 & 0 & 0 & \frac{\bar{k}_3-k_3}{2} & 0 \\
 0 & 0 & 0 & \frac{k_4+\bar{k}_4}{2}   & 0 & 0 & 0 & \frac{\bar{k}_4-k_4}{2} \\
 \frac{\bar{k}_1-k_1}{2}& 0 & 0 & 0 &\frac{k_1+\bar{k}_1}{2}  & 0 & 0 & 0 \\
 0 &  \frac{\bar{k}_2-k_2}{2} & 0 & 0 & 0 & \frac{k_2+\bar{k}_2}{2}  & 0 & 0 \\
 0 & 0 & \frac{\bar{k}_3-k_3}{2} & 0 & 0 & 0 &\frac{k_3+\bar{k}_3}{2}  & 0 \\
 0 & 0 & 0 &  \frac{\bar{k}_4-k_4}{2}& 0 & 0 & 0 & \frac{k_4+\bar{k}_4}{2}  \\
\end{array}
\right)
\end{align}
with
\begin{align}
k_1&=\Big({\frac{g_1g_2g_3g_4}{\bar{g}_1\bar{g}_2\bar{g}_3\bar{g}_4}}\Big)^{1/4}\,,\qquad\,\,
k_2=\Big({\frac{\bar{g}_1g_2\bar{g}_3g_4}{g_1\bar{g}_2g_3\bar{g}_4}}\Big)^{1/4}\,,\nn
k_3&=\Big({\frac{\bar{g}_1\bar{g}_2g_3g_4}{g_1g_2\bar{g}_3\bar{g}_4}}\Big)^{1/4}\,,\qquad
k_4=\Big({\frac{g_1\bar{g}_2\bar{g}_3g_4}{\bar{g}_1g_2g_3\bar{g}_4}}\Big)^{1/4}\,,
\end{align}
and
\begin{align}
g_1&=1+\tanh\left(c/2\right)z^1,\qquad g_2=1-\tanh\left(c/2\right)z^2\,,\nn
 g_3&= 1-\tanh\left(c/2\right)z^3,\qquad g_4= 1+\tanh\left(c/2\right)z^4\,.
\end{align}
The action on the supersymmetry parameters $\varepsilon$ can be seen by diagonalising the $W$-tensor $W_{ab}$ of $D=5$ gauged
supergravity \eqref{eq:Wtens} and restricting $\varepsilon^a$ to lie within the space spanned by the eigenvectors of $W_{ab}$ with eigenvalues $e^{\mathcal{K}/2}\overline{\mathcal{W}}$ (1st) and $e^{\mathcal{K}/2}\mathcal{W}$ (5th). In this basis the $USp(8)$ transformation is found to be
\begin{align}
\hat{\mathcal{H}}=\text{diag}\left(k_1,k_2,k_3,k_4,\bar{k}_1,\bar{k}_2,\bar{k}_3,\bar{k}_4\right)\,.
\end{align}

The dilaton shift action can also be seen as a K\"ahler transformation acting in the $D=5$ theory, as noted in \cite{Arav:2020obl}. 
Under
$\varphi\rightarrow\varphi+c$ we have
$\mathcal{K}\to \mathcal{K}+f+\bar f$ and $\mathcal{W}\to e^{-f}\mathcal{W}$ with $f=f(z^A)$ given by
\begin{align}
e^f=\cosh^4(c/2)g_1g_2g_3g_4\,.
\end{align}
Under this transformation
the preserved supersymmetries of the BPS equations transform as 
$\varepsilon_1\to e^{(f-\bar f)/4}\varepsilon_1$ and $\varepsilon_2\to e^{-(f-\bar f)/4}\varepsilon_2$
i.e.
$\varepsilon_1\to k_1\varepsilon_1$ and $\varepsilon_2\to \bar k_1\varepsilon_2$. This shows
that the dilaton shift is realised by an $SL(2,\mathbb{R})$ transformation that is also acting as
an $SL(2,\mathbb{R})$ transformation on the preserved supersymmetries. This allows us to conclude that
the S-folding procedure will preserve the supersymmetry of the $D=5$ solutions as noted in the text.

\subsection{The $\mathcal{N}=4$ one-sided Janus solution in type IIB}\label{connection}
Here we show that the one-sided Janus solution \eqref{eq:one-sided Janus solutions proper distance}, after being uplifted to $D=10$, can be cast
into the form of the general $AdS_4$ solutions of type IIB which preserve $\mathcal{N}=4$ supersymmetry 
\cite{DHoker:2007zhm,DHoker:2007hhe}.

In \cite{DHoker:2007zhm,DHoker:2007hhe} they consider the type IIB Einstein metric written in the form
\begin{equation}
\dd s^2 = f_4^2\dd s^2_{AdS_4} + f_1^2 \dd\Omega_2^2 +f_2^2\dd\tilde{\Omega}_2^2+ds^2(\Sigma)\,,
\end{equation}
where $ds^2(\Sigma)$ is the metric on a Riemann surface. 
Introducing a complex coordinate $w$ on $\Sigma$ we write 
\begin{align}
ds^2(\Sigma)= 4\rho^2\dd w\dd \bar{w}\,,
\end{align}
where
$\rho$  as well as $f_1,f_2,f_4$ are functions of $w,\bar w$.
To specify a solution in the language of \cite{DHoker:2007zhm}, it is sufficient to provide two harmonic functions on the Riemann surface, $h_1$, $h_2$.
To do so, as in \cite{VanRaamsdonk:2020djx}, one can introduce the real functions
\begin{align}
W&\equiv \partial_w h_1\partial_{\bar{w}}h_2+\partial_w h_2\partial_{\bar{w}}h_1\,,\nn
N_1&\equiv 2h_1h_2\left|\partial_w h_1\right|^2-h_1^2W\,,\nn
N_2&\equiv 2h_1h_2\left|\partial_w h_2\right|^2-h_2^2W\,.
\end{align}
Then, for example, the $D=10$ dilaton $\Phi$ is given by
\begin{align}
e^{2\Phi}&=\frac{N_2}{N_1}\,,
\end{align}
while the metric functions have the form
\begin{align}
\rho^8&=\frac{W^2}{h_1^4h_2^4}N_1N_2\,,\qquad\qquad
f_1^2=2e^{\frac{\Phi}{2}}h_1^2\sqrt{-\frac{W}{N_1}}\,,\nn
f_2^2&=2e^{-\frac{\Phi}{2}}h_2^2\sqrt{-\frac{W}{N_2}}\,,\qquad
f_4^2=2e^{-\frac{\Phi}{2}}\sqrt{-\frac{N_2}{W}}\,.
\end{align}

To connect with the uplifted one sided Janus solution \eqref{eq:one-sided Janus solutions proper distance} we take
the Riemann surface to be an infinite strip and write
\begin{align}\label{eq:coordinate_transformation_Gutperle}
w=\frac{\bar r }{L}+i\psi\,, \end{align}
with $-\infty<\bar r <\infty$ and $\psi\in [0,\pi/2]$.
We then take the harmonic functions to be
\begin{align}
h_1&=-i\frac{ e^{-\varphi_{(s)}}L^2}{2 \sqrt{2}} \left(\sinh w-\sinh\bar{w}\right)=\frac{e^{-\varphi_{(s)}}L^2}{\sqrt{2}} \cosh \frac{\bar r}{L} \sin \psi\,,\nn
h_2&=\frac{ e^{\varphi_{(s)}}L^2}{4 \sqrt{2}} \left(e^{w}+e^{\bar{w}}\right)=\frac{ e^{\varphi_{(s)}}L^2}{2 \sqrt{2}} e^{\bar  r/L}\cos \psi\,,
\end{align}
and hence
\begin{align}
W&=-\frac{L^4}{16}\sin 2\psi\,,\nn
N_1&=\frac{e^{-2\varphi_{(s)}}L^8}{256}  \sin 2\psi  \left(1+e^{-2{\bar r}/L}\right) \left(2+2 e^{2{\bar r}/L} \cos ^2\psi +e^{4{\bar r}/L}-\cos 2 \psi \right)\,,\nn
N_2&=\frac{e^{2\varphi_{(s)}}L^8}{256} e^{2{\bar r}/L} \sin 2 \psi  \left(2+e^{2{\bar r}/L}+\cos 2 \psi \right)\,.
\end{align}

With a little effort we can show that this agrees with the uplift of \eqref{eq:one-sided Janus solutions proper distance}
after using the results\footnote{In fact to get an exact match with the metric and also
for the two-form and four-form potentials in \eqref{coordsforn4} we should relabel
$\tilde \theta\to \pi-\theta$, $\tilde \xi\to \xi+\pi$ as well as
$\theta\to \tilde \theta$, $\xi\to \tilde \xi$, so that 
$d\Omega_2\leftrightarrow d\tilde\Omega_2$ and 
$\text{vol}_{\tilde S^2}\to -\text{vol}_{S^2}$
and
$\text{vol}_{S^2}\to \text{vol}_{\tilde S^2}$.} in section \ref{so3so3model}. For example, in both cases the $D=10$ dilaton is given by
\begin{align}
\label{eq:dilaton_Gutperle}
e^{2\Phi}=\frac{e^{4\varphi_{(s)}}e^{4{\bar r}/L} \left(2+e^{2{\bar r}/L}+\cos 2 \psi \right)}{\left(1+e^{2{\bar r}/L}\right) \left(2+2 e^{2{\bar r}/L} \cos ^2\psi +e^{4{\bar r}/L}-\cos 2 \psi \right)}\,.
\end{align}
Notice that
as $\bar r\to\infty$, where the solution approaches the $AdS_5$ vacuum, we have $e^{2\Phi}\to e^{4\varphi_{(s)}}$
while as $\bar r\to-\infty$ we have $e^{2\Phi}\to 0$.

\section{Holographic Renormalisation}\label{appren}
The holographic renormalisation for the 10-scalar truncation was discussed in detail in \cite{Arav:2020obl} (see also the closely related discussion 
in \cite{Bobev:2016nua}). The counter term action required to remove all divergences was given in (B.7) of \cite{Arav:2020obl}. In
addition a set of finite counter terms was given in (B.8) of \cite{Arav:2020obl}, invariant under the discrete symmetries \eqref{genz2}-\eqref{gens33},
which depends on 14 constant ``$\delta$-coefficients" (in particular it was assumed that they are independent of sources for $\varphi$). 
By analysing the conditions for configurations
preserving $ISO(2,1)$ symmetry to have a local energy density that is a total spatial derivative, it was shown, in the notation of \cite{Arav:2020obl} 
that
\begin{align}\label{susyscheme}
\delta_{4(1)}
=-\frac{1}{4}+2\delta_{\beta},\quad
\delta_{4(3)}=\frac{3}{4}+2\delta_{\beta},\quad
\delta_{\partial\phi^2(1)}&=2\delta_{\alpha}\,.
\end{align}
The finite counter terms combined with this condition are consistent with a renormalisation scheme preserving $\mathcal{N}=1$ supersymmetry of the $d=4$ boundary theory.

It is of interest to determine a scheme that is consistent with the full $\mathcal{N}=4$ supersymmetry of the boundary theory. While the full analysis is left for future work, here we make a simple observation that further constrains the $\delta$-coefficients. The finite counterterms considered above are invariant under the discrete symmetries \eqref{genz2}-\eqref{gens33} which preserve the superpotential and hence
preserve the supercharge associated with the $\mathcal{N}=1$ supersymmetry that is being considered in the BPS equations. One can check 
that the action is also invariant under the additional discrete symmetries
\begin{align}\label{essfour}
z^1\leftrightarrow \bar z^4, \,\,\,\,z^2\leftrightarrow z^3;\,\,\,\, &\Leftrightarrow\,\,\,\,
\phi_1\leftrightarrow \phi_4,\,\,\,\, \phi_2\leftrightarrow \phi_3,\,\,\,\, (\alpha_2,\alpha_3)\to -(\alpha_2,\alpha_3)\,,\nn
z^1\leftrightarrow -\bar z^2, \,\,\,\, z^3\leftrightarrow -z^4;\,\,\,\, &\Leftrightarrow\,\,\,\,
\phi_2\leftrightarrow \phi_4,\,\,\,\, \phi_1\leftrightarrow \phi_3,\,\,\,\, (\alpha_1,\alpha_3)\to -(\alpha_1,\alpha_3)\,,\nn
z^1\leftrightarrow -\bar z^3, \,\,\,\, z^2\leftrightarrow -z^4;\,\,\,\, &\Leftrightarrow\,\,\,\,
\phi_3\leftrightarrow \phi_4,\,\,\,\, \phi_1\leftrightarrow \phi_2,\,\,\,\, (\alpha_1,\alpha_2)\to -(\alpha_1,\alpha_2)\,.
\end{align}
These symmetries do not preserve the BPS equations but instead transform the supercharges into each other; they are the
generalisation of (3.3) of \cite{Khavaev:2000gb} to include the $\alpha_i$ scalars.
For a scheme preserving 
$\mathcal{N}=4$ supersymmetry we should therefore impose that the finite counterterms are also invariant under these discrete symmetries
and hence we should also impose in (B.8) of \cite{Arav:2020obl}:
 \begin{align}\label{newschemeconds}
\delta_{4(2)}=\delta_{4(1)},\qquad
\delta_{4(4)}=\delta_{4(3)},\qquad
\delta_{\partial\phi^2(2)}=\delta_{\partial\phi^2(1)},\qquad
\delta_{R\phi^2(2)}=\delta_{R\phi^2(1)}\,.
\end{align}

To illustrate the impact of these conditions, we now consider the equal mass, $\mathcal {N}=1^*$ $SO(3)$ invariant truncation which depends on four scalars $\phi_1=\phi_2=\phi_3$, $\phi_4$,
$\alpha_1=\alpha_2=\alpha_3$ and $\varphi$, with $\beta_1=\beta_2=0$.
With a $D=5$ metric of the form
\begin{align}\label{iso21metansapp}
ds^2 &= e^{2A(\bar r,x)}(dt^2 - dy_1^2 - dy_2^2) - e^{2V(\bar r,x)} dx^2 - d\bar r^2\,,
\end{align}
with all scalar fields functions of $(\bar r,x)$ only, we can use the schematic expansion as we approach the boundary at $\bar r\to\infty$: 
\begin{align}\label{asexp10sceqmass}
A&=\frac{\bar r}{L}+\Omega+\cdots+{A}_{(v)}e^{-4\bar r/L}+\cdots\,,\nn
V&=\frac{\bar r}{L}+\Omega+f+\cdots+{V}_{(v)}e^{-4\bar r/L}+\cdots\,,\nn
\phi_2=\phi_3=\phi_1&=\phi_{1,(s)}e^{-\bar r/L}+\dots+{\phi}_{1,(v)}e^{-3\bar r/L}+\cdots\,,\nn
\phi_4&=\phi_{4,(s)}e^{-\bar r/L}+\dots+{\phi}_{4,(v)}e^{-3\bar r/L}+\cdots\,,\nn
\alpha_2=\alpha_3=\alpha_1&=\alpha_{1,(s)}\frac{\bar r}{L}e^{-2\bar r/L}+{\alpha}_{1,(v)}e^{-2\bar r/L}+\cdots\,,\nn
\varphi&=\varphi_{(s)}+\cdots+{\varphi}_{(v)}e^{-4\bar r/L}+\cdots\,,
\end{align}
where $\phi_{1,(s)}$, $\phi_{4,(s)}$, $\alpha_{1,(s)}$, $\varphi_{(s)}$ determine the source terms for the scalar operators in \eqref{opfieldmapz}
and, as in \cite{Arav:2020obl}, we focus on $\varphi_{(s)}=0$.
Now using \eqref{susyscheme} and \eqref{newschemeconds} in (B.37) of \cite{Arav:2020obl} we obtain\footnote{Note that we have corrected a typo in 
(B.37) of  \cite{Arav:2020obl}: the sign of the third term on the right hand side of $\langle\mathcal{O}_{\phi_4}\rangle$ should be + and not -.}
\begin{align}\label{equalmassalphvevnfour}
\langle\mathcal{O}_{\alpha_1}\rangle
&=\langle\mathcal{O}_{\alpha_2}\rangle=\langle\mathcal{O}_{\alpha_3}\rangle=\frac{1}{4\pi GL}\Big({\alpha}_{1,(v)}-2\delta_{\alpha}\alpha_{1,(s)}\Big)\,,\nn
\langle\mathcal{O}_{\phi_1}\rangle
&=\langle\mathcal{O}_{\phi_2}\rangle=\langle\mathcal{O}_{\phi_3}\rangle
=\frac{1}{2\pi GL}\Big(\phi_{1,(v)}+\frac{5}{6}\phi_{1,(s)}^3-\frac{9-2\delta_{4(5)}}{3}\phi_{1,(s)}^2\phi_{4,(s)}
\nn&\qquad-\frac{1+16\delta_{\beta}}{6}\phi_{1,(s)}\phi_{4,(s)}^2+\frac{L^2}{4}(1+4\delta_{\alpha})\Box\phi_{1,(s)}-\frac{L^2}{24}(1+2\delta_{R\phi^2(1)})R\phi_{1,(s)}\Big)\,,\nn
\langle\mathcal{O}_{\phi_4}\rangle&=\frac{1}{2\pi GL}\Big(\phi_{4,(v)}-\frac{9-2\delta_{4(5)}}{3}\phi_{1,(s)}^3-\frac{1+16\delta_{\beta}}{2}\phi_{1,(s)}^2\phi_{4,(s)}+\frac{7+32\delta_{\beta}}{6}\phi_{4,(s)}^3\nn
&\phantom{=\frac{1}{2\pi GL}\Big(}+\frac{L^2}{4}(1+4\delta_{\alpha})\Box\phi_{4,(s)}-\frac{L^2}{24}(1+2\delta_{R\phi^2(1)})R\phi_{4,(s)}\Big)\,,\nn
\langle\mathcal{O}_{\varphi}\rangle&=\frac{1}{\pi GL}\Big({\varphi}_{(v)}-\frac{3}{4}(\alpha_{1,(s)}-4{\alpha}_{1,(v)})(\phi_{1,(s)}^2-\phi_{1,(s)}\phi_{4,(s)})\Big)\,,
\end{align}

We now consider the further truncation to the $SO(3)\times SO(3)$ invariant truncation which depends on three scalars $\phi_1=\phi_2=\phi_3=-\phi_4$, $\alpha_1=\alpha_2=\alpha_3$ and $\varphi$. Correspondingly we should set $\phi_{4,(s)}=-\phi_{1,(s)},~ \phi_{4(v)}=-\phi_{1(v)}$ and then 
\eqref{equalmassalphvevnfour} becomes
\begin{align}\label{equalmassalphvevsimp}
\langle\mathcal{O}_{\alpha_1}\rangle
&=\langle\mathcal{O}_{\alpha_2}\rangle=\langle\mathcal{O}_{\alpha_3}\rangle=\frac{1}{4\pi GL}\Big({\alpha}_{1,(v)}-2\delta_{\alpha}\alpha_{1,(s)}\Big)\,,\nn
\langle\mathcal{O}_{\phi_1}\rangle
&=\langle\mathcal{O}_{\phi_2}\rangle=\langle\mathcal{O}_{\phi_3}\rangle
=-\langle\mathcal{O}_{\phi_4}\rangle=\frac{1}{2\pi GL}\Big(\phi_{1,(v)}+\frac{11-2\delta_{4(5)}-8\delta_{\beta}}{3}\phi_{1,(s)}^3
\nn&\qquad\qquad\qquad\qquad+\frac{L^2}{4}(1+4\delta_{\alpha})\Box\phi_{1,(s)}-\frac{L^2}{24}(1+2\delta_{R\phi^2(1)})R\phi_{1,(s)}\Big)\,,\nn
\langle\mathcal{O}_{\varphi}\rangle&=\frac{1}{\pi GL}\Big({\varphi}_{(v)}-\frac{3}{2}(\alpha_{1,(s)}-4{\alpha}_{1,(v)})\phi_{1,(s)}^2\Big)\,.
\end{align}
We emphasise that if we had not imposed \eqref{newschemeconds}, then we would not have obtained 
the equality $\langle\mathcal{O}_{\phi_1}\rangle
=\langle\mathcal{O}_{\phi_2}\rangle=\langle\mathcal{O}_{\phi_3}\rangle
=-\langle\mathcal{O}_{\phi_4}\rangle$.

This refinement of the RG scheme does not play a direct role for the one point functions in section \ref{nfourstuff}
since the source terms for the scalars all vanish  (after a possible shift of the dilaton). However, it does impact upon other observables.
We also note here that to get  the expectation values as given in \eqref{vevsonesided}, one should follow the discussion in section C.3 of \cite{Arav:2020obl}.


\providecommand{\href}[2]{#2}\begingroup\raggedright\endgroup

\end{document}